\newcommand{\lenstronomy}{\texttt{Lenstronomy}}

\documentclass[twocolumn, tighten]{openjournal}
\usepackage{url}
\usepackage{xcolor}
\definecolor{xlinkcolor}{cmyk}{1,1,0,0}
\usepackage[
 bookmarks=true, 
 pdfnewwindow=true,      
 colorlinks=true,    
 linkcolor=xlinkcolor,     
 citecolor=xlinkcolor,     
 filecolor=xlinkcolor,  
 urlcolor=xlinkcolor,      
 final=true,
]{hyperref}

\graphicspath{{./}{figures/}}

\makeatletter
\renewcommand\@makecaption[2]{%
  \par
  \vskip\abovecaptionskip
  \begingroup
    \footnotesize\rmfamily
    \begingroup
      \samepage
      \flushing
      \let\footnote\@footnotemark@gobble
      \ifnum\pdfstrcmp{\@captype}{table}=0
        \@make@capt@title{\textsc{Table \thetable}}{#2}%
      \else
        \ifnum\pdfstrcmp{\@captype}{figure}=0
          \@make@capt@title{\textsc{Figure \thefigure}}{#2}%
        \else
          \@make@capt@title{#1}{#2}%
        \fi
      \fi\par
    \endgroup
  \endgroup
  \vskip\belowcaptionskip
}
\makeatother

\usepackage{etoolbox}
\AtBeginEnvironment{tabular}{\renewcommand{\arraystretch}{1.5}}
\AtBeginEnvironment{tabularx}{\renewcommand{\arraystretch}{1.5}}
\usepackage{amsmath}
\usepackage{orcidlink}

\newcommand{\Hzero}{\ensuremath{\mathrm{H_0}}}

\shorttitle{Modeling of Doubly Imaged Quasars}
\shortauthors{Brady et al.}

\begin{document}

\title{\textbf{TDCOSMO XXVI: Uniform Lens Modeling of Eight Doubly Imaged Quasars}}

\author{
    Ryan Brady$^1$\orcidlink{0009-0003-1338-6336}
}

\author{Xiang-Yu Huang$^1$\orcidlink{0000-0001-7113-0599}
}

\author{
Simon Birrer$^1$\orcidlink{0000-0003-3195-5507}
}

\author{
Anowar J. Shajib$^{2, 3, 4}$\orcidlink{0000-0002-5558-888X}
}

\author{Nafis Sadik Nihal$^4$\orcidlink{0009-0009-6956-0675}
}

\author{Cameron Lemon$^5$\orcidlink{0000-0003-2456-9317}
}

\author{ Martin Millon$^{6, 7}$\orcidlink{0000-0001-7051-497X}
}

\author{Veronica Motta$^{8}$\orcidlink{0000-0003-4446-7465}
}

\author{Dominique Sluse$^9$\orcidlink{0000-0001-6116-2095}
}

\author{Frederic Courbin$^{10, 11, 12}$\orcidlink{0000-0003-0758-6510}
}

\affiliation{$^1$Department of Physics and Astronomy, Stony Brook University, Stony Brook, NY 11794-3800, USA}

\affiliation{$^2$Department of Astronomy \& Astrophysics, University of Chicago, Chicago, IL 60637, USA}

\affiliation{$^3$Kavli Institute for Cosmological Physics, University of Chicago, Chicago, IL 60637, USA}

\affiliation{$^4$Center for Astronomy, Space Science and Astrophysics, Independent University, Bangladesh, Dhaka 1229, Bangladesh}

\affiliation{$^5$Department of Physics, Oskar Klein Centre, Stockholm University, SE-106 91, Stockholm, Sweden}

\affiliation{$^6$Institute for Particle Physics and Astrophysics, ETH Zurich,
Wolfgang-Pauli-Strasse 27, CH-8093 Zurich, Switzerland}

\affiliation{$^7$D\'epartement de Physique Th\'eorique, Universit\'e de Gen\`eve, 24 quai Ernest-Ansermet, CH-1211 Gen\`eve 4, Switzerland}

\affiliation{$^8$Instituto de F\'{\i}sica y Astronom\'{\i}a, Universidad de Valpara\'{\i}so, Avda. Gran Breta\~na 1111, Valpara\'{\i}so, Chile}

\affiliation{$^9$STAR Institute, University of Li{\`e}ge, Quartier Agora, All\'ee du six Ao\^ut 19c, 4000 Li\`ege, Belgium}

\affiliation{$^{10}$ICC-UB Institut de Ciències del Cosmos, Universitat de Barcelona, Martí Franquès, 1, 08028 Barcelona, Spain}

\affiliation{$^{11}$Institut Català de Recerca i Estudis Avançats (ICREA), Pg. Lluís Companys 23, 08010 Barcelona, Spain}

\affiliation{$^{12}$Institut dEstudis Espacials de Catalunya (IEEC), Edifici RDIT, Campus UPC, Castelldefels,
08860 Barcelona, Spain}

\begin{abstract}
We present the first uniform gravitational lens modeling analysis of eight doubly imaged quasars from multi-band observations with the Hubble Space Telescope. Previous time-delay cosmography analyses by the TDCOSMO Collaboration have primarily relied on quadruply imaged quasars, while doubly imaged systems, despite being more abundant, remain underutilized due to their fewer geometric constraints. Using an open-source \lenstronomy~framework, we reconstruct the lensing systems with a pipeline tailored for doubles. Comparing our results to the literature, the modeled Einstein radii agree at an average of $1.5\sigma$, which is expected given data and modeling heterogeneity, while modeled image separations differ from Gaia DR2 measurements with an r.m.s of only 3.6 mas. We find a strong correlation between Fermat potential precision and the surface brightness of the spatially extended host arcs, establishing that arc surface brightness is the primary driver of mass model precision in doubly imaged systems. To further quantify the information contributed by the lensed arcs, we performed a conjugate point analysis that uses only the quasar image positions to constrain the lens mass profiles. The resulting posteriors are substantially broader than those from full image modeling, and a strong anti-correlation between mass parameter hypervolume and arc magnitude additionally confirms that arc brightness determines the degree to which the lens mass profile can be constrained in doubles. A hierarchical cosmographic analysis incorporating time-delay measurements and stellar kinematics to infer $\Hzero$ will be presented in a subsequent publication. The uniform pipeline and arc surface brightness trends established here will significantly accelerate the construction of time-delay cosmography samples from the large lens populations expected from LSST, Roman, and Euclid.


\end{abstract}

\section{\textbf{Introduction}}
\label{sec:intro}

The $\Lambda$ Cold Dark Matter ($\Lambda$CDM) cosmological model serves as the prevailing framework to describe the Universe on the largest scales. Within this paradigm, the Universe is composed predominantly of cold dark matter and dark energy, the latter manifesting as a cosmological constant ($\Lambda$) responsible for its accelerated expansion \citep{weinberg:2013}. A cornerstone of this model is the Hubble constant (\Hzero), which quantifies the present-day rate of said expansion. Precise determination of \Hzero~is therefore essential not only for anchoring the $\Lambda$CDM model but also for testing its internal consistency across independent cosmological probes.

Despite the successes of $\Lambda$CDM in explaining a wide range of cosmological observations, a persistent and statistically significant discrepancy has appeared in recent years between measurements of $\Hzero$ derived from observations of the early Universe and those from the local Universe. Measurements of the cosmic microwave background (CMB) anisotropies by the \textit{Planck} satellite, when interpreted within the $\Lambda$CDM framework, yield a value of $\Hzero = 67.4 \pm 0.5$ km s$^{-1}$ Mpc$^{-1}$ \citep{planck:2018}. In contrast, direct distance ladder measurements, such as those conducted by the SH0ES collaboration using Cepheid-calibrated Type Ia supernovae, report a significantly higher value of $\Hzero = 73.04 \pm 1.04$ km s$^{-1}$ Mpc$^{-1}$ \citep{riess:2022}. Comprehensive reviews (e.g., \citealp{diValentino:2021, hu:2023, perivolaropoulos:2024, divalentino:2025}) detail how this discrepancy, commonly known as the ``Hubble tension," now exceeds the $5\sigma$ level and may point either to unrecognized systematics within one or more measurement techniques or to new physics beyond the standard $\Lambda$CDM framework.

Time-delay cosmography (TDC) of strongly lensed quasars has emerged as an independent and complementary method for determining $\Hzero$ that circumvents many of the systematics inherent in other techniques \citep{treu:2022, birrer:2024}. First proposed by \citet{refsdal:1964}, TDC relies on the principle that multiple images of a background source, lensed by a foreground galaxy, arrive at the observer at different times due to differences in both the geometric path length and the gravitational potential traversed by the light rays. The measured time delays between these images, when combined with the corresponding Fermat potential differences between images (i.e., the effective arrival-time surface differences as inferred from lens modeling), yield an estimate of the so-called ``time-delay distance'' ($\mathrm{D_{\Delta t})}$.  This quantity encapsulates a combination of angular diameter distances between the observer, lens, and source, and is inversely proportional to the Hubble constant. As such, precise measurements of $\mathrm{D_{\Delta t}}$ translate directly into constraints on $\Hzero$.

However, the inference of $\mathrm{D_{\Delta t}}$ from lens modeling is subject to the mass-sheet transform (MST) and subsequent mass-sheet degeneracy \citep{falco:1985}, a degeneracy in which the surface mass density can be rescaled by a factor $\lambda$ while preserving the lensing observables, such as the image positions and flux ratios, and altering the time delays. Under this transformation, the inferred time-delay distance scales as $\mathrm{D_{\Delta t}} \propto 1/\lambda$, leading to a direct degeneracy with $\Hzero$. This effect can be decomposed into contributions from mass along the line of sight, $\mathrm{\kappa_{ext}}$, and an internal component associated with the lens galaxy itself, parameterized by $\lambda_{\rm int}$. The latter reflects uncertainties in the radial mass profile of the lens and cannot be broken by lensing data alone, necessitating additional constraints from stellar kinematics or well-informed priors to infer $\Hzero$ \citep{teodori:2022, birrer:2024, tdcosmo:xix}. As this work focuses not on cosmological results but on a scalable modeling framework, we do not consider sampling $\lambda_\text{int}$. For future analyses aimed at constraining $\Hzero$, this radial mass slope uncertainty will be marginalized over. For a further discussion of lensing degeneracies, see, e.g., \citet{saha:2000, saha:2006}.

The H0LiCOW collaboration has applied the TDC framework to a sample of five well-characterized quadruply imaged quasar systems and one doubly imaged system (henceforth denoted quads and doubles, respectively), reporting values of $\Hzero$ consistent with the local distance ladder measurements, finding $\Hzero = 73.3^{+1.7}_{-1.8}$ km s$^{-1}$ Mpc$^{-1}$ \citep{wong:2020}. To improve upon the precision of TDC and to further probe systematics, the TDCOSMO collaboration was formed \citep{tdcosmo:i}. The most recent TDCOSMO 2025 analysis \citep{TDCOSMO:2025} represents a significant advance in this program through hierarchical Bayesian analyses. In this approach, individual lens systems are allowed to share statistical information through population-level hyperparameters while retaining system-specific freedom, thereby reducing sensitivity to modeling assumptions for any one lens. \citet{TDCOSMO:2025} incorporated high quality stellar kinematics data from JWST NIRSpec, VLT MUSE, and Keck KCWI observations, along with the aforementioned enhanced modeling techniques, to probe $\Hzero$ using a slightly expanded sample of seven quads and one double. This work reports $\Hzero = 71.6^{+3.9}_{-3.3}$ km s$^{-1}$ Mpc$^{-1}$ for their time-delay sample combined with Pantheon+ supernova \citep{pantheon} constraints in flat $\Lambda$CDM, or $\Hzero = 74.3^{+3.1}_{-3.7}$ km s$^{-1}$ Mpc$^{-1}$ when including improved external lens samples from SLACS and SL2S with spectroscopy (\citealt{bolton:2006, more:2012}, respectively). These measurements remain consistent with local distance ladder determinations while achieving improved precision through better data and Bayesian methodology. These results also show that the observed tension in $\Hzero$ persists across independent methods, and emphasizes the importance of TDC as an arbitrator in the broader effort to resolve this discrepancy.

While quads provide richer constraints per system, there exists an approximately 4-to-1 ratio of doubles vs. quads in the Universe \citep{luhtaru:2021}. In addition to being far more common, doubles tend to exhibit simpler lens geometries, making them especially attractive for statistical studies \citep{oguri:2010}. Though they offer fewer lensing observables, their typically longer time delays can yield competitive precision in time delay measurements \citep{liao:2017}. The potential of doubles for $\Hzero$ measurements has been demonstrated previously in individual systems. Most notably, \citet{Birrer:2019} measured $\Hzero=72.5^{+2.1}_{-2.3}$ km s$^{-1}$ Mpc$^{-1}$ via an analysis of the doubly imaged quasar SDSS 1206+4332; while the quasar itself is doubly imaged, the system also exhibits a quadruply imaged extended host galaxy that provides additional lensing constraints. In that work, the internal MST was also assumed to be known as $\lambda_\text{int} = 1$. This result was obtained using detailed lens modeling combined with stellar kinematics and line-of-sight corrections, demonstrating that doubles can achieve tight constraints on $\Hzero$ when high quality imaging and spectroscopic data are available. More recently, \citet{paic:2025} presented the first full TDCOSMO analysis of a doubly imaged system, HE 1104-1805, combining 17 years of photometric monitoring, high quality MUSE integral field spectroscopy, and Hubble Space Telescope (HST) imaging. This analysis yielded $\Hzero = 64.2^{+5.8}_{-5.0}$ km s$^{-1}$ Mpc$^{-1} \times \lambda_\text{int}$ while achieving 8.5\% precision. These recent successes prove that with modern observational capabilities, improved modeling frameworks, and the hierarchical Bayesian approach, doubles can provide cosmological constraints with precision complementing that of quads.

\begin{figure*}[t]
    \centering
    \includegraphics[width=1\textwidth]{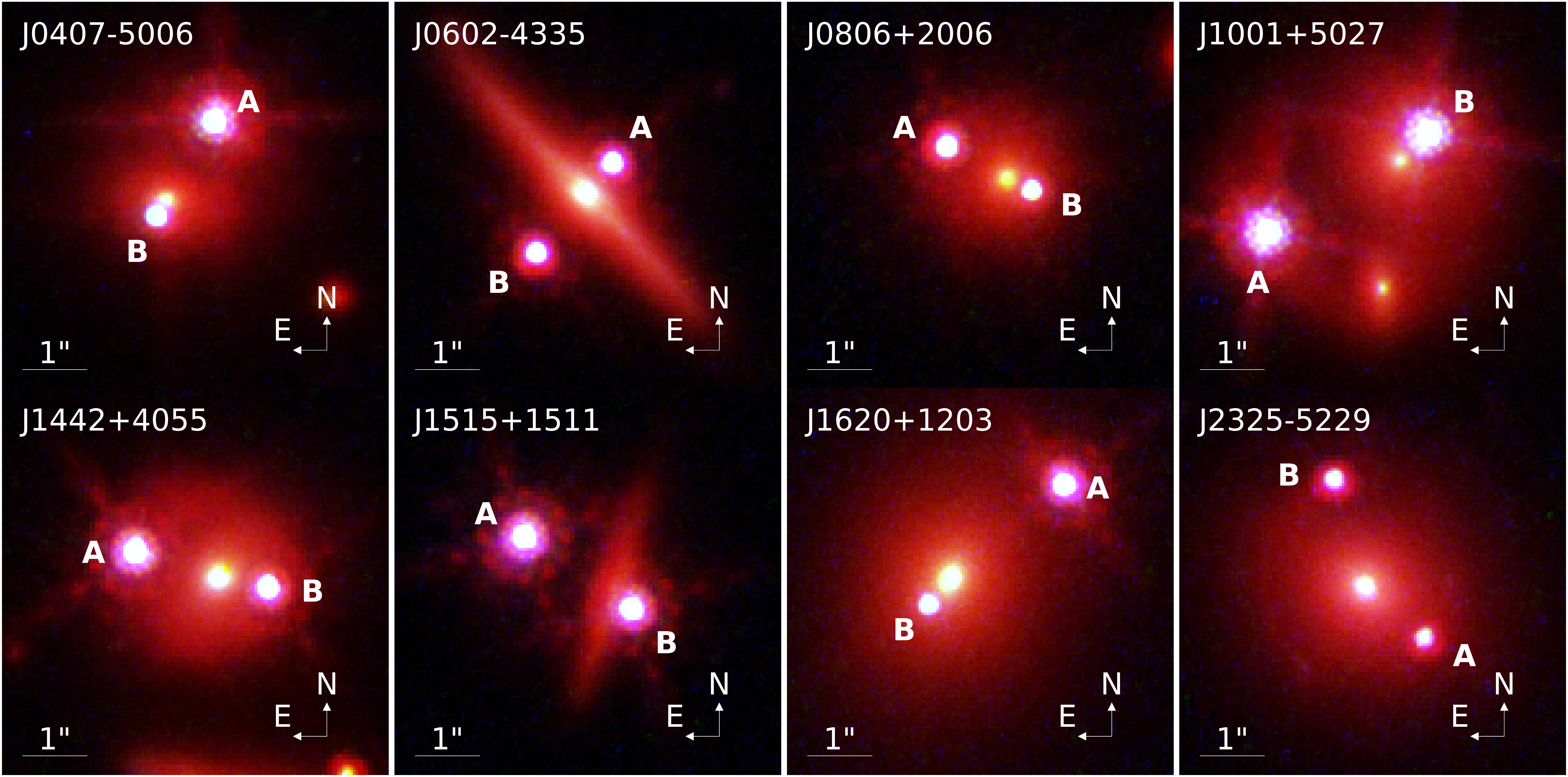}
    \caption{Composite red-green-blue (RGB) images of the eight doubly imaged quasar systems observed during program HST GO-17199 (PI: Lemon). Each figure presents an image constructed from the HST observations, with F160W data mapped to the red channel, F814W data to the green channel, and F475X data to the blue channel. To enhance visualization, the intensity scaling of each band is adjusted individually. Image labels are assigned in accordance with the respective previous time-delay literature.}
    \label{fig:rgbs}
\end{figure*}

While the ultimate goal of TDC is a precise and accurate determination of $\Hzero$, the cornerstone of this technique lies in the construction of reliable gravitational lens models to deduce the Fermat potential differences between images. In this work, we take a first step toward establishing a standardized, scalable modeling foundation needed for cosmography with doubles. With a sample of eight doubles observed by the Hubble Space Telescope, we utilize the \lenstronomy~software \citep{birrer:2018, lenstronomy_two} to develop a uniform modeling pipeline to extract the key lensing parameters, such as the Einstein radii and lensed image positions, that will ultimately be used in conjunction with measured time delays to infer $\Hzero$. This paper therefore represents the first stage of a hierarchical approach, where we isolate the lens modeling to ensure the accuracy and internal consistency of the lens mass parameters before incorporating the time-delay measurements and propagating uncertainties into cosmographic inference. This strategy mirrors the approach of recent modeling efforts such as STRIDES \citep{schmidt:2023}, which emphasized automation and uniformity to handle large lens samples efficiently. With the advent of large-scale surveys yielding thousands of lens candidates, a dedicated modeling framework for doubles is essential for achieving percent-level precision on $\Hzero$.

In Section \ref{sec:hst_sampls}, we present the discoveries and relevant cosmological information for the eight doubles in the study. Section \ref{sec:modeling_procedure} details the model components and the \lenstronomy~modeling pipeline. The best-fit mass and light profiles of all eight systems are presented in Section \ref{sec:results}, alongside modeled astrometry and photometry. To complement the full image modeling, we performed a conjugate point analysis for each system. In Section \ref{sec:conjugate}, we compare the conjugate point inferences to the full image modeling results and quantify the information gain provided by modeling the full surface brightness distribution of the lensed arcs. We summarize our findings and discuss future hierarchical outlooks in Section \ref{sec:discussion}.

\section{\textbf{HST Samples}}
\label{sec:hst_sampls}

As part of HST program GO-17199 (PI: Lemon), each lens was observed using three filters with the Wide Field Camera 3 (WFC3): F160W for infrared (IR) data, and F814W and F475X for ultraviolet-visible (UVIS) data. These systems were chosen due to the existence of previous time delay measurements from long-term monitoring campaigns. Throughout this work, image labels are assigned as defined by their respective labeling from the previous time-delay literature. To enhance data sampling and to avoid saturating the quasar images, a 4-point dither pattern was utilized for IR observations, while a 2-point dither pattern was applied for UVIS exposures, with both long and short exposures taken at each dither position. For data reduction, alignment, and combination of exposures in each filter, we utilized the Python package \texttt{AstroDrizzle} \citep{avila:2015}. The final reduced images have a pixel scale of $0\farcs08$/pixel for IR exposures, \textcolor{black}{which was chosen to improve sampling relative to the native detector pixel scale of 0\farcs13 while avoiding unnecessary oversampling and the associated increase in correlated noise. The UVIS exposures remained at $0\farcs04$/pixel to avoid introducing additional correlated noise through further resampling.}  RGB composites of the eight systems of study are shown in Figure \ref{fig:rgbs}. Further observational information is presented in Appendix \ref{append::c}.

\subsection{Notes on Individual Doubles}
\label{subsec:notes_on_doubles}
This section provides a brief description of the eight doubles studied in this work. 
\subsubsection{J0407-5006}
\label{subsubsec:J0407}

This lens was identified from the STRIDES 2016 followup campaign \citep{treu:2018} by \citet{anguita:2018}. Spectroscopic observations were obtained using the New Technology Telescope (NTT) equipped with EFOSC2, establishing a source redshift of $\mathrm{z_s} = 1.515$, while the redshift of the lensing galaxy remains unmeasured. \citet{tdcosmo:ii} determined the time delay to be $\rm \Delta t_{AB}=-128.4^{+3.5}_{-3.8}$ days from monitoring data acquired at the Max Planck Institute for Astrophysics
 2.2 m telescope at La Silla observatory.

\subsubsection{J0602-4335} \label{subsubsec:J0602}

J0602-4335 was discovered by \citet{dawes:2023} in the Dark Energy Spectroscopic Instrument (DESI) Legacy Imaging Surveys \citep{desi_legacy_imaging}. With archival EFOSC2 spectroscopic data, \citet{dux:2025} determined the source redshift to be $\mathrm{z_s}=2.92$, with the lens redshift not yet measured. \citet{dux:2025} also report a time delay of $\rm \Delta t_{AB} = 23.6\pm 2.1$ days from $r$-band monitoring
with the Max-Planck-Gesellschaft telescope.

\subsubsection{J0806+2006} \label{subsubsec:J0806}

This system was identified from the Sloan Digital Sky Survey (SDSS) spectroscopic quasar sample \citep{richards:2002} by \citet{inada:2006}. Followup imaging in optical and near-infrared bands using the University of Hawaii 88-inch telescope and the Keck I telescope determined that the lens galaxy lies at a redshift of $\mathrm{\mathrm{\mathrm{z_l}}} = 0.573$. Spectroscopic observations with Keck II confirmed a source of $\mathrm{z_s} = 1.540$. \citet{bekov:2024}, via long-term monitoring at the Maidanak Observatory, measured a time delay of $\rm \Delta t_{AB} = -53.0 \pm 6.0$ days.

\subsubsection{J1001+5027} \label{subsubsec:J1001}
This system was discovered by \citet{oguri:2005} through spectroscopic analysis conducted with the Double Imaging Spectrograph at the Astrophysical Research Consortium 3.5m telescope at Apache Point Observatory. Followup spectroscopy using Gemini/GMOS by \citet{inada:2012} confirmed a lens redshift of $\mathrm{\mathrm{\mathrm{z_l}}}=0.415$ and report a host quasar redshift of $\mathrm{\mathrm{z_s}}=1.841$ from archival SDSS data. Our HST data suggests that \textcolor{black}{a} perturbing galaxy lays $\sim$$2\farcs0$ away from the main lens, with the redshift of said perturber unknown. \citet{kumar:2013} found a time delay of $\rm \Delta t_{AB} = -119.3 \pm 3.3$ days from long-term observations conducted with the Mercator Telescope, the Maidanak Observatory, and the Himalayan Chandra Telescope.

\subsubsection{J1442+4055} \label{subsubsec:J1442}

J1442+4055 was discovered as a double in the SDSS-III/BOSS Quasar Lens Survey by \citet{more:2016}, who reported that the quasar source lies at $\rm z_s = 2.575$ as a result of the SDSS pipeline \citep{paris:2014}. Followup observations by \citet{shalyapin:2019}, using Gran Telescopio Canarias (GTC) and Liverpool Telescope (LT) spectroscopic observations, further confirmed the lens redshift to be $\mathrm{z_l} = 0.284$, while also quantifying the redshift of a nearby ($<5\farcs0$) galaxy to be $\rm z_{G2}=0.323 \pm0.051$. The aforementioned work also determined the time delay to be $\rm \Delta t_{AB} = -25.0 \pm 1.5$ days from the monitoring conducted by the LT.

\subsubsection{J1515+1511} \label{subsubsec:J1515}

This system was discovered as a double by \citet{inada:2014} through the SDSS Quasar Lens Search (SQLS). Followup imaging and spectroscopy with telescopes including the TNG and Gemini North found a lens redshift of $\mathrm{z_l} = 0.742$. Subsequent analyses by \citet{shalyapin:2017} report a source redshift of $\mathrm{z_s} = 2.049$ based upon updated SDSS/BOSS data, while also employing a 2.6 year $r$-band monitoring campaign using the LT and spectroscopy with the GTC to measure a secondary galaxy $\sim$$16\farcs0$ away at $\mathrm{z_\text{G2}} = 0.541$, which may contribute external shear to the system.  Combining the aforementioned photometric data with three seasons of monitoring taken with the ECAM instrument at the \textit{Euler} Swiss Telescope, \citet{cosmograil:xix} found the time delay to be \textcolor{black}{$\rm \Delta t_{AB} = -210.5^{+5.5}_{-5.7}$ days.}

\subsubsection{J1620+1203} \label{subsubsec:J1620}

J1620+1203 was identified as a double as part of the SDSS SQLS by \citet{kayo:2010}. Followup imaging using the Faint Object Camera and Spectrograph (FOCAS) on the Subaru Telescope determined that the source quasar has a redshift of $\mathrm{z_s} = 1.158$, while the lensing galaxy redshift was determined as $\mathrm{\mathrm{\mathrm{z_l}}} = 0.398$. From long-term photometric monitoring at the Leonhard Euler 1.2 m Swiss Telescope, \citet{cosmograil:xix} determined the time delay to be $\rm \Delta t_{AB} = -171.5 \pm 8.7$ days.

\subsubsection{J2325-5229} \label{subsubsec:J2325}

J2325-5229 was discovered as a double in the VISTA Hemisphere Survey (VHS) and the Dark Energy Survey (DES) using a morphology-independent machine learning technique \citep{ostrovski:2017}. The lens redshift was determined to be $\mathrm{\mathrm{\mathrm{z_l}}} = 0.400$. Spectroscopic followup observations conducted using NTT with EFOSC2, as well as archival data from the Anglo-Australian Telescope (AAT), confirmed the quasar source redshift to be $\mathrm{z_s} = 2.739$. \citet{tdcosmo:ii} determined the time delay to be $\rm \Delta t_{AB}=43.8^{+4.5}_{-4.0}$ days from long-term monitoring conducted at La Silla observatory.

\section{\textbf{Lens Modeling Procedure}}
\label{sec:modeling_procedure}

\textcolor{black}{Throughout this work, we employ a uniform framework in which all systems are modeled using the same lens mass parameterization, point spread function (PSF) treatment, optimization strategy, and inference procedure. Ancillary model components, such as additional S\'ersic profiles for the lens light, treatment of perturbers, or additional host galaxy complexity are introduced only when the baseline model leaves clear, coherent residual structures in the imaging data that indicate the adopted model is inadequate. Across the eight systems in the dataset, only three required minor modifications to the baseline framework. Specifically, J0602-4335 required two lens light components in all three data bands to model its edge-on morphology, J1001+5027 included an additional mass and light component for a nearby perturber, and J2325-5229 required additional complexity in the F160W source model to reproduce its extended arcs.}

All gravitational lens modeling was conducted with the Python-based package \texttt{Lenstronomy}. The entire pipeline, including mask construction, PSF handling, fitting notebooks, and subsequent analyses, can be found at \href{https://github.com/brady-ryan/doubles_modeling/tree/main}{https://github.com/brady-ryan/doubles\_modeling/tree/main}. A summary of the model components is presented in Table \ref{tab:priors}.

\begin{figure}[t]
    \centering
    \includegraphics[width=.5\textwidth]{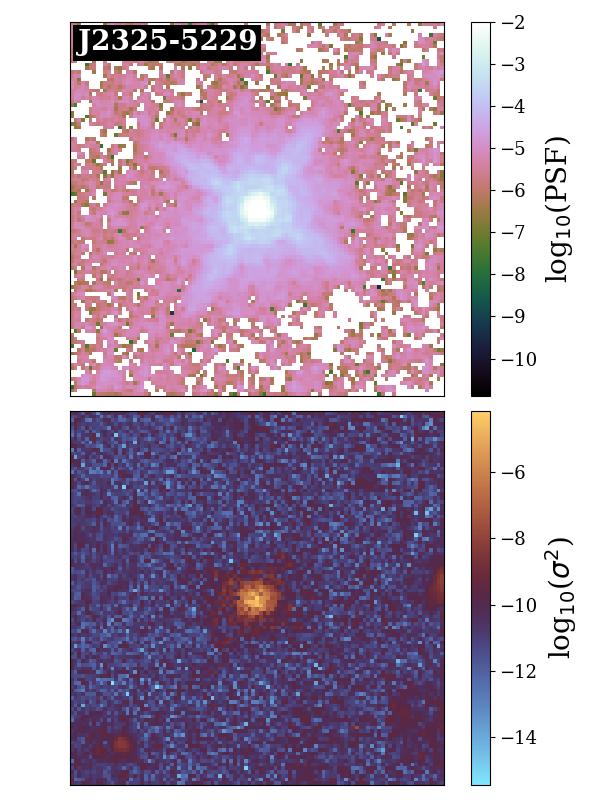}
    \caption{Initial Point Spread Function (PSF) visualization for J2325-5229 in the F160W band. Precise inference of astrometry and photometry requires accurate PSF modeling and a realistic quantification of PSF uncertainties. \textbf{Top.} Initial IR PSF as inferred with STARRED from isolated stars in the science images. All PSFs are supersampled by a factor of three, allowing the PSF to be evaluated on a finer grid and better capture the small-scale structure of the point sources. \textbf{Bottom.} Per-pixel PSF variance map calculated with \texttt{psfr}, used to appropriately weight the likelihood in the IR modeling.}
    \label{fig:psfs}
\end{figure}

\begin{table*}[t]
\label{tab:priors}
\centering
\caption{Summary of prior distributions adopted for the lens mass, lens light, and source light models used in the image modeling. The main deflector mass is modeled as an isothermal ellipsoid with broad, uninformative priors on the Einstein radius and ellipticity components, and includes an external shear term with a conservative upper bound. Lens and source surface brightness are described by elliptical S\'ersic profiles with wide priors on effective radius, S\'ersic  index, and ellipticity. For J1001+5027, which contains a nearby perturber, additional mass and light components are included with identical priors. For the edge-on system J0602–4335, two S\'ersic  components are used to model the lens light in all three data bands, with the first and second components adopting ellipticity bounds $\mathcal{U}(-0.2, 0.2)$ and $\mathcal{U}(-0.8, 0.8)$ to capture the central core and extended wings, respectively. Uniform background terms are fit independently in all three bands to account for potential inaccurate background subtraction.}
\begin{tabular*}{\textwidth}{@{\extracolsep{\fill}}cccc}\hline
Component & Parameter & Prior & Notes \\
\hline
\multicolumn{4}{c}{\textbf{Mass Components}} \\
\hline
Main Lens (SIE) & $\theta_\mathrm{E}$ [arcsec] & $\mathcal{U}(0.1,\,4.0)$ & Einstein radius. Additional SIE for J1001+5027 \\
          & $e_{\rm mass, 1}$ & $\mathcal{U}(-0.4,\,0.4)$ & Ellipticity component$^{a}$ \\
          & $e_{\rm mass, 2}$ & $\mathcal{U}(-0.4,\,0.4)$ & Ellipticity component$^{a}$ \\
\hline
External Shear & $|\gamma_{\rm ext}|$ & $\mathcal{U}(0,\,0.15)$ & Shear magnitude \\
\hline
\multicolumn{4}{c}{\textbf{Light Components}} \\
\hline
Main Lens & $R_{\rm eff}$ [arcsec] & $\mathcal{U}(0.08,\,5.0)$ & Effective radius \\
          & $n_{\text{S\'ersic}}$ & $\mathcal{U}(0.5,\,6.0)$ & S\'ersic index \\
          & $e_{\rm L, 1}$ & $\mathcal{U}(-0.5,\,0.5)$ & Ellipticity Component.$^{a}$ Modified for J0602 \\
          & $e_{\rm L, 2}$ & $\mathcal{U}(-0.5,\,0.5)$ & Ellipticity Component.$^{a}$ Modified for J0602 \\
\hline
Extended Source & $R_{\rm eff}$ [arcsec] & $\mathcal{U}(0.08,\,1.5)$ & Effective radius \\
       & $n_{\text{S\'ersic}}$ & $\mathcal{U}(0.5,\,4.0)$ & S\'ersic Index \\
       & $e_{\rm source, 1}$ & $\mathcal{U}(-0.5,\,0.5)$ & Ellipticity Component \\
       & $e_{\rm source, 2}$ & $\mathcal{U}(-0.5,\,0.5)$ & Ellipticity Component \\
       & SHAPLET & Fixed & $n_{\max}=5$ (J2325 only) \\
\hline
\end{tabular*}

\vspace{2mm}
\raggedright
$^{a}$In addition to the uniform priors on ellipticity components, we impose a Gaussian prior $\mathcal{N}(0,\,10^\circ)$ on the position angle offset between the mass and F160W lens light, i.e., $|\phi_{\rm mass} - \phi_{\rm L}|$.
\end{table*}

\subsection{Model components and priors}
\label{subsec:components}
\textit{Mass Treatment - }The results of the SLACS survey found that elliptical galaxies are well described by near-isothermal mass profiles \citep{gavazzi:2007, auger:2010}. Therefore, the main deflecting galaxies in our study were modeled as Singular Isothermal Ellipsoids (SIE). \textcolor{black}{Unlike previous modeling studies in this series which utilize a Power-Law Elliptical Mass Distribution to describe the lens mass profile (e.g. \citealt{tdcosmo:ix, tdcosmox}), we intentionally fix the slope of the mass profile to this isothermal case. The systems analyzed within this study are characterized by only two lensed images and exhibit faint, less extended source galaxy arcs, limiting the imaging constraints on the radial mass profile. As a result, allowing the slope to vary would increase parameter uncertainties without significantly improving the information recovered from the data. Since this work does not infer time-delay distances or cosmological parameters, fixing the slope to the isothermal value provides a consistent and comparable modeling framework.} We acknowledge that, for cosmographic applications, deviations from isothermality can bias time-delay distances and magnifications, however, for the purposes of this work, the SIE approximation is sufficient. Future cosmographic analyses using these systems \textcolor{black}{will allow for a free power-law slope in conjunction with additional constraints from stellar kinematics (Huang et al. - in prep)}. The convergence at image-plane position $(\theta_1,\theta_2)$ of an SIE is given by
\begin{equation}
    \kappa(\theta_1,\theta_2) = \frac{\theta_\text{E}}{2} \frac{1}{\sqrt{q_\text{mass} \theta_1^2 + \theta_2^2/q_\text{mass}}},
\end{equation}
where $\theta_\text{E}$ is the Einstein radius and $q_\text{mass}$ is the minor/major axis ratio of the ellipsoid based upon the modeled mass ellipticity components ($e_{\text{mass}, 1}$, $e_{\text{mass}, 2}$).
For J1001+5027, an additional SIE component was implemented to model the mass distribution of the perturbing galaxy.

Any additional linear distortions due to line-of-sight perturbers were modeled as an external shear, with strength
\begin{equation}
    \gamma_{\text{ext}} = \sqrt{\gamma_{\text{ext,1}}^2 + \gamma_{\text{ext,2}}^2},
\end{equation}
where $\rm \gamma_{ext, 1}$ and $\rm \gamma_{ext, 2}$ are the two Cartesian components of the external shear that describe its amplitude and orientation. Recent works have demonstrated that the external shear can also absorb contributions from mass distribution complexities that are not fully described by simple parametric profiles \citep{etherington:2024}. For galaxy-scale lenses in field and group environments, physically motivated external shear from line-of-sight structure typically ranges from 1-3\%, rarely exceeding 5\% \citep{keeton:1997, treu:2009}. To allow for potentially enhanced environmental shear in our sample while suppressing the nonphysical large values that absorb model complexity, we impose a prior of $\gamma_\text{ext}\leq 0.15$. This upper limit is approximately three times the typical field expectation, providing sufficient flexibility for systems in moderately rich environments while preventing unphysical shear amplitudes that would indicate model failure rather than genuine external perturbations.

\textit{Lens Light - }The lens galaxy light was parameterized with two components. Firstly, we applied a uniform background whose amplitude varies in all three bands to describe potential inaccurate background subtractions. Secondly, the main lens light was described via elliptical S\'ersic profiles of the form
\begin{equation}
    I(R) = I_\text{eff} \exp\left( -b_n \left[ \left( \frac{R}{R_{\rm{eff}}} \right)^{\frac{1}{n}} - 1 \right] \right),
\end{equation}
where $R = \sqrt{q_\text{L}\theta_1^2 + \theta_2^2/q_L}$, $\theta_1$ and $\theta_2$ are the angular coordinates relative to the profile center, $q_\text{L}$ is the axis ratio of the light ellipsoid, $I_\text{eff}$ is the intensity at the effective radius $R_\text{eff}$, $n$ is the Sérsic index, and $b_n$ is a constant chosen so that $R_\text{eff}$ encloses half of the total light. The ellipticity was parameterized in terms of its components $(e_{\text{L,1}},e_{\text{L,2}})$, from which the lens light position angle follows as
\begin{equation}
\label{eq:ellip}
    \phi_{\text{L}} = \tfrac{1}{2}\arctan(e_{\text{L,2}}, e_{\text{L,1}}).
\end{equation}
The adopted lens light model varies by system. \textcolor{black}{The baseline lens light model consists of a single elliptical S\'ersic profile. Additional elliptical S\'ersic components were introduced only when a single component left significant residuals in the lens galaxy light, and were retained only if they produced a clear improvement in the fit.} For J0602-4335, we use two elliptical S\'ersic components per filter to reproduce the edge-on light distribution, consisting of a compact central component and a more extended, flattened component that captures the elongated outer light profile. For J1001+5027, we use one elliptical S\'ersic profile to describe the main lens galaxy and an additional elliptical S\'ersic profile to model the light from the nearby perturber in each band. For the remaining systems, the IR models require two elliptical S\'ersic profiles to capture both the central concentration and extended components of the lens galaxy light, while a single elliptical S\'ersic component is sufficient for each UVIS model.

Previous studies have shown that mass-light misalignments are typically $\lesssim 10$ degrees in systems with low external shear (\citealt{treu:2009, Sluse:2012, Shajib:2019, schmidt:2023}). Therefore, we imposed a Gaussian prior on the alignment between the position angles of the deflector mass distribution, $\phi_{\rm mass}$, and the IR lens light, $\phi_{\rm L}$, requiring the offset $|\phi_{\rm mass} - \phi_{\rm L}|$ to follow a distribution with a standard deviation of 10 degrees, reflecting a more conservative boundary. In addition, the mass distribution and lens light centroids were required to be common across all three bands.

\textit{Extended Source Light - }For all systems, the quasar host galaxy was modeled with a single elliptical S\'ersic profile. For J2325-5229, we additionally included a shapelets component \citep{birrer:2015} for the source in the IR band to improve the fit to the system’s complex arc structure in the image plane. The shapelets basis, composed of 2D Gauss-Hermite polynomials, allows localized arc features to be captured more accurately than with a Sérsic profile alone. We adopted a maximum polynomial order of $\mathrm{n_{max}=5}$, corresponding to
\begin{equation}
    \mathrm{N_{shapelets}=\frac{(n_{max}+1)(n_{max}+2)}{2}},
\end{equation}
shapelet coefficients, which provided sufficient flexibility to reproduce the arc morphology in the IR data.

\textit{PSF Treatment - } \textcolor{black}{The quasar images are modeled as point sources using a PSF reconstructed independently in each imaging band with STARRED \citep{millon:2024}. For each band, several isolated stars were selected from the stacked science image. STARRED jointly fits all stellar cutouts with a common PSF model while simultaneously optimizing the centroid and flux of each star. The differing sub-pixel centroids of the stars provide additional information about the PSF on scales smaller than the detector pixel, while a sparsity-promoting regularization in a wavelet basis suppresses noise and stabilizes the reconstruction. The resulting PSF is reconstructed on a grid supersampled by a factor of three, reducing pixelization biases when convolving the modeled image with the PSF during the lens modeling process. Because no additional stars were available in the field of view of J1001+5027, the UVIS PSFs for this system were reconstructed using the two quasar images together with one isolated field star. On average, six stars were used to reconstruct the PSF for each system in each imaging band.}

To account for uncertainties in the PSF reconstruction, we incorporated a PSF variance map computed with \texttt{psfr} \citep{psfr} for the IR imaging only. This variance map quantifies the uncertainty in the reconstructed PSF arising from both the finite number of reference stars and the regularization imposed during the deconvolution process. The primary role of this variance map is to prevent inaccuracies in the PSF model from driving unphysical adjustments in other model components in an effort to artificially improve the likelihood. By effectively increasing the uncertainty in PSF-dominated pixels where the PSF is less well constrained, the variance map downweights these regions and acts as a safeguard against incorrect fits driven by PSF mismodeling. \textcolor{black}{The variance map itself is constructed by comparing the reconstructed PSF to each reference star, subtracting the expected noise contribution, and measuring the remaining pixel-by-pixel residual scatter.} For the UVIS data, no PSF variance map was applied, as we do not observe significant or structured residuals at the locations of the point sources that would indicate PSF-driven modeling pathologies. The sole exception is J0407-5006, for which a PSF variance map was included in all bands because of the small separation of Image B from the lens. The final PSF and associated variance map for the IR band of J2325-5229 is shown in Figure~\ref{fig:psfs}. 

Lastly, we adopt an astrometric uncertainty of 4 mas. This choice represents a compromise between \citet{tdcosmox} and \citet{schmidt:2023}, who report r.m.s offsets of 2 mas and 6 mas, respectively, based on HST \lenstronomy~model comparisons with Gaia DR3 \citep{2021A&A...649A...1G}. \textcolor{black}{In our modeling, the positional uncertainty is assigned to each measured quasar image position when evaluating the astrometric likelihood. This uncertainty sets the tolerance for deviations between the observed and modeled image positions, allowing the optimizer and MCMC sampler to explore a realistic range of solutions while preventing the posterior from becoming artificially over-constrained by the measured astrometry.}

\subsection{Fitting sequence}
\label{subsec:fitting_seq}

\begin{figure}[t]
    \centering
    \includegraphics[width=.5\textwidth]{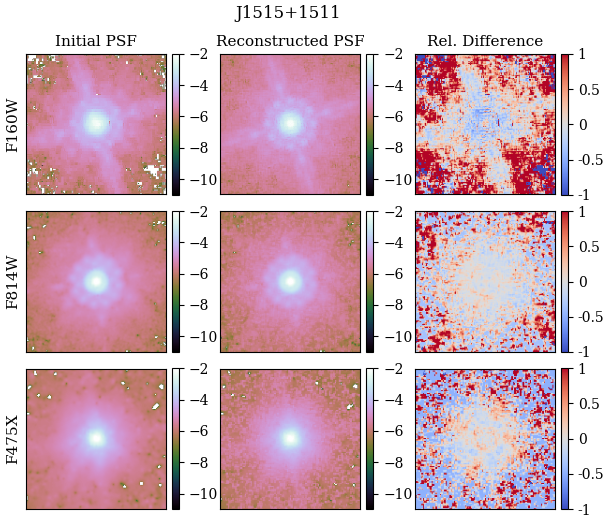}
    \caption{\textcolor{black}{Comparison of the initial PSF models across all three data bands with the final PSF calculated by \texttt{Lenstronomy}'s iterative PSF refinement routine for the system J1515+1511. The field of view has been reduced to the central half of the PSF to emphasize the core, as the outer regions are dominated by noise. The last column displays relative differences between the final and the initial PSFs, normalized by the initial PSF estimate.}}
    \label{fig:psf_iter}
\end{figure}

The sequence proceeded as follows. To suppress unrelated light, we applied an outer circular boundary removing nearby contaminating sources and background noise. An initial Particle Swarm Optimization (PSO; \citealt{kennedy:1995}) was run on the F160W band alone to establish a baseline model fit. The choice of F160W as the reference band was motivated by its superior signal-to-noise ratio and clearer detection of the lens galaxy and extended host light. The PSF at this initial stage is only a rough estimate and can bias the recovered quasar images if not properly accounted for. To address this, we performed iterative PSF reconstruction, where the lens light components from the current best-fit model were subtracted, isolating the quasar residuals. These residuals are stacked using median combination with 90 degree rotational symmetry to generate an improved empirical PSF, reducing sensitivity to outliers or asymmetries. This updated PSF is then re-applied to the forward model in a cycle of PSO optimization followed by PSF iteration, repeated five times to progressively refine both the model parameters and PSF. Because the PSF is derived from the model residuals themselves, this iterative scheme helps correct for inaccuracies in the initial PSF that could otherwise lead to local minima in the parameter space. \textcolor{black}{Figure \ref{fig:psf_iter} presents a comparison of the initial PSF estimate computed with STARRED to the final PSF determined from the \lenstronomy~iterative reconstruction for the system J1515+1511. The reconstruction refines both the PSF core and the extended diffraction structure, with the most noticeable relative differences occurring in the diffraction features and low-flux wings. Similar behavior is observed across the full sample.}


This initial PSF reconstruction is performed prior to introducing the source model in order to stabilize the point source astrometry and constrain the PSF core. After introducing the extended source component, the overall model was re-optimized through alternating PSO and PSF iteration cycles. The F814W band was then aligned to the IR reference frame using \lenstronomy's iterative alignment routine, which matches the coordinate frames of different filters based on the astrometric positions of the lensed quasar images. Following alignment, we performed joint PSO optimization of both F160W and F814W bands, with PSF refinement applied to both filters. The F475X band was subsequently aligned and incorporated, enabling a full three-band joint fit.

This final stage involved an extended optimization campaign with progressively decreasing PSO parameter width scales (100\%, then 60\%, then 30\%) to converge on the global minimum, interspersed with PSF iteration steps. The fitting sequence concluded with Markov Chain Monte Carlo (MCMC) sampling using \texttt{emcee} \citep{emcee}, ensuring thorough exploration of the parameter space and full marginalization over model uncertainties. Each system was sampled for at least 15,000 steps, yielding a minimum of $5.25\times10^6$ total samples per system.

\begin{table*}[htb]
\centering
\caption{Median lens light parameters computed from the best-fit model. The associated uncertainties are statistical and were calculated using the 84th and 16th percentiles. The position angle $\phi_{\rm L}$ is measured in degrees north of East. For all systems except J1001+5027, the F160W band is modeled using two elliptical S\'ersic lens light components. For J1001+5027, each band includes one elliptical S\'ersic component for the main lens and one for the perturber. For systems with two F160W lens light components, an additional “Composite” row reports the effective radius $R_{\rm eff}$, axis ratio $q_{\rm L}$, and position angle $\phi_{\rm L}$ inferred from a double-component photometric fit (the S\'ersic index is not defined for the composite model and is therefore omitted). Additionally, for J0602–4335, all three bands are modeled with double-component elliptical Sérsic profiles to capture both the central core and extended wings of the lens light distribution. Following \citet{Shajib:2019, shajib:2021, adnan:2025}, we added a 2\% systematic floor in quadrature to $R_\text{eff}$.}
\label{tab:lens_light_params}
\begin{tabular*}{\textwidth}{@{\extracolsep{\fill}}lccccc}
\hline
Lens System & Filter & $R_{\text{eff}}$ [arcsec] & $n_{\text{S\'ersic}}$ & $q_\text{L}$ & $\phi_{\text{L}}$ [deg] \\
\hline
J0407-5006 & F160W & $0.81^{+0.04}_{-0.04}$ & $5.40^{+0.22}_{-0.27}$ & $0.844^{+0.012}_{-0.012}$ & $137.8^{+2.7}_{-2.6}$ \\
 & F160W & $0.83^{+0.05}_{-0.04}$ & $5.94^{+0.04}_{-0.09}$ & $0.865^{+0.011}_{-0.011}$ & $143.9^{+3.3}_{-3.1}$ \\
 & F160W Composite & $0.20^{+0.01}_{-0.01}$ & $--$ & $0.523^{+0.079}_{-0.006}$ & $17.8^{+0.7}_{-0.5}$ \\
 & F814W & $0.46^{+0.03}_{-0.03}$ & $4.63^{+0.23}_{-0.19}$ & $0.661^{+0.008}_{-0.008}$ & $5.6^{+0.8}_{-0.8}$ \\
 & F475X & $0.55^{+0.10}_{-0.09}$ & $5.73^{+0.20}_{-0.35}$ & $0.784^{+0.058}_{-0.054}$ & $12.4^{+161.2}_{-8.2}$ \\
\hline
J0602-4335 & F160W & $0.16^{+0.01}_{-0.01}$ & $2.33^{+0.03}_{-0.03}$ & $0.663^{+0.004}_{-0.002}$ & $38.9^{+0.3}_{-0.3}$ \\
 & F160W & $0.57^{+0.01}_{-0.01}$ & $0.83^{+0.01}_{-0.01}$ & $0.127^{+0.001}_{-0.001}$ & $46.3^{+0.1}_{-0.1}$ \\
 & F160W Composite & $0.37^{+0.01}_{-0.01}$ & $--$ & $0.379^{+0.002}_{-0.002}$ & $44.9^{+0.1}_{-0.1}$ \\
 & F814W & $0.22^{+0.01}_{-0.01}$ & $1.78^{+0.02}_{-0.03}$ & $0.732^{+0.007}_{-0.007}$ & $40.0^{+0.7}_{-0.7}$ \\
 & F814W & $0.62^{+0.01}_{-0.01}$ & $0.85^{+0.02}_{-0.02}$ & $0.112^{+0.001}_{-0.001}$ & $45.7^{+0.3}_{-0.3}$ \\
 & F814W Composite & $0.45^{+0.01}_{-0.01}$ & $--$ & $0.394^{+0.004}_{-0.004}$ & $44.8^{+0.4}_{-0.4}$ \\
 & F475X & $0.19^{+0.01}_{-0.01}$ & $1.15^{+0.04}_{-0.04}$ & $0.880^{+0.019}_{-0.019}$ & $28.3^{+4.6}_{-5.0}$ \\
 & F475X & $0.66^{+0.02}_{-0.02}$ & $0.74^{+0.04}_{-0.03}$ & $0.111^{+0.001}_{-0.001}$ & $45.5^{+0.1}_{-0.1}$ \\
 & F475X Composite & $0.53^{+0.01}_{-0.01}$ & $--$ & $0.398^{+0.010}_{-0.010}$ & $44.7^{+0.4}_{-0.4}$ \\
\hline
J0806+2006 & F160W & $0.09^{+0.01}_{-0.01}$ & $5.68^{+0.24}_{-0.48}$ & $0.954^{+0.017}_{-0.017}$ & $111.0^{+9.9}_{-10.0}$ \\
 & F160W & $0.87^{+0.04}_{-0.03}$ & $1.80^{+0.18}_{-0.17}$ & $0.903^{+0.012}_{-0.012}$ & $121.4^{+2.9}_{-2.9}$ \\
 & F160W Composite & $0.37^{+0.01}_{-0.01}$ & $--$ & $0.950^{+0.010}_{-0.022}$ & $120.1^{+2.2}_{-2.2}$ \\
 & F814W & $0.43^{+0.02}_{-0.02}$ & $3.63^{+0.12}_{-0.11}$ & $0.926^{+0.009}_{-0.009}$ & $104.9^{+3.8}_{-3.7}$ \\
 & F475X & $0.18^{+0.01}_{-0.01}$ & $0.93^{+0.10}_{-0.09}$ & $0.713^{+0.043}_{-0.041}$ & $164.6^{+5.1}_{-4.9}$ \\
\hline
J1001+5027 & F160W & $1.06^{+0.03}_{-0.03}$ & $3.02^{+0.03}_{-0.02}$ & $0.825^{+0.003}_{-0.003}$ & $83.2^{+0.6}_{-0.6}$ \\
 & F814W & $1.50^{+0.13}_{-0.11}$ & $4.25^{+0.16}_{-0.15}$ & $0.807^{+0.008}_{-0.008}$ & $75.8^{+1.3}_{-1.3}$ \\
 & F475X & $4.78^{+0.19}_{-0.35}$ & $5.59^{+0.26}_{-0.28}$ & $0.372^{+0.024}_{-0.022}$ & $71.6^{+1.4}_{-1.4}$ \\
\hline
J1442+4055 & F160W & $0.08^{+0.01}_{-0.01}$ & $2.40^{+0.16}_{-0.21}$ & $0.863^{+0.010}_{-0.018}$ & $115.8^{+3.1}_{-1.8}$ \\
 & F160W & $0.08^{+0.01}_{-0.01}$ & $5.87^{+0.10}_{-0.19}$ & $0.842^{+0.008}_{-0.013}$ & $122.0^{+2.1}_{-1.3}$ \\
 & F160W Composite & $0.19^{+0.01}_{-0.01}$ & $--$ & $0.827^{+0.005}_{-0.009}$ & $125.0^{+1.4}_{-1.0}$ \\
 & F814W & $0.92^{+0.03}_{-0.03}$ & $5.96^{+0.03}_{-0.06}$ & $0.808^{+0.005}_{-0.006}$ & $127.5^{+1.3}_{-0.8}$ \\
 & F475X & $0.87^{+0.06}_{-0.11}$ & $5.86^{+0.10}_{-0.27}$ & $0.897^{+0.017}_{-0.017}$ & $140.0^{+4.9}_{-4.7}$ \\
\hline
J1515+1511 & F160W & $0.55^{+0.02}_{-0.02}$ & $1.50^{+0.04}_{-0.04}$ & $0.260^{+0.005}_{-0.004}$ & $106.2^{+0.2}_{-0.2}$ \\
 & F160W & $0.54^{+0.02}_{-0.02}$ & $1.50^{+0.04}_{-0.04}$ & $0.257^{+0.005}_{-0.003}$ & $106.2^{+0.2}_{-0.2}$ \\
 & F160W Composite & $0.42^{+0.01}_{-0.01}$ & $--$ & $0.217^{+0.001}_{-0.001}$ & $106.9^{+0.1}_{-0.1}$ \\
 & F814W & $0.46^{+0.02}_{-0.02}$ & $1.35^{+0.07}_{-0.07}$ & $0.220^{+0.004}_{-0.004}$ & $109.4^{+0.3}_{-0.3}$ \\
 & F475X & $1.61^{+1.78}_{-0.98}$ & $3.24^{+1.17}_{-1.26}$ & $0.306^{+0.048}_{-0.049}$ & $119.6^{+6.3}_{-5.4}$ \\
\hline
J1620+1203 & F160W & $0.18^{+0.01}_{-0.01}$ & $1.55^{+0.23}_{-0.20}$ & $0.479^{+0.044}_{-0.034}$ & $130.6^{+0.8}_{-0.8}$ \\
 & F160W & $2.72^{+0.19}_{-0.22}$ & $5.47^{+0.34}_{-0.48}$ & $0.796^{+0.002}_{-0.002}$ & $114.2^{+0.3}_{-0.3}$ \\
 & F160W Composite & $0.67^{+0.01}_{-0.01}$ & $--$ & $0.773^{+0.001}_{-0.001}$ & $117.7^{+0.2}_{-0.2}$ \\
 & F814W & $0.95^{+0.03}_{-0.03}$ & $3.79^{+0.06}_{-0.06}$ & $0.777^{+0.004}_{-0.004}$ & $113.3^{+0.6}_{-0.6}$ \\
 & F475X & $0.65^{+0.03}_{-0.03}$ & $5.97^{+0.02}_{-0.05}$ & $0.703^{+0.017}_{-0.016}$ & $107.1^{+1.8}_{-1.9}$ \\
\hline
J2325-5229 & F160W & $0.79^{+0.02}_{-0.02}$ & $1.81^{+0.04}_{-0.04}$ & $0.797^{+0.002}_{-0.002}$ & $35.1^{+0.4}_{-0.4}$ \\
 & F160W & $0.08^{+0.01}_{-0.01}$ & $5.96^{+0.03}_{-0.06}$ & $0.851^{+0.006}_{-0.006}$ & $41.2^{+1.3}_{-1.3}$ \\
 & F160W Composite & $0.46^{+0.01}_{-0.01}$ & $--$ & $0.807^{+0.001}_{-0.002}$ & $35.8^{+0.3}_{-0.2}$ \\
 & F814W & $1.44^{+0.10}_{-0.09}$ & $4.81^{+0.13}_{-0.13}$ & $0.838^{+0.005}_{-0.005}$ & $34.5^{+0.9}_{-0.9}$ \\
 & F475X & $2.93^{+0.19}_{-0.22}$ & $5.92^{+0.06}_{-0.12}$ & $0.891^{+0.015}_{-0.015}$ & $32.2^{+4.5}_{-4.3}$ \\
\hline
\end{tabular*}
\end{table*}

\begin{figure*}[t]
    \centering
    \includegraphics[width=.89\textwidth]{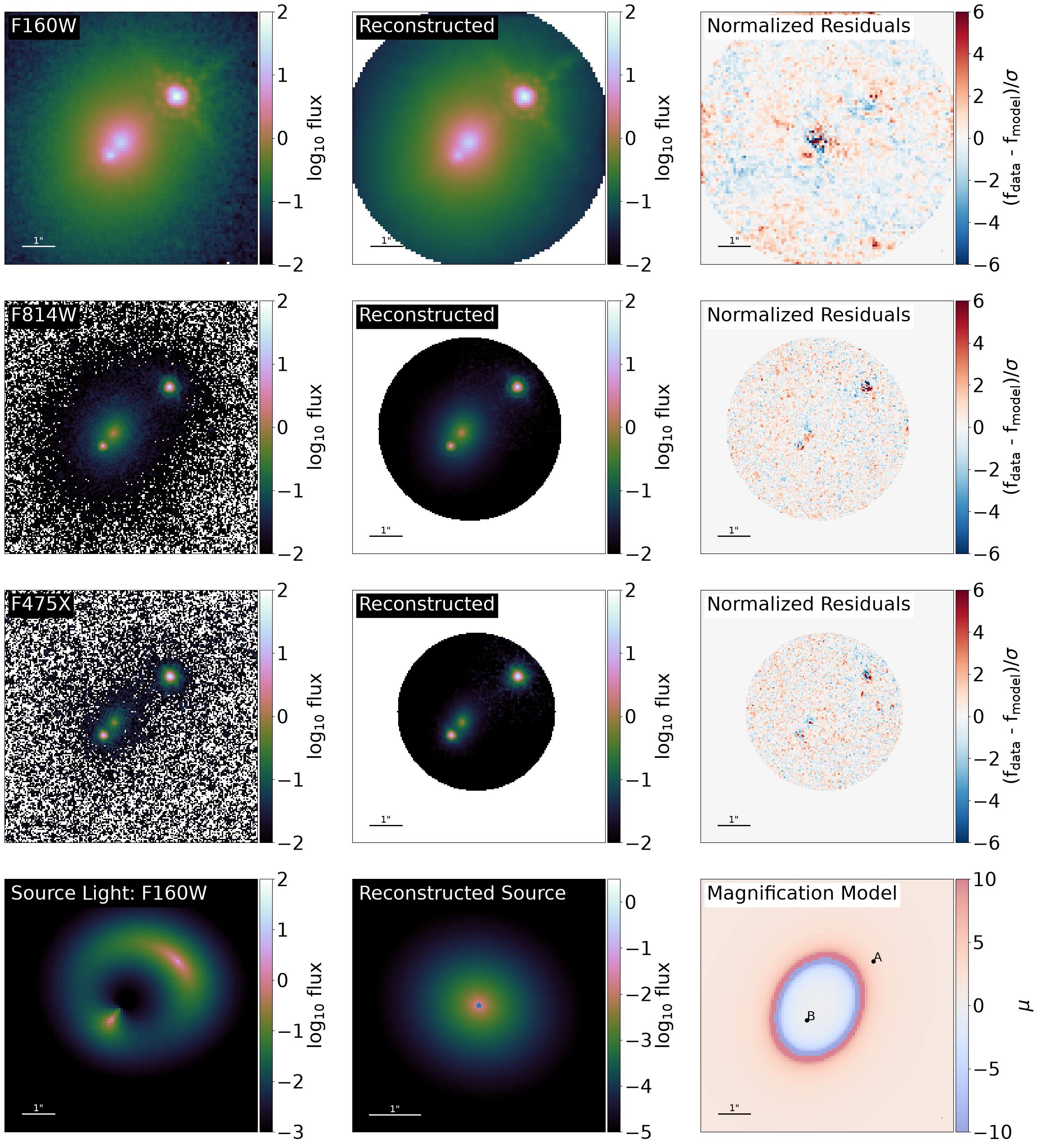}
    \caption{\lenstronomy~model visualization for the system J1620+1203. \textbf{Top Row.} HST IR science image cutout, followed by the IR best-fit reconstruction, followed by residuals of the fit normalized by the pixel noise level. \textbf{Second Row.} The same components as the top row, but for the F814W data. \textbf{Third Row.} The same components as the top row, but for the F475X data. \textbf{Bottom Row.} Reconstructed source arcs modeled in the image plane for the IR data, followed by the reconstructed source in the source plane. Here, the central blue star denotes the position of the quasar. The last panel showcases the magnification model displaying spatial variations in magnification.}
    \label{fig:j1620}
\end{figure*}

\section{\textbf{Full Image Modeling Results}}
\label{sec:results}

\begin{table*}[htb]
\centering
\caption{Median lens mass parameters and subsequent predicted Fermat potential differences between the quasar images computed from our uniform SIE+shear mass modeling best-fits. The associated uncertainties are statistical and were calculated using the 84th and 16th percentiles. The position angles $\phi_\text{mass}$ and $\phi_\text{ext}$ are measured in degrees north of East.}
\label{tab:lens_mass_params}
\begin{tabular*}{\textwidth}{@{\extracolsep{\fill}}lcccccccc}
\hline
Lens System & $\theta_\text{E}$ [arcsec] & $q_{\text{mass}}$ & $\phi_{\text{mass}}$ [deg] & $\gamma_{\text{ext}}$ & $\phi_{\text{ext}}$ [deg] & $\Delta \Phi_\text{AB}$ [arcsec$^2$] \\
\hline
J0407-5006 & $0.787^{+0.006}_{-0.006}$ & $0.80^{+0.04}_{-0.04}$ & $121.2^{+6.9}_{-5.4}$ & $0.113^{+0.008}_{-0.005}$ & $123.5^{+3.6}_{-3.9}$ & $-0.862^{+0.006}_{-0.009}$ \\ [3pt]
J0602-4335 & $0.914^{+0.014}_{-0.019}$ & $0.59^{+0.06}_{-0.05}$ & $53.7^{+3.5}_{-4.1}$ & $0.093^{+0.023}_{-0.029}$ & $66.6^{+7.3}_{-5.9}$ & $0.548^{+0.012}_{-0.011}$ \\ [3pt]
J0806+2006 & $0.782^{+0.019}_{-0.018}$ & $0.88^{+0.06}_{-0.07}$ & $115.4^{+8.2}_{-8.5}$ & $0.078^{+0.033}_{-0.030}$ & $110.3^{+5.0}_{-6.2}$ & $-0.493^{+0.013}_{-0.018}$ \\ [3pt]
J1001+5027 & $1.395^{+0.004}_{-0.004}$ & $0.73^{+0.01}_{-0.01}$ & $36.6^{+1.4}_{-1.4}$ & $0.123^{+0.005}_{-0.005}$ & $20.0^{+1.0}_{-0.9}$ & $-2.363^{+0.011}_{-0.011}$ \\ [3pt]
J1442+4055 & $1.032^{+0.007}_{-0.008}$ & $0.80^{+0.03}_{-0.03}$ & $96.9^{+4.3}_{-4.3}$ & $0.033^{+0.010}_{-0.010}$ & $58.4^{+9.8}_{-9.7}$ & $-0.607^{+0.008}_{-0.006}$ \\ [3pt]
J1515+1511 & $0.896^{+0.033}_{-0.019}$ & $0.87^{+0.06}_{-0.08}$ & $109.2^{+7.4}_{-7.1}$ & $0.096^{+0.021}_{-0.036}$ & $27.8^{+5.7}_{-3.9}$ & $-1.117^{+0.024}_{-0.043}$ \\ [3pt]
J1620+1203 & $1.431^{+0.020}_{-0.021}$ & $0.76^{+0.05}_{-0.04}$ & $121.6^{+4.1}_{-4.3}$ & $0.041^{+0.019}_{-0.015}$ & $146.9^{+16.8}_{-13.9}$ & $-2.397^{+0.050}_{-0.059}$ \\ [3pt]
J2325-5229 & $1.466^{+0.002}_{-0.002}$ & $0.94^{+0.02}_{-0.02}$ & $34.5^{+5.4}_{-5.4}$ & $0.040^{+0.004}_{-0.004}$ & $97.1^{+2.9}_{-3.4}$ & $0.698^{+0.005}_{-0.005}$ \\ [3pt]
\hline
\end{tabular*}
\end{table*}

\begin{table*}[htb]
\centering
\caption{Astrometric positions of the two quasar images as infered from the best-fit models. The absolute coordinates ($\alpha$ and $\delta$) of the system are taken directly from the HST header values \texttt{RA\_TARG} and \texttt{DEC\_TARG}. Offsets for quasar images A and B are measured relative to the best-fit lens light centroid. The image separation is the distance between the two quasar images. The total uncertainty in relative astrometry is dominated by systematic errors from sub-pixel PSF reconstruction, which we estimate to be 4 mas following \citet{tdcosmox} and \citet{schmidt:2023}.}
\label{tab:positions}
\begin{tabular*}{\textwidth}{@{\extracolsep{\fill}}lcccccccc}
\hline
Lens System & \multicolumn{2}{c}{Location} & \multicolumn{2}{c}{Image A} & \multicolumn{2}{c}{Image B} & Image \\
Name & $\alpha$ & $\delta$ & $\Delta \alpha$ & $\Delta \delta$ & $\Delta \alpha$ & $\Delta \delta$ & Separation \\
 & [hh:mm:ss] & [hh:mm:ss] & [arcsec] & [arcsec] & [arcsec] & [arcsec] & [arcsec] \\
\hline
J0407-5006 & 04:07:10.272 & -50:06:01.510 & -0.721 & 1.225 & 0.137 & -0.248 & 1.704 \\
J0602-4335 & 06:02:16.100 & -43:35:40.200 & -0.412 & 0.470 & 0.759 & -0.931 & 1.826 \\
J0806+2006 & 08:06:23.680 & 20:06:31.460 & 0.924 & 0.500 & -0.396 & -0.178 & 1.484 \\
J1001+5027 & 10:01:28.500 & 50:27:57.420 & 2.042 & -1.119 & -0.443 & 0.436 & 2.932 \\
J1442+4055 & 14:42:54.700 & 40:55:35.620 & 1.289 & 0.410 & -0.762 & -0.162 & 2.129 \\
J1515+1511 & 15:15:38.540 & 15:11:35.120 & 1.370 & 0.863 & -0.298 & -0.254 & 2.008 \\
J1620+1203 & 16:20:26.234 & 12:03:40.680 & -1.771 & 1.455 & 0.335 & -0.409 & 2.813 \\
J2325-5229 & 23:25:41.210 & -52:29:15.000 & -0.917 & -0.792 & 0.489 & 1.651 & 2.819 \\
\hline
\end{tabular*}
\end{table*}

In this section, we present the results for each system from the pipeline and provide an overarching comparison to the literature. Table \ref{tab:lens_light_params} presents the lens light model parameters as inferred from the best-fit lens light reconstructions. All systems display the expected chromatic trend between the F160W and F814W bands, with $R_\text{eff}$ increasing toward shorter wavelengths as bluer light traces the younger stellar populations that inhabit the outer regions of the galaxy. Five of the eight systems continue this trend from F814W to F475X. In the remaining cases, the trend likely weakens due to the lower signal-to-noise ratio of the lens galaxy in this bluest band.

Table \ref{tab:lens_mass_params} presents the best-fit lens mass model parameters alongside the predicted Fermat potential differences between image pairs. For clarity, we define the Fermat potential difference as
\begin{equation}
    \Delta \Phi_\text{AB}=\frac{(\theta_\text{A}-\beta)^2}{2}-\frac{(\theta_\text{B}-\beta)^2}{2}-\psi_\text{A}+\psi_\text{B},
\end{equation}
such that $\Delta \Phi_\text{AB}$ is negative when the information from Image A arrives before Image B. Here, $\theta_\text{i}$ are the respective quasar images, $\beta$ is the source position, and $\psi_\text{i}$ is the deflection potential at position $\theta_\text{i}$ as inferred from the best-fit lens mass model.

\begin{figure}[t]
    \centering
    \includegraphics[width=.499\textwidth]{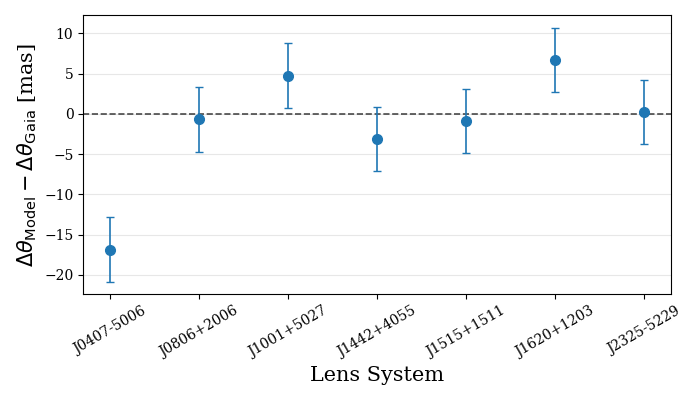}
    \caption{Comparison of our multi-band modeled image separation to Gaia Data Release 2 values. There is no Gaia data available for J0602-4335, and thus it is not included. The vertical error bars represent our estimated 4 mas uncertainties in the HST image modeling. The dashed black line indicates a model with no image separation offset from Gaia DR2. Excluding J0407-5006, we find an r.m.s image separation difference of 3.6 mas. We note J0407-5006 as an outlier due to the close proximity of Image B to the lens centroid ($\sim$$0\farcs28$). At this separation, the lensed image and lens galaxy are difficult to deblend in Gaia due to its comparatively coarser angular resolution, which may introduce systematic offsets in the reported position.}
    \label{fig:gaia_astrom}
\end{figure}

\begin{figure}[t]
    \centering
    \includegraphics[width=.499\textwidth]{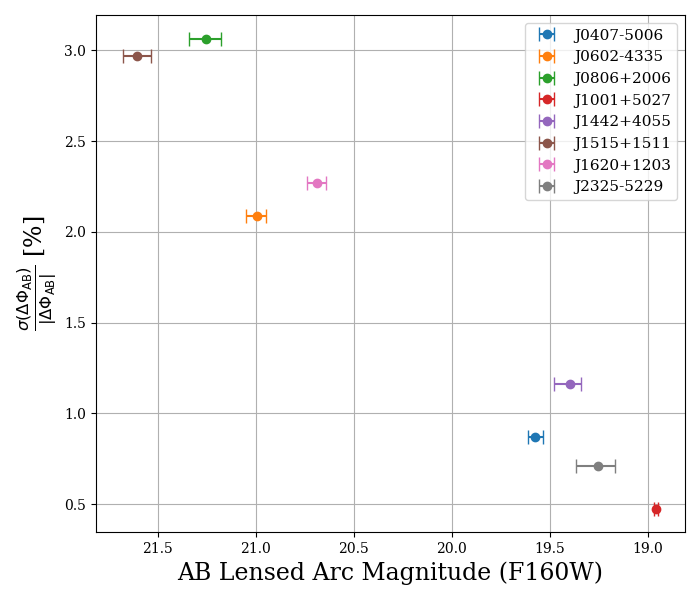}
    \caption{Comparison of Fermat potential precision with calculated AB Lensed Arc Magnitude. Each point correlates to an individual system, with the horizontal error bars denoting the uncertainty in the modeled arc magnitude. We find a Pearson correlation of $r = 0.97 $ $(p = 0.0005)$ between arc brightness and Fermat potential precision, indicating that improved arc brightness directly translates into tighter constraints on the Fermat potential, and therefore on  cosmological inferences that utilize time delays.}
    \label{fig:fermat_vs_mag}
\end{figure}

\begin{figure*}[t]
    \centering
    \includegraphics[width=.99\textwidth]{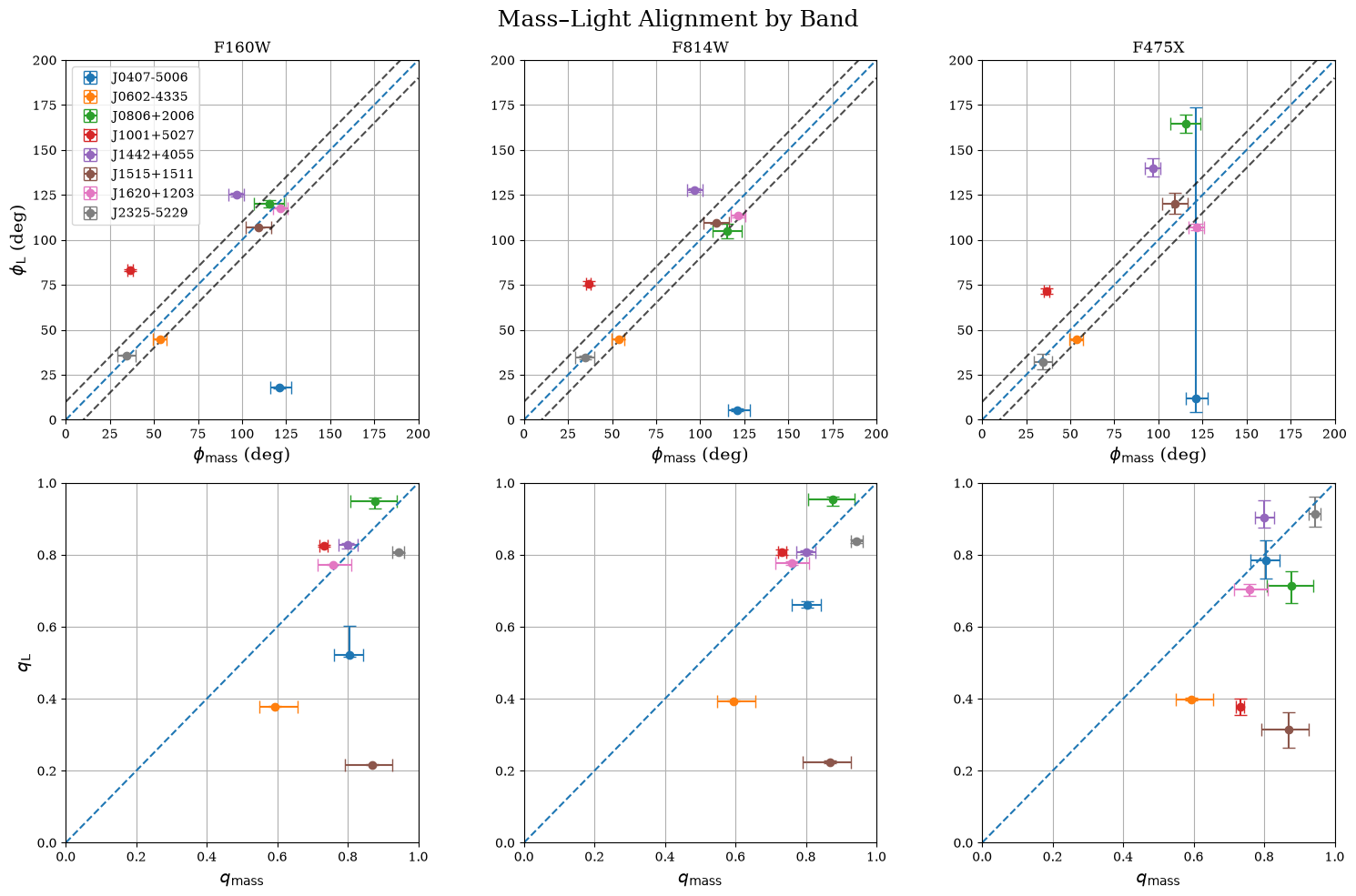}
    \caption{Mass and Lens Light profile shape parameters by band, from left to right: F160W, F814W, and F475X. For all plots, the error bars denote the $\pm1\sigma$ credibility regions. \textbf{Top Row.} Comparison of mass profile position angle with lens light profile position angle. As the F160W data contains the highest signal-to-noise, we imposed a Gaussian mass-light position angle alignment prior with a standard deviation of 10 degrees to this data band. The dashed blue lines indicate profiles of equal mass-light position angles, while the dashed black lines denote $\pm10$ degrees from equality. \textbf{Bottom Row.} Comparison of mass major/minor axis ratio with light major/minor axis ratio. The dashed blue lines denote models of equal mass-light axis ratios.}
    \label{fig:mass-light}
\end{figure*}

Table \ref{tab:positions} lists the model-derived astrometry. \textcolor{black}{As described in Section \ref{sec:modeling_procedure}, the model was fit assuming a 4 mas uncertainty on each quasar image position, and this uncertainty is adopted for the astrometric measurements reported here. The lens-image separations vary within reason across the sample, ranging from 0\farcs28 for image B in J0407-5006 to 2\farcs33 for image A in J1001+5027. Likewise, the image-image separations range from 1\farcs48 in J0806+2006 to 2\farcs93 in J1001+5027, with an average separation of 2\farcs21 across the sample.}

Table \ref{tab:mags} showcases the calculated image, lens, and host photometry from the models, for which we add in quadrature a WFC3 AB magnitude zeropoint uncertainty of $\pm$0.018 mag (see \citeauthor{bajaj:2020} \citeyear{bajaj:2020} for more details). For the flux ratios, the zeropoint uncertainty cancels exactly and does not contribute. To estimate photometric uncertainties arising from blending between the lensed arcs and the PSF, \textcolor{black}{we re-modeled J1442+4055 three times using three independently reconstructed PSFs for each data band.} This system was chosen because the extended arc magnitude, $\rm m_{arc}=19.40$ mag, lies in the brighter half of the sample and therefore provides a representative baseline for the impact of arc blending and PSF mismodeling on image fluxes. \textcolor{black}{The image fluxes recovered from the three models were compared, and the standard deviation divided by the mean flux was adopted as the fractional systematic uncertainty.} The resulting fractional uncertainties are 8\% in F160W, 3\% in F814W, and 1\% in F475X. \textcolor{black}{Because this exercise was intended to quantify the typical uncertainty introduced by blending between the lensed arcs and the PSF, rather than derive uncertainties specific to a single lens system, the resulting band-dependent fractional uncertainties were adopted as systematic floors for the image fluxes and flux ratios throughout the sample.} These systematic uncertainties should be regarded as approximate estimates, and future work should explore additional sources of systematic error in the photometry.

In order to quantify in which systems we detected host quasar light with confidence, we performed a suite of computations using the same structure as Section~\ref{subsec:fitting_seq}, but without the inclusion of an extended source model. From the no-source and source-included models, we computed the Bayesian Inference Criterion (BIC) as:
\begin{equation}
    \mathrm{BIC}=k\ln(n)-2\ln(\mathcal{L}),
\end{equation}
where $k$ is the number of model parameters, $n$ is the number of pixels utilized in the model, and $\mathcal{L}$ denotes the maximized likelihood of the model. We assess the statistical preference for the inclusion of a host component via $\Delta\mathrm{BIC} \equiv \mathrm{BIC_{\rm no-source}} - \mathrm{BIC_{\rm source}}$, such that positive values indicate a preference for the source-included model. Following \citet{kess_raftery:1995}, values of $\Delta\mathrm{BIC} \gtrsim 10$ are considered strong statistical evidence. In all eight systems, the inclusion of a source model is strongly favored with $\Delta \rm BIC > 300$.

\begin{table*}[htb]
\centering
\caption{Median AB magnitudes of the lens light, quasar images, and lensed/delensed host galaxy, alongside image flux ratios, across all three data bands. Uncertainties reflect both statistical errors from the best-fit flux models (derived from the 84th and 16th percentiles) and a zeropoint calibration uncertainty of $\pm$0.018 mag from HST WFC3. For the quasar image magnitudes and fluxes, an additional systematic uncertainty due to PSF mismodeling and blending with the extended lensed arcs has been included, estimated from multiple PSF realizations on models of J1442+4055. Lens and source light magnitudes were computed by integrating the respective light profiles over the full image grid. Entries denoted by $\text{``N/A"}$ indicate systems and filters in which flux from the host quasar was not measured. }
\label{tab:mags}
\begin{tabular*}{\textwidth}{@{\extracolsep{\fill}}lccccccc}
\hline
Lens System & Filter & Lens & Image A & Image B & Host (lensed) & Host (delensed) & $f_\text{A}/f_\text{B}$ \\
\hline
J0407-5006 & F160W & $18.17^{+0.02}_{-0.02}$ & $17.68^{+0.09}_{-0.09}$ & $18.81^{+0.09}_{-0.09}$ & $19.57^{+0.04}_{-0.04}$ & $20.94^{+0.06}_{-0.05}$ & $2.84^{+0.32}_{-0.32}$ \\ [3pt] 
 & F814W & $20.15^{+0.03}_{-0.03}$ & $18.06^{+0.04}_{-0.04}$ & $19.69^{+0.04}_{-0.04}$ & $21.53^{+0.05}_{-0.05}$ & $22.88^{+0.05}_{-0.06}$ & $4.48^{+0.20}_{-0.20}$ \\ [3pt] 
 & F475X & $22.92^{+0.10}_{-0.08}$ & $18.64^{+0.02}_{-0.02}$ & $19.85^{+0.02}_{-0.02}$ & $23.72^{+0.08}_{-0.07}$ & $25.07^{+0.08}_{-0.07}$ & $3.04^{+0.05}_{-0.05}$ \\ [3pt] 
\hline
J0602-4335 & F160W & $17.33^{+0.02}_{-0.02}$ & $19.26^{+0.09}_{-0.09}$ & $19.05^{+0.09}_{-0.09}$ & $20.99^{+0.06}_{-0.05}$ & $22.98^{+0.07}_{-0.06}$ & $0.82^{+0.09}_{-0.09}$ \\ [3pt] 
 & F814W & $18.84^{+0.02}_{-0.02}$ & $19.37^{+0.04}_{-0.04}$ & $19.57^{+0.04}_{-0.04}$ & N/A & N/A & $1.20^{+0.05}_{-0.05}$ \\ [3pt] 
 & F475X & $20.64^{+0.02}_{-0.02}$ & $19.61^{+0.02}_{-0.02}$ & $19.80^{+0.02}_{-0.02}$ & N/A & N/A & $1.19^{+0.02}_{-0.02}$ \\ [3pt] 
\hline
J0806+2006 & F160W & $18.42^{+0.02}_{-0.03}$ & $18.82^{+0.09}_{-0.09}$ & $19.73^{+0.09}_{-0.09}$ & $21.26^{+0.09}_{-0.08}$ & $22.99^{+0.27}_{-0.18}$ & $2.31^{+0.27}_{-0.26}$ \\ [3pt] 
 & F814W & $19.85^{+0.03}_{-0.03}$ & $19.22^{+0.04}_{-0.04}$ & $20.03^{+0.04}_{-0.04}$ & $23.02^{+0.10}_{-0.09}$ & $24.80^{+0.19}_{-0.17}$ & $2.10^{+0.09}_{-0.09}$ \\ [3pt] 
 & F475X & $22.94^{+0.04}_{-0.04}$ & $19.83^{+0.02}_{-0.02}$ & $20.54^{+0.02}_{-0.02}$ & $23.94^{+0.17}_{-0.15}$ & $25.68^{+0.23}_{-0.22}$ & $1.93^{+0.03}_{-0.03}$ \\ [3pt] 
\hline
J1001+5027 & F160W & $18.12^{+0.02}_{-0.02}$ & $18.11^{+0.09}_{-0.09}$ & $17.94^{+0.09}_{-0.09}$ & $18.96^{+0.02}_{-0.02}$ & $20.71^{+0.02}_{-0.02}$ & $0.85^{+0.10}_{-0.10}$ \\ [3pt] 
 & F814W & $19.29^{+0.03}_{-0.03}$ & $17.82^{+0.04}_{-0.04}$ & $17.67^{+0.04}_{-0.04}$ & $21.01^{+0.03}_{-0.03}$ & $22.78^{+0.03}_{-0.03}$ & $0.87^{+0.04}_{-0.04}$ \\ [3pt] 
 & F475X & $21.61^{+0.05}_{-0.04}$ & $18.07^{+0.02}_{-0.02}$ & $18.15^{+0.02}_{-0.02}$ & N/A & N/A & $1.08^{+0.02}_{-0.02}$ \\ [3pt] 
\hline
J1442+4055 & F160W & $17.91^{+0.02}_{-0.03}$ & $17.90^{+0.09}_{-0.09}$ & $18.51^{+0.09}_{-0.09}$ & $19.40^{+0.09}_{-0.06}$ & $21.26^{+0.15}_{-0.09}$ & $1.75^{+0.20}_{-0.20}$ \\ [3pt] 
 & F814W & $18.83^{+0.02}_{-0.02}$ & $18.16^{+0.04}_{-0.04}$ & $18.99^{+0.04}_{-0.04}$ & $23.78^{+0.19}_{-0.23}$ & $25.65^{+0.21}_{-0.21}$ & $2.14^{+0.09}_{-0.09}$ \\ [3pt] 
 & F475X & $20.72^{+0.03}_{-0.05}$ & $18.88^{+0.02}_{-0.02}$ & $19.82^{+0.02}_{-0.02}$ & $23.63^{+0.17}_{-0.11}$ & $25.51^{+0.21}_{-0.12}$ & $2.37^{+0.03}_{-0.03}$ \\ [3pt] 
\hline
J1515+1511 & F160W & $19.59^{+0.02}_{-0.02}$ & $18.06^{+0.09}_{-0.09}$ & $18.36^{+0.09}_{-0.09}$ & $21.61^{+0.08}_{-0.07}$ & $22.75^{+0.09}_{-0.08}$ & $1.31^{+0.15}_{-0.15}$ \\ [3pt] 
 & F814W & $21.40^{+0.04}_{-0.03}$ & $18.17^{+0.04}_{-0.04}$ & $18.57^{+0.04}_{-0.04}$ & $24.74^{+0.30}_{-0.25}$ & $25.89^{+0.31}_{-0.26}$ & $1.44^{+0.06}_{-0.06}$ \\ [3pt] 
 & F475X & $23.61^{+0.40}_{-0.39}$ & $18.77^{+0.02}_{-0.02}$ & $19.22^{+0.02}_{-0.02}$ & $23.06^{+0.08}_{-0.07}$ & $24.21^{+0.08}_{-0.08}$ & $1.51^{+0.02}_{-0.02}$ \\ [3pt] 
\hline
J1620+1203 & F160W & $17.37^{+0.02}_{-0.02}$ & $18.76^{+0.09}_{-0.09}$ & $20.81^{+0.09}_{-0.09}$ & $20.69^{+0.06}_{-0.05}$ & $22.09^{+0.08}_{-0.07}$ & $6.63^{+0.78}_{-0.78}$ \\ [3pt] 
 & F814W & $18.78^{+0.02}_{-0.02}$ & $18.99^{+0.04}_{-0.04}$ & $20.47^{+0.04}_{-0.04}$ & $22.84^{+0.05}_{-0.05}$ & $24.23^{+0.07}_{-0.07}$ & $3.90^{+0.17}_{-0.17}$ \\ [3pt] 
 & F475X & $21.21^{+0.03}_{-0.03}$ & $19.39^{+0.02}_{-0.02}$ & $20.84^{+0.02}_{-0.02}$ & N/A & N/A & $3.80^{+0.06}_{-0.06}$ \\ [3pt] 
\hline
J2325-5229 & F160W & $17.73^{+0.02}_{-0.02}$ & $20.47^{+0.09}_{-0.09}$ & $20.15^{+0.09}_{-0.09}$ & $19.25^{+0.12}_{-0.08}$ & $20.63^{+0.21}_{-0.14}$ & $0.74^{+0.08}_{-0.08}$ \\ [3pt] 
 & F814W & $18.79^{+0.03}_{-0.03}$ & $21.01^{+0.04}_{-0.04}$ & $20.61^{+0.04}_{-0.04}$ & N/A & N/A & $0.69^{+0.03}_{-0.03}$ \\ [3pt] 
 & F475X & $20.83^{+0.03}_{-0.03}$ & $21.95^{+0.02}_{-0.02}$ & $21.32^{+0.02}_{-0.02}$ & N/A & N/A & $0.56^{+0.01}_{-0.01}$ \\ [3pt] 
\hline
\end{tabular*}
\end{table*}

Our best-fit lens model for J1620+1203 is shown in Figure \ref{fig:j1620}. The first column displays the observed HST imaging in the F160W, F814W, and F475X bands, while the second column shows the corresponding model reconstructions generated by our pipeline. The third column presents the residuals of the fit normalized by the pixel noise level, providing a diagnostic of the model fit quality. Across all three bands, the residual maps are largely consistent with noise, indicating that the model accurately reproduces the observed surface brightness distribution of both the lens galaxy and the quasar images. In the two UVIS bands, localized residuals are visible near the quasar image cores, likely induced by sub-pixel uncertainties in the PSF modeling. The bottom left panel shows the reconstructed lensed host emission in the image plane for the F160W band, for which we detect extended arc structure. The extended host arc provides further constraints that couple the quasar image configuration to the underlying source position, thereby providing additional information to better describe the lens mass distribution. The bottom central panel shows the corresponding reconstruction of the source in the source plane. Finally, the bottom right panel presents the lensing magnification map, with the labeled quasar image positions overlaid. Model visualizations for the remaining systems are presented in Appendix \ref{append::b}.

When comparing our results for all systems to the literature, we find an average difference in $\theta_\text{E}$ of $1.5\sigma$. This slight scatter is anticipated given that prior measurements were obtained with heterogeneous instruments, mass parameterizations, and fitting methodologies, making exact agreement unlikely even for well-constrained systems. The largest discrepancy comes from J0407-5006, whose measurement differs from \citet{anguita:2018} by 4.5$\sigma$. This system exhibits the smallest image-lens separation within the sample, with Image B laying only $0\farcs28$ from the lens centroid. At such small separations, ground-based observations, such as those conducted in the aforementioned study, likely struggle to fully deblend the quasar image from the lens galaxy, in turn biasing the inferred astrometry and mass model parameters. When we exclude J0407-5006 from this comparison, our difference in $\theta_\text{E}$ drops to 0.92$\sigma$, supporting the reliability of our modeling pipeline and the consistency of our measurements with previous work. A full system-by-system comparison to the literature is presented in Appendix \ref{sec:appendA}.

Figure \ref{fig:gaia_astrom} displays how our best-fit image separations deviate from Gaia DR2 values for the seven systems with available Gaia data \citep{gaia:dr2}, with no measurement available for J0602-4335. We find an r.m.s image separation difference of 3.6 mas, which is comfortably within our estimated astrometric uncertainty of 4.0 mas. This sub-pixel level agreement between our HST-based models and independent Gaia astrometry serves as an additional external validation of our pipeline, confirming that the point source positions recovered by our full image modeling are not systematically biased by PSF mismodeling or lens light subtraction residuals in the iterative PSF reconstruction routine. Notably, the largest Gaia residual occurs for J0407-5006, which also has the smallest image-lens separation in our sample of $0\farcs28$ between Image B and the lens centroid. In such close configurations, Gaia’s effective angular resolution ($\sim$$0\farcs40$, \citeauthor{gaia:dr2} \citeyear{gaia:dr2}) exceeds the scale of the image-lens separation, making accurate astrometry more challenging compared to the $0\farcs04$ pixels of the HST UVIS detector. Therefore, we note this system as an outlier and do not include it in our r.m.s image separation difference calculation. Most notably, our work reports the first Einstein radius and image separation measurements for J0602-4335, finding $\theta_\text{E}=0\farcs914^{+0.014}_{-0.019}$ and $\Delta \theta=1\farcs826$, respectively.

\begin{figure*}[t]
    \centering
    \includegraphics[width=0.97\textwidth]{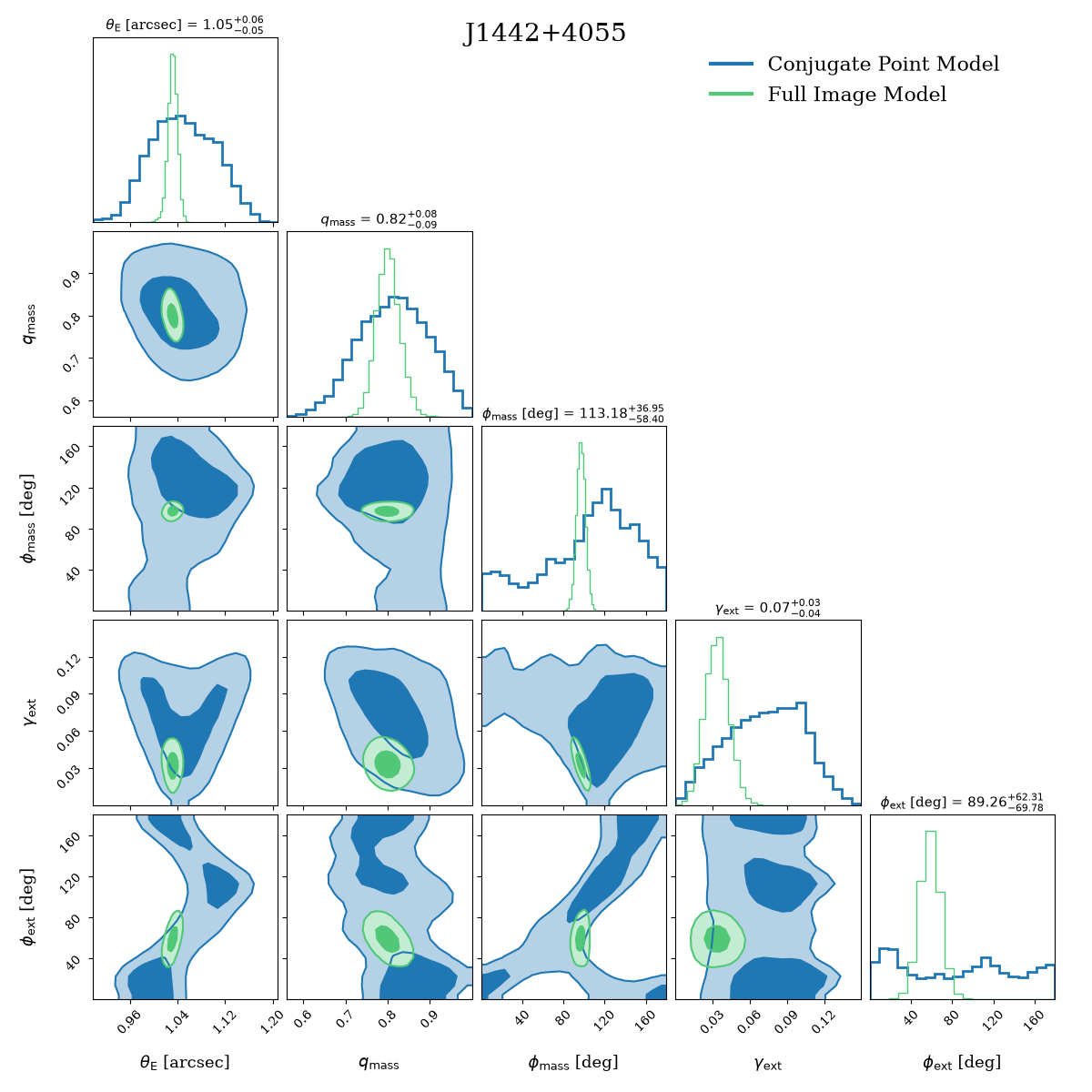}
    \caption{Overlayed mass posterior probability distributions for the conjugate point model (blue) and the full image model (green) for the system J1442+4055. Contours reflect the 1- and 2$\sigma$ credible regions, and the quoted parameter values correspond to the conjugate point posteriors. While the full image posteriors fall cleanly within the broader conjugate point contours, the ladder method is substantially less informative and displays well-known degeneracies associated with the mass shape, mass position angle, and external shear.}
    \label{fig:conj_posteriors}
\end{figure*}

Figure \ref{fig:fermat_vs_mag} displays the relationship between modeled Fermat potential precision and calculated AB lensed arc magnitude across the sample. We find a Pearson correlation coefficient of $r = 0.97 $ $(p = 0.0005)$, indicating a strong relation in which brighter arcs yield tighter Fermat potential constraints. The system with the brightest arc, J1001+5027 at $\rm m_{arc} = 18.96$ mag, achieves a Fermat potential precision of \textcolor{black}{$\frac{\sigma_{\Delta\Phi}}{\Delta \Phi} = 0.47\%$}, while the faintest detected arc, J1515+1511 at $\rm m_{arc} = 21.6$ mag, yields \textcolor{black}{$\frac{\sigma_{\Delta\Phi}}{\Delta \Phi} = 2.99\%$}. This dependence is particularly relevant for doubles, as extended arc structure provides one of the few sources of spatial information capable of constraining the lens mass model beyond the two point source positions. We emphasize that this trend is derived under the assumption of an SIE mass model. Adopting a more flexible power-law profile would introduce additional freedom in the radial slope, and thus impact the precision on the Fermat potential. As shown by \citet{suyu:2012}, degeneracies between the radial mass profile slope and the time-delay distance in doubles can lead to order $\sim$15\% variations if the slope is not tightly constrained. The structure of the extended arcs helps to break these degeneracies, justifying the need for future deep observations in order to further reduce the error budget in cosmographic studies.

Figure \ref{fig:mass-light} compares the mass and light shape parameters across bands. We first consider the position angle (top row). In the F160W band, where the lens light is best constrained, five of eight systems lie within the imposed $\pm10$ degree prior, indicating generally good alignment between the mass and light distributions. Notable outliers include J0407-5006 and J1442+4055. In these cases, nearby companion galaxies visible in Figure \ref{fig:rgbs} may perturb the lensing potential. Including such perturbers may improve the agreement between mass and light position angles, as demonstrated by \citet{Sluse:2012}, suggesting that the observed discrepancies may reflect limitations of the simplified SIE+shear lens model rather than intrinsically unusual galaxy structure. While the F814W and F475X bands show broadly similar trends, the light parameters are strongly correlated across bands and the mass model is constrained jointly using all datasets. As such, these comparisons do not constitute independent evidence for mass-light alignment. Moreover, the bluest F475X band exhibits increased scatter, consistent with the fundamentally lower signal of the lens galaxy light, which leads to less reliable position angle recovery. In the bottom row, we compare the axis ratios. Most systems cluster near the equality line in F160W, with both mass and light distributions typically round ($q \gtrsim 0.8$). The primary exceptions are J0602-4335 and J1515+1511, which show consistently low light axis ratios across all bands, reflecting their elongated morphologies seen in Figure \ref{fig:rgbs}. As with the position angles, uncertainties in $q_{\rm L}$ increase in the bluer bands due to reduced lens light signal. Overall, the F160W results indicate that mass and light are broadly well aligned in both orientation and ellipticity for this dataset, with deviations attributable to potential environmental perturbations.

\section{\textbf{Conjugate Point Analysis}}
\label{sec:conjugate}

To further assess the constraining power of the extended lensed host galaxy arcs relative to the point source positions alone, we performed a conjugate point analysis for each system in our sample. Furthermore, this comparison also quantifies the additional information provided by high resolution, multi-band HST imaging beyond the positional constraints that could be achieved with lower resolution data. By comparing these results to the full image modeling presented in Section \ref{sec:results}, we can quantify the degree to which the extended arc morphology improves constraints on the lens parameters, or, conversely, whether point source astrometry alone is sufficient for characterization of the lens.
Much like full image modeling, the conjugate point method relies on solving the lens equation,
\begin{equation}
{\beta} = {\theta}_\text{i} - {\alpha}({\theta}_\text{i}; {\eta}),
\label{eq:lenseq}
\end{equation}
where ${\theta}_\text{i}$ are the observed image positions of the point sources, ${\beta}$ is the corresponding unlensed source position in the source plane, ${\alpha}$
 is the deflection angle field, and ${\eta}$ denotes the set of lens model parameters. For any given choice of ${\eta}$, a physically consistent lens model must map all observed image positions ${\theta}_\text{i}$ to a common source position ${\beta}.$

\begin{figure}[t]
    \centering
    \includegraphics[width=.49\textwidth]{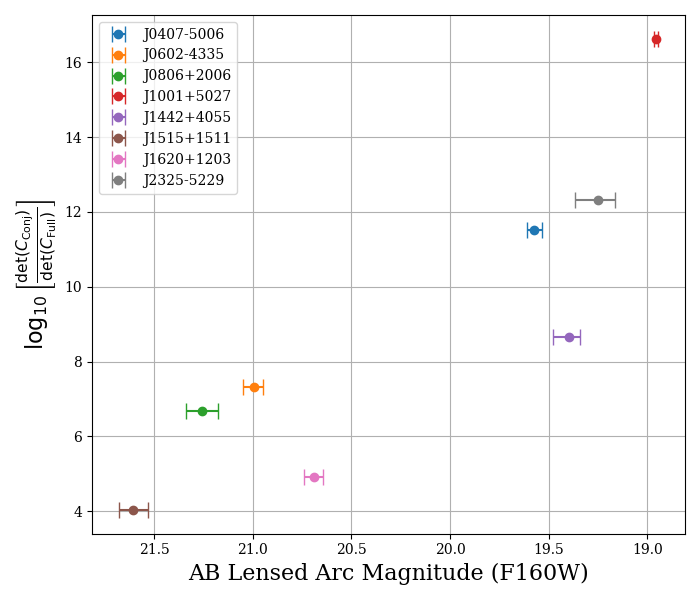}
    \caption{Comparison of mass posterior hypervolume between conjugate point and full image modeling results with AB lensed arc magnitude. The hypervolume of the mass parameter posterior is approximated as the determinant of the corresponding covariance matrix. Each point correlates to an individual system, with y-axis values denoting the logarithm of the difference between mass parameter covariance matrix determinants. The x-axis error bars represent the uncertainty in the calculated arc magnitude. We find that the conjugate point mass posterior volume is larger than that of the full image modeling by an average of $10^9$ times, with a Pearson correlation coefficient of $r = -0.87 $ $(p = 0.005)$ existing between the differences and the arc magnitude. As such, this relationship indicates that brighter lensed arcs lead to significantly tighter constraints on the lens mass distribution.}
    \label{fig:conj_full_vol}
\end{figure}

This minimalist approach is computationally efficient and particularly attractive for large statistical samples, as it only requires image astrometry and eliminates the cost of source reconstruction and pixel-by-pixel surface brightness fitting. However, it also discards potentially valuable constraints encoded in the arc geometry. Because doubles only have two lensed images, and therefore less lens constraints than quads, the question is raised of whether the arc information becomes critical for breaking degeneracies in the lens model, particularly those involving the ellipticity and external shear. To address this question systematically, we adopted the best-fit quasar image positions measured from our full image modeling as observational truths with an astrometric uncertainty of $\sigma=0\farcs004$. We retained the same mass parameterization used in the full modeling, which included an SIE with external shear. To further enforce consistency between methods, the center of the conjugate point mass distribution was allowed to vary at at the same width as the full image modeling centroid posterior. All other lens parameters (e.g. Einstein radius $\theta_\text{E}$, axis ratio $q_{\rm mass}$, position angle $\phi_{\rm mass}$, and external shear $\gamma_{\rm ext}$) were allowed to vary freely during the fit. Optimization was conducted via PSO, while posterior sampling was performed using the \lenstronomy~MCMC framework with the same methodology as described in Section \ref{sec:results}.

Figure \ref{fig:conj_posteriors} illustrates this comparison for the system J1442+4055, with the conjugate point posteriors (blue) consistently encompassing the full image modeling posteriors (green) as narrow, well-constrained islands within the broader conjugate point distributions. The most pronounced degeneracies in the conjugate point models appear in the joint distributions of $q_\text{mass}, \phi_\text{mass}$, and $ \gamma_\text{ext}$, where the contours are broad and exhibit the well-known mass-shear degeneracy \citep{keeton:1997}. The extended arc surface brightness distribution effectively constricts this degeneracy by providing additional constraints on the lensing potential through the morphology and curvature of the arc \citep{suyu:2009}. More generally, the conjugate point method recovers Einstein radii across the entire sample consistent with full image modeling and shows no evidence of systematic bias, with a mean fractional offset of $-0.41\%$ and a system-to-system scatter of $7.0\%$.

Figure \ref{fig:conj_full_vol} further quantifies this information gain across the full sample by comparing the mass parameter posterior hypervolumes between the two methods, approximated as the determinant of the respective covariance matrices, and the calculated AB lensed arc magnitude. We find that the conjugate point posterior volume is larger than that of the full image modeling by an average factor of $10^9$ times. Systems with brighter and more extended arcs, such as J1001+5027 and J2325-5229, show the largest relative information gain from full image modeling, consistent with the Fermat potential precision trends discussed in Section \ref{sec:results}. While the full image modeling framework naturally introduces additional constraints through the simultaneous modeling of the lens light distribution (e.g., shared centroids and mass-light alignment priors), the observed correlation between arc brightness and mass parameter hypervolume ($r=-0.87$ and $p=0.005$) suggests that the dominant driver of improved constraints is the information content of the extended arc surface brightness itself. This result has practical implications for future cosmography with doubles in that, while conjugate point models are computationally attractive, the systematic underconstraint of the mass shape parameters would propagate directly into uncertainties on the Fermat potential and time-delay distance, potentially limiting the cosmographic precision achievable in such an inference.

\section{\textbf{Summary}}
\label{sec:discussion}

In this work, we have presented the first uniform multi-band gravitational lens modeling analysis of eight doubly imaged quasars observed with the Hubble Space Telescope. Using a consistent \lenstronomy~pipeline, we reconstructed the mass and light distributions of each deflector, measured image astrometry and photometry, and created a foundation to determine Fermat potential differences for future time-delay cosmography analyses. The uniform treatment of all systems enables direct comparison across the sample and establishes a scalable framework for modeling doubles in preparation for hierarchical Bayesian inference. Notably, the Einstein radii derived from our models agree with the previous literature within $1.5\sigma$, which is anticipated given that previous models were developed using various mass modeling prescriptions and various ground-based datasets which lack the astrometric resolution and host arc detection that our multi-band HST data provides. As an additional validator of the pipeline, our modeled image separations have an r.m.s difference from Gaia DR2 values of only 3.6 mas.

A central result of this study is the quantitative demonstration of the constraining power of extended arc surface brightness information. We find a strong correlation between lensed arc brightness and Fermat potential precision ($r = 0.97$), indicating that systems with higher signal-to-noise extended arcs yield substantially tighter constraints on the lensing potential. Although the full image modeling includes additional constraints from the joining of the lens mass and light centroids, the strong dependence of precision on arc brightness demonstrates that the primary source of tightening arises from the detailed structure of the extended arcs themselves. We note that this trend is established via an SIE mass modeling framework, and that freeing the radial power-law slope of the lens naturally lessens the Fermat potential precision. Nonetheless, this scaling relation provides a physically intuitive predictor for which doubles are most promising for precision cosmography and a justification for future deep observations. 

In order to further quantify the information gain provided by modeling the full surface brightness distribution of the lens and extended host arcs, we conducted a conjugate point analysis that utilizes only the best-fit quasar image positions from the full image modeling to constrain the lens mass profiles. Through this systematic comparison, we find that full image modeling reduces the mass parameter posterior hypervolume by an average factor of $10^9$ relative to positional constraints alone. In addition, we observe a strong anti-correlation between the mass parameter hypervolume and magnitude of the lensed arcs ($r=-0.87$). The substantial tightening of mass parameters reflects the ability of the extended arcs to constrain the degeneracies between mass ellipticity and external shear that remain largely ungoverned in doubles when only two quasar image positions are available.

As large imaging surveys expand the known population of doubles, the development of uniform modeling pipelines will be necessary for incorporating these systems into hierarchical Bayesian analyses aimed at constraining $\Hzero$. Our results demonstrate that, under the assumption of an SIE mass model, when high resolution imaging resolves the extended host galaxy arcs, doubly imaged systems can yield well-constrained mass models and precise Fermat potential predictions suitable for such cosmographic inferences. This strong dependence of Fermat potential precision on arc brightness directly motivates deeper and higher resolution followup observations. More broadly, the standardized outputs of our pipeline are designed to serve as the foundational inputs to a hierarchical Bayesian framework aimed at population-level time-delay cosmography, in which shared physical parameters across the ensemble can be constrained simultaneously to mitigate systematics that may limit isolated lens analyses. Such an approach, currently in development (Huang et al. - in prep), would allow a statistically significant sample of doubles to place precise and independent constraints on the Hubble constant.

\hspace{.2in}

\begin{acknowledgements} \textit{Acknowledgements.} The \lenstronomy~software is open source and freely available at \href{https://github.com/lenstronomy/lenstronomy}{https://github.com/lenstronomy/lenstronomy}. All notebooks detailing the lens modeling procedure and analysis are available at \href{https://github.com/brady-ryan/doubles_modeling}{https://github.com/brady-ryan/doubles\_modeling}. Individual system modeling results, for data analysis reproduction purposes, can be found at \citet{brady_2026_19421439}.

We thank all the friends of the TDCOSMO collaboration for useful feedback that improved this manuscript. Namely, we thank Dominique Sluse, Eric Paic, and Tommaso Treu for their careful assessment of this work as internal TDCOSMO reviewers.

This research is based on observations made with the NASA/ESA Hubble Space Telescope obtained from the Space Telescope Science Institute, which is operated by the Association of Universities for Research in Astronomy, Inc. These observations are associated with program HST-GO-17199. 

R.B. acknowledges support from the NASA New York Space Grant Consortium under award 80NSSC20M0096.

R.B, X-Y.H, and S.B. acknowledge support through HST-GO-17199.

C.L. acknowledges funding from the European Union’s Horizon Europe research and innovation programme under the Marie Sklodovska-Curie grant agreement No. 101105725.

M.M. acknowledges support by the SNSF (Swiss National Science Foundation) through return CH grant P5R5PT\_225598 and Ambizione grant PZ00P2\_223738

V.M. acknowledges support from ANID FONDECYT Regular grant number 1231418, and Centro de Astrof\'{\i}sica de Valpara\'{\i}so CIDI 21.

D.S. acknowledges the support of the Fonds de la Recherche Scientifique-FNRS, Belgium, under grant No. 4.4503.1.

F.C. acknowledges support by the Agencia Estatal de Investigación (AEI), Ministerio de Ciencia, Innovación y Universidades, Spain, under the project PID2024‑155455NB‑I00, within the 2024 Call for Knowledge Generation Projects.

\end{acknowledgements}

\textit{Facilities:} Stony Brook University SeaWulf

\textit{Software:} AstroDrizzle \citep{avila:2015}, emcee \citep{emcee}, Jupyter \citep{jupyter}, \lenstronomy~\citep{birrer:2018, lenstronomy_two}, matplotlib \citep{Hunter:2007}, Numpy \citep{numpy2}, psfr \citep{psfr}, SExtractor \citep{sextractor}, STARRED \citep{millon:2024}.

\bibliography{sources} 

@article{anguita:2018,
    author = {Anguita, T and Schechter, P L and Kuropatkin, N and Morgan, N D and Ostrovski, F and Abramson, L E and Agnello, A and Apostolovski, Y and Fassnacht, C D and Hsueh, J W and Motta, V and Rojas, K and Rusu, C E and Treu, T and Williams, P and Auger, M and Buckley-Geer, E and Lin, H and McMahon, R and Abbott, T M C and Allam, S and Annis, J and Bernstein, R A and Bertin, E and Brooks, D and Burke, D L and Carnero Rosell, A and Carrasco-Kind, M and Carretero, J and Cunha, C E and D’Andrea, C B and De Vicente, J and DePoy, D L and Desai, S and Diehl, H T and Doel, P and Flaugher, B and García-Bellido, J and Gerdes, D W and Gruen, D and Gruendl, R A and Gschwend, J and Hartley, W G and Hollowood, D L and Honscheid, K and James, D J and Kuehn, K and Lima, M and Maia, M A G and Miquel, R and Plazas, A A and Sanchez, E and Scarpine, V and Smith, M and Soares-Santos, M and Sobreira, F and Suchyta, E and Tarle, G and Walker, A R},
    title = {The STRong lensing Insights into the Dark Energy Survey (STRIDES) 2016 follow-up campaign – II. New quasar lenses from double component fitting},
    journal = {Monthly Notices of the Royal Astronomical Society},
    volume = {480},
    number = {4},
    pages = {5017-5028},
    year = {2018},
    month = {08},
    issn = {0035-8711},
    doi = {10.1093/mnras/sty2172},
    url = {https://doi.org/10.1093/mnras/sty2172},
    eprint = {https://academic.oup.com/mnras/article-pdf/480/4/5017/25595899/sty2172.pdf},
}

@article{treu:2018,
    author = {Treu, T and Agnello, A and Baumer, M A and Birrer, S and Buckley-Geer, E J and Courbin, F and Kim, Y J and Lin, H and Marshall, P J and Nord, B and Schechter, P L and Sivakumar, P R and Abramson, L E and Anguita, T and Apostolovski, Y and Auger, M W and Chan, J H H and Chen, G C F and Collett, T E and Fassnacht, C D and Hsueh, J-W and Lemon, C and McMahon, R G and Motta, V and Ostrovski, F and Rojas, K and Rusu, C E and Williams, P and Frieman, J and Meylan, G and Suyu, S H and Abbott, T M C and Abdalla, F B and Allam, S and Annis, J and Avila, S and Banerji, M and Brooks, D and Rosell, A Carnero and Kind, M Carrasco and Carretero, J and Castander, F J and D’Andrea, C B and da Costa, L N and De Vicente, J and Doel, P and Eifler, T F and Flaugher, B and Fosalba, P and García-Bellido, J and Goldstein, D A and Gruen, D and Gruendl, R A and Gutierrez, G and Hartley, W G and Hollowood, D and Honscheid, K and James, D J and Kuehn, K and Kuropatkin, N and Lima, M and Maia, M A G and Martini, P and Menanteau, F and Miquel, R and Plazas, A A and Romer, A K and Sanchez, E and Scarpine, V and Schindler, R and Schubnell, M and Sevilla-Noarbe, I and Smith, M and Smith, R C and Soares-Santos, M and Sobreira, F and Suchyta, E and Swanson, M E C and Tarle, G and Thomas, D and Tucker, D L and Walker, A R},
    title = {The STRong lensing Insights into the Dark Energy Survey (STRIDES) 2016 follow-up campaign – I. Overview and classification of candidates selected by two techniques},
    journal = {Monthly Notices of the Royal Astronomical Society},
    volume = {481},
    number = {1},
    pages = {1041-1054},
    year = {2018},
    month = {08},
    issn = {0035-8711},
    doi = {10.1093/mnras/sty2329},
    url = {https://doi.org/10.1093/mnras/sty2329},
    eprint = {https://academic.oup.com/mnras/article-pdf/481/1/1041/25699627/sty2329.pdf},
}

@ARTICLE{inada:2006,
       author = {{Inada}, Naohisa and {Oguri}, Masamune and {Becker}, Robert H. and {White}, Richard L. and {Gregg}, Michael D. and {Schechter}, Paul L. and {Kawano}, Yozo and {Kochanek}, Christopher S. and {Richards}, Gordon T. and {Schneider}, Donald P. and {Barentine}, J.~C. and {Brewington}, Howard J. and {Brinkmann}, J. and {Harvanek}, Michael and {Kleinman}, S.~J. and {Krzesinski}, Jurek and {Long}, Dan and {Neilsen}, Jr., Eric H. and {Nitta}, Atsuko and {Snedden}, Stephanie A. and {York}, Donald G.},
        title = "{SDSS J0806+2006 and SDSS J1353+1138: Two New Gravitationally Lensed Quasars from the Sloan Digital Sky Survey}",
      journal = {\aj},
         year = 2006,
        month = apr,
       volume = {131},
       number = {4},
        pages = {1934-1941},
          doi = {10.1086/500591},
archivePrefix = {arXiv},
       eprint = {astro-ph/0512239},
 primaryClass = {astro-ph},
       adsurl = {https://ui.adsabs.harvard.edu/abs/2006AJ....131.1934I}
}

@ARTICLE{shalyapin:2019,
       author = {{Shalyapin}, Vyacheslav N. and {Goicoechea}, Luis J.},
        title = "{Gravitationally Lensed Quasar SDSS J1442+4055: Redshifts of Lensing Galaxies, Time Delay, Microlensing Variability, and Intervening Metal System at z {\ensuremath{\sim}} 2}",
      journal = {\apj},
         year = 2019,
        month = mar,
       volume = {873},
       number = {2},
          eid = {117},
        pages = {117},
          doi = {10.3847/1538-4357/ab08f0},
archivePrefix = {arXiv},
       eprint = {1903.06505},
 primaryClass = {astro-ph.GA},
       adsurl = {https://ui.adsabs.harvard.edu/abs/2019ApJ...873..117S}
}

@ARTICLE{shalyapin:2017,
       author = {{Shalyapin}, Vyacheslav N. and {Goicoechea}, Luis J.},
        title = "{Doubly Imaged Quasar SDSS J1515+1511: Time Delay and Lensing Galaxies}",
      journal = {\apj},
         year = 2017,
        month = feb,
       volume = {836},
       number = {1},
          eid = {14},
        pages = {14},
          doi = {10.3847/1538-4357/836/1/14},
archivePrefix = {arXiv},
       eprint = {1701.04272},
 primaryClass = {astro-ph.GA},
       adsurl = {https://ui.adsabs.harvard.edu/abs/2017ApJ...836...14S}
}

@ARTICLE{kayo:2010,
       author = {{Kayo}, Issha and {Inada}, Naohisa and {Oguri}, Masamune and {Morokuma}, Tomoki and {Hall}, Patrick B. and {Kochanek}, Christopher S. and {Schneider}, Donald P.},
        title = "{Eight New Quasar Lenses from the Sloan Digital Sky Survey Quasar Lens Search}",
      journal = {\aj},
         year = 2010,
        month = apr,
       volume = {139},
       number = {4},
        pages = {1614-1621},
          doi = {10.1088/0004-6256/139/4/1614},
archivePrefix = {arXiv},
       eprint = {0912.1462},
 primaryClass = {astro-ph.CO},
       adsurl = {https://ui.adsabs.harvard.edu/abs/2010AJ....139.1614K}
}

@ARTICLE{ostrovski:2017,
       author = {{Ostrovski}, Fernanda and {McMahon}, Richard G. and {Connolly}, Andrew J. and {Lemon}, Cameron A. and {Auger}, Matthew W. and {Banerji}, Manda and {Hung}, Johnathan M. and {Koposov}, Sergey E. and {Lidman}, Christopher E. and {Reed}, Sophie L. and {Allam}, Sahar and {Benoit-L{\'e}vy}, Aur{\'e}lien and {Bertin}, Emmanuel and {Brooks}, David and {Buckley-Geer}, Elizabeth and {Carnero Rosell}, Aurelio and {Carrasco Kind}, Matias and {Carretero}, Jorge and {Cunha}, Carlos E. and {da Costa}, Luiz N. and {Desai}, Shantanu and {Diehl}, H. Thomas and {Dietrich}, J{\"o}rg P. and {Evrard}, August E. and {Finley}, David A. and {Flaugher}, Brenna and {Fosalba}, Pablo and {Frieman}, Josh and {Gerdes}, David W. and {Goldstein}, Daniel A. and {Gruen}, Daniel and {Gruendl}, Robert A. and {Gutierrez}, Gaston and {Honscheid}, Klaus and {James}, David J. and {Kuehn}, Kyler and {Kuropatkin}, Nikolay and {Lima}, Marcos and {Lin}, Huan and {Maia}, Marcio A.~G. and {Marshall}, Jennifer L. and {Martini}, Paul and {Melchior}, Peter and {Miquel}, Ramon and {Ogando}, Ricardo and {Plazas Malag{\'o}n}, Andr{\'e}s and {Reil}, Kevin and {Romer}, Kathy and {Sanchez}, Eusebio and {Santiago}, Basilio and {Scarpine}, Vic and {Sevilla-Noarbe}, Ignacio and {Soares-Santos}, Marcelle and {Sobreira}, Flavia and {Suchyta}, Eric and {Tarle}, Gregory and {Thomas}, Daniel and {Tucker}, Douglas L. and {Walker}, Alistair R.},
        title = "{VDES J2325-5229 a z = 2.7 gravitationally lensed quasar discovered using morphology-independent supervised machine learning}",
      journal = {\mnras},
         year = 2017,
        month = mar,
       volume = {465},
       number = {4},
        pages = {4325-4334},
          doi = {10.1093/mnras/stw2958},
archivePrefix = {arXiv},
       eprint = {1607.01391},
 primaryClass = {astro-ph.GA},
       adsurl = {https://ui.adsabs.harvard.edu/abs/2017MNRAS.465.4325O}
}

@article{birrer:2018,
title = {lenstronomy: Multi-purpose gravitational lens modelling software package},
journal = {Physics of the Dark Universe},
volume = {22},
pages = {189-201},
year = {2018},
issn = {2212-6864},
doi = {https://doi.org/10.1016/j.dark.2018.11.002},
url = {https://www.sciencedirect.com/science/article/pii/S2212686418301869},
author = {Simon Birrer and Adam Amara}
}

@MISC{avila:2015,
       author = {{Avila}, Roberto and {Koekemoer}, Anton and {Mack}, Jennifer and {Fruchter}, Andrew},
        title = "{Optimizing pixfrac in Astrodrizzle: An example from the Hubble Frontier Fields}",
 howpublished = {Instrument Science Report WFC3 2015-04, 17 pages},
         year = 2015,
        month = mar,
        pages = {4},
       adsurl = {https://ui.adsabs.harvard.edu/abs/2015wfc..rept....4A}
}

@ARTICLE{millon:2024,
       author = {{Millon}, Martin and {Michalewicz}, Kevin and {Dux}, Fr{\'e}d{\'e}ric and {Courbin}, Fr{\'e}d{\'e}ric and {Marshall}, Philip J.},
        title = "{Image Deconvolution and Point-spread Function Reconstruction with STARRED: A Wavelet-based Two-channel Method Optimized for Light-curve Extraction}",
      journal = {\aj},
         year = 2024,
        month = aug,
       volume = {168},
       number = {2},
          eid = {55},
        pages = {55},
          doi = {10.3847/1538-3881/ad4da7},
archivePrefix = {arXiv},
       eprint = {2402.08725},
 primaryClass = {astro-ph.IM},
       adsurl = {https://ui.adsabs.harvard.edu/abs/2024AJ....168...55M}
}

@ARTICLE{planck:2018,
       author = {{Planck Collaboration} and {Aghanim}, N. and {Akrami}, Y. and {Ashdown}, M. and {Aumont}, J. and {Baccigalupi}, C. and {Ballardini}, M. and {Banday}, A.~J. and {Barreiro}, R.~B. and {Bartolo}, N. and {Basak}, S. and {Battye}, R. and {Benabed}, K. and {Bernard}, J. -P. and {Bersanelli}, M. and {Bielewicz}, P. and {Bock}, J.~J. and {Bond}, J.~R. and {Borrill}, J. and {Bouchet}, F.~R. and {Boulanger}, F. and {Bucher}, M. and {Burigana}, C. and {Butler}, R.~C. and {Calabrese}, E. and {Cardoso}, J. -F. and {Carron}, J. and {Challinor}, A. and {Chiang}, H.~C. and {Chluba}, J. and {Colombo}, L.~P.~L. and {Combet}, C. and {Contreras}, D. and {Crill}, B.~P. and {Cuttaia}, F. and {de Bernardis}, P. and {de Zotti}, G. and {Delabrouille}, J. and {Delouis}, J. -M. and {Di Valentino}, E. and {Diego}, J.~M. and {Dor{\'e}}, O. and {Douspis}, M. and {Ducout}, A. and {Dupac}, X. and {Dusini}, S. and {Efstathiou}, G. and {Elsner}, F. and {En{\ss}lin}, T.~A. and {Eriksen}, H.~K. and {Fantaye}, Y. and {Farhang}, M. and {Fergusson}, J. and {Fernandez-Cobos}, R. and {Finelli}, F. and {Forastieri}, F. and {Frailis}, M. and {Fraisse}, A.~A. and {Franceschi}, E. and {Frolov}, A. and {Galeotta}, S. and {Galli}, S. and {Ganga}, K. and {G{\'e}nova-Santos}, R.~T. and {Gerbino}, M. and {Ghosh}, T. and {Gonz{\'a}lez-Nuevo}, J. and {G{\'o}rski}, K.~M. and {Gratton}, S. and {Gruppuso}, A. and {Gudmundsson}, J.~E. and {Hamann}, J. and {Handley}, W. and {Hansen}, F.~K. and {Herranz}, D. and {Hildebrandt}, S.~R. and {Hivon}, E. and {Huang}, Z. and {Jaffe}, A.~H. and {Jones}, W.~C. and {Karakci}, A. and {Keih{\"a}nen}, E. and {Keskitalo}, R. and {Kiiveri}, K. and {Kim}, J. and {Kisner}, T.~S. and {Knox}, L. and {Krachmalnicoff}, N. and {Kunz}, M. and {Kurki-Suonio}, H. and {Lagache}, G. and {Lamarre}, J. -M. and {Lasenby}, A. and {Lattanzi}, M. and {Lawrence}, C.~R. and {Le Jeune}, M. and {Lemos}, P. and {Lesgourgues}, J. and {Levrier}, F. and {Lewis}, A. and {Liguori}, M. and {Lilje}, P.~B. and {Lilley}, M. and {Lindholm}, V. and {L{\'o}pez-Caniego}, M. and {Lubin}, P.~M. and {Ma}, Y. -Z. and {Mac{\'\i}as-P{\'e}rez}, J.~F. and {Maggio}, G. and {Maino}, D. and {Mandolesi}, N. and {Mangilli}, A. and {Marcos-Caballero}, A. and {Maris}, M. and {Martin}, P.~G. and {Martinelli}, M. and {Mart{\'\i}nez-Gonz{\'a}lez}, E. and {Matarrese}, S. and {Mauri}, N. and {McEwen}, J.~D. and {Meinhold}, P.~R. and {Melchiorri}, A. and {Mennella}, A. and {Migliaccio}, M. and {Millea}, M. and {Mitra}, S. and {Miville-Desch{\^e}nes}, M. -A. and {Molinari}, D. and {Montier}, L. and {Morgante}, G. and {Moss}, A. and {Natoli}, P. and {N{\o}rgaard-Nielsen}, H.~U. and {Pagano}, L. and {Paoletti}, D. and {Partridge}, B. and {Patanchon}, G. and {Peiris}, H.~V. and {Perrotta}, F. and {Pettorino}, V. and {Piacentini}, F. and {Polastri}, L. and {Polenta}, G. and {Puget}, J. -L. and {Rachen}, J.~P. and {Reinecke}, M. and {Remazeilles}, M. and {Renzi}, A. and {Rocha}, G. and {Rosset}, C. and {Roudier}, G. and {Rubi{\~n}o-Mart{\'\i}n}, J.~A. and {Ruiz-Granados}, B. and {Salvati}, L. and {Sandri}, M. and {Savelainen}, M. and {Scott}, D. and {Shellard}, E.~P.~S. and {Sirignano}, C. and {Sirri}, G. and {Spencer}, L.~D. and {Sunyaev}, R. and {Suur-Uski}, A. -S. and {Tauber}, J.~A. and {Tavagnacco}, D. and {Tenti}, M. and {Toffolatti}, L. and {Tomasi}, M. and {Trombetti}, T. and {Valenziano}, L. and {Valiviita}, J. and {Van Tent}, B. and {Vibert}, L. and {Vielva}, P. and {Villa}, F. and {Vittorio}, N. and {Wandelt}, B.~D. and {Wehus}, I.~K. and {White}, M. and {White}, S.~D.~M. and {Zacchei}, A. and {Zonca}, A.},
        title = "{Planck 2018 results. VI. Cosmological parameters}",
      journal = {\aap},
         year = 2020,
        month = sep,
       volume = {641},
          eid = {A6},
        pages = {A6},
          doi = {10.1051/0004-6361/201833910},
archivePrefix = {arXiv},
       eprint = {1807.06209},
 primaryClass = {astro-ph.CO},
       adsurl = {https://ui.adsabs.harvard.edu/abs/2020A&A...641A...6P}
}

@ARTICLE{riess:2022,
       author = {{Riess}, Adam G. and {Yuan}, Wenlong and {Macri}, Lucas M. and {Scolnic}, Dan and {Brout}, Dillon and {Casertano}, Stefano and {Jones}, David O. and {Murakami}, Yukei and {Anand}, Gagandeep S. and {Breuval}, Louise and {Brink}, Thomas G. and {Filippenko}, Alexei V. and {Hoffmann}, Samantha and {Jha}, Saurabh W. and {D'arcy Kenworthy}, W. and {Mackenty}, John and {Stahl}, Benjamin E. and {Zheng}, WeiKang},
        title = "{A Comprehensive Measurement of the Local Value of the Hubble Constant with 1 km s$^{-1}$ Mpc$^{-1}$ Uncertainty from the Hubble Space Telescope and the SH0ES Team}",
      journal = {\apjl},
         year = 2022,
        month = jul,
       volume = {934},
       number = {1},
          eid = {L7},
        pages = {L7},
          doi = {10.3847/2041-8213/ac5c5b},
archivePrefix = {arXiv},
       eprint = {2112.04510},
 primaryClass = {astro-ph.CO},
       adsurl = {https://ui.adsabs.harvard.edu/abs/2022ApJ...934L...7R}
}

@ARTICLE{divalentino:2021,
       author = {{Di Valentino}, Eleonora and {Mena}, Olga and {Pan}, Supriya and {Visinelli}, Luca and {Yang}, Weiqiang and {Melchiorri}, Alessandro and {Mota}, David F. and {Riess}, Adam G. and {Silk}, Joseph},
        title = "{In the realm of the Hubble tension-a review of solutions}",
      journal = {Classical and Quantum Gravity},
         year = 2021,
        month = jul,
       volume = {38},
       number = {15},
          eid = {153001},
        pages = {153001},
          doi = {10.1088/1361-6382/ac086d},
archivePrefix = {arXiv},
       eprint = {2103.01183},
 primaryClass = {astro-ph.CO},
       adsurl = {https://ui.adsabs.harvard.edu/abs/2021CQGra..38o3001D}
}

@ARTICLE{refsdal:1964,
       author = {{Refsdal}, S.},
        title = "{On the possibility of determining Hubble's parameter and the masses of galaxies from the gravitational lens effect}",
      journal = {\mnras},
         year = 1964,
        month = jan,
       volume = {128},
        pages = {307},
          doi = {10.1093/mnras/128.4.307},
       adsurl = {https://ui.adsabs.harvard.edu/abs/1964MNRAS.128..307R}
}

@ARTICLE{wong:2020,
       author = {{Wong}, Kenneth C. and {Suyu}, Sherry H. and {Chen}, Geoff C. -F. and {Rusu}, Cristian E. and {Millon}, Martin and {Sluse}, Dominique and {Bonvin}, Vivien and {Fassnacht}, Christopher D. and {Taubenberger}, Stefan and {Auger}, Matthew W. and {Birrer}, Simon and {Chan}, James H.~H. and {Courbin}, Frederic and {Hilbert}, Stefan and {Tihhonova}, Olga and {Treu}, Tommaso and {Agnello}, Adriano and {Ding}, Xuheng and {Jee}, Inh and {Komatsu}, Eiichiro and {Shajib}, Anowar J. and {Sonnenfeld}, Alessandro and {Blandford}, Roger D. and {Koopmans}, L{\'e}on V.~E. and {Marshall}, Philip J. and {Meylan}, Georges},
        title = "{H0LiCOW - XIII. A 2.4 per cent measurement of H$_{0}$ from lensed quasars: 5.3{\ensuremath{\sigma}} tension between early- and late-Universe probes}",
      journal = {\mnras},
         year = 2020,
        month = oct,
       volume = {498},
       number = {1},
        pages = {1420-1439},
          doi = {10.1093/mnras/stz3094},
archivePrefix = {arXiv},
       eprint = {1907.04869},
 primaryClass = {astro-ph.CO},
       adsurl = {https://ui.adsabs.harvard.edu/abs/2020MNRAS.498.1420W}
}

@ARTICLE{emcee,
       author = {{Foreman-Mackey}, Daniel and {Hogg}, David W. and {Lang}, Dustin and {Goodman}, Jonathan},
        title = "{emcee: The MCMC Hammer}",
      journal = {\pasp},
         year = 2013,
        month = mar,
       volume = {125},
       number = {925},
        pages = {306},
          doi = {10.1086/670067},
archivePrefix = {arXiv},
       eprint = {1202.3665},
 primaryClass = {astro-ph.IM},
       adsurl = {https://ui.adsabs.harvard.edu/abs/2013PASP..125..306F}
}

@article{Hunter:2007,
 author = {Hunter, John D.},
 title = {Matplotlib: A 2D Graphics Environment},
 journal = {Comput. Sci. Eng.},
 issue_date = {May 2007},
 volume = {9},
 number = {3},
 month = may,
 year = {2007},
 issn = {1521-9615},
 pages = {90-95},
 numpages = {6},
 url = {https://doi.org/10.1109/mcse.2007.55},
 doi = {10.1109/mcse.2007.55},
 acmid = {1251845},
 publisher = {Institute of Electrical and Electronics Engineers (IEEE)},
 address = {Piscataway, NJ, USA},
 source = {Crossref},
}

@article{numpy2,
 author = {van der Walt, Stéfan and Colbert, S Chris and Varoquaux, Gaël},
 title = {The NumPy Array: A Structure for Efficient Numerical Computation},
 journal = {Comput. Sci. Eng.},
 volume = {13},
 number = {2},
 pages = {22-30},
 year = {2011},
 doi = {10.1109/mcse.2011.37},
 url = {https://doi.org/10.1109/mcse.2011.37},
 eprint = {https://aip.scitation.org/doi/pdf/10.1109/MCSE.2011.37},
 source = {Crossref},
 publisher = {Institute of Electrical and Electronics Engineers (IEEE)},
 issn = {1521-9615},
 month = mar,
}

@conference{jupyter, 
Title = {Jupyter Notebooks -- a publishing format for reproducible computational workflows}, 
Author = {Thomas Kluyver and Benjamin Ragan-Kelley and Fernando P{\'e}rez and Brian Granger and Matthias Bussonnier and Jonathan Frederic and Kyle Kelley and Jessica Hamrick and Jason Grout and Sylvain Corlay and Paul Ivanov and Dami{\'a}n Avila and Safia Abdalla and Carol Willing}, 
Booktitle = {Positioning and Power in Academic Publishing: Players, Agents and Agendas}, 
Editor = {F. Loizides and B. Schmidt}, 
Organization = {IOS Press}, 
Pages = {87 - 90}, 
Year = {2016} 
}

@ARTICLE{sextractor,
       author = {{Bertin}, E. and {Arnouts}, S.},
        title = "{SExtractor: Software for source extraction.}",
      journal = {\aaps},
         year = 1996,
        month = jun,
       volume = {117},
        pages = {393-404},
          doi = {10.1051/aas:1996164},
       adsurl = {https://ui.adsabs.harvard.edu/abs/1996A&AS..117..393B}
}

@ARTICLE{Birrer:2019,
       author = {{Birrer}, S. and {Treu}, T. and {Rusu}, C.~E. and {Bonvin}, V. and {Fassnacht}, C.~D. and {Chan}, J.~H.~H. and {Agnello}, A. and {Shajib}, A.~J. and {Chen}, G.~C. -F. and {Auger}, M. and {Courbin}, F. and {Hilbert}, S. and {Sluse}, D. and {Suyu}, S.~H. and {Wong}, K.~C. and {Marshall}, P. and {Lemaux}, B.~C. and {Meylan}, G.},
        title = "{H0LiCOW - IX. Cosmographic analysis of the doubly imaged quasar SDSS 1206+4332 and a new measurement of the Hubble constant}",
      journal = {\mnras},
         year = 2019,
        month = apr,
       volume = {484},
       number = {4},
        pages = {4726-4753},
          doi = {10.1093/mnras/stz200},
archivePrefix = {arXiv},
       eprint = {1809.01274},
 primaryClass = {astro-ph.CO},
       adsurl = {https://ui.adsabs.harvard.edu/abs/2019MNRAS.484.4726B}
}

@ARTICLE{birrer:2024,
       author = {{Birrer}, S. and {Millon}, M. and {Sluse}, D. and {Shajib}, A.~J. and {Courbin}, F. and {Erickson}, S. and {Koopmans}, L.~V.~E. and {Suyu}, S.~H. and {Treu}, T.},
        title = "{Time-Delay Cosmography: Measuring the Hubble Constant and Other Cosmological Parameters with Strong Gravitational Lensing}",
      journal = {\ssr},
         year = 2024,
        month = aug,
       volume = {220},
       number = {5},
          eid = {48},
        pages = {48},
          doi = {10.1007/s11214-024-01079-w},
archivePrefix = {arXiv},
       eprint = {2210.10833},
 primaryClass = {astro-ph.CO},
       adsurl = {https://ui.adsabs.harvard.edu/abs/2024SSRv..220...48B}
}

@ARTICLE{schmidt:2023,
       author = {{Schmidt}, T. and {Treu}, T. and {Birrer}, S. and {Shajib}, A.~J. and {Lemon}, C. and {Millon}, M. and {Sluse}, D. and {Agnello}, A. and {Anguita}, T. and {Auger-Williams}, M.~W. and {McMahon}, R.~G. and {Motta}, V. and {Schechter}, P. and {Spiniello}, C. and {Kayo}, I. and {Courbin}, F. and {Ertl}, S. and {Fassnacht}, C.~D. and {Frieman}, J.~A. and {More}, A. and {Schuldt}, S. and {Suyu}, S.~H. and {Aguena}, M. and {Andrade-Oliveira}, F. and {Annis}, J. and {Bacon}, D. and {Bertin}, E. and {Brooks}, D. and {Burke}, D.~L. and {Carnero Rosell}, A. and {Carrasco Kind}, M. and {Carretero}, J. and {Conselice}, C. and {Costanzi}, M. and {da Costa}, L.~N. and {Pereira}, M.~E.~S. and {De Vicente}, J. and {Desai}, S. and {Doel}, P. and {Everett}, S. and {Ferrero}, I. and {Friedel}, D. and {Garc{\'\i}a-Bellido}, J. and {Gaztanaga}, E. and {Gruen}, D. and {Gruendl}, R.~A. and {Gschwend}, J. and {Gutierrez}, G. and {Hinton}, S.~R. and {Hollowood}, D.~L. and {Honscheid}, K. and {James}, D.~J. and {Kuehn}, K. and {Lahav}, O. and {Menanteau}, F. and {Miquel}, R. and {Palmese}, A. and {Paz-Chinch{\'o}n}, F. and {Pieres}, A. and {Plazas Malag{\'o}n}, A.~A. and {Prat}, J. and {Rodriguez-Monroy}, M. and {Romer}, A.~K. and {Sanchez}, E. and {Scarpine}, V. and {Sevilla-Noarbe}, I. and {Smith}, M. and {Suchyta}, E. and {Tarle}, G. and {To}, C. and {Varga}, T.~N. and {DES Collaboration}},
        title = "{STRIDES: automated uniform models for 30 quadruply imaged quasars}",
      journal = {\mnras},
         year = 2023,
        month = jan,
       volume = {518},
       number = {1},
        pages = {1260-1300},
          doi = {10.1093/mnras/stac2235},
archivePrefix = {arXiv},
       eprint = {2206.04696},
 primaryClass = {astro-ph.CO},
       adsurl = {https://ui.adsabs.harvard.edu/abs/2023MNRAS.518.1260S}
}

@ARTICLE{bekov:2024,
       author = {{Bekov}, D. Kh. and {Akhunov}, T.~A. and {Burkhonov}, O.~A. and {Alimova}, N.~R.},
        title = "{Light Curves of Lensed Components and Time Delay Measurements in the Binary Gravtationally Lensed Quasars SDSS J2124<inline-formula id=``IEq1''><tex-math id=``IEq1\_TeX''>+</tex-math></inline-formula>1632 and SDSS J0806<inline-formula id=``IEq2''><tex-math id=``IEq2\_TeX''>+</tex-math></inline-formula>2006}",
      journal = {Astrophysical Bulletin},
         year = 2024,
        month = mar,
       volume = {79},
       number = {1},
        pages = {15-24},
          doi = {10.1134/S1990341323600278},
       adsurl = {https://ui.adsabs.harvard.edu/abs/2024AstBu..79...15B}
}

@ARTICLE{liao:2017,
       author = {{Liao}, Kai and {Treu}, Tommaso and {Marshall}, Phil and {Fassnacht}, Christopher D. and {Rumbaugh}, Nick and {Dobler}, Gregory and {Aghamousa}, Amir and {Bonvin}, Vivien and {Courbin}, Frederic and {Hojjati}, Alireza and {Jackson}, Neal and {Kashyap}, Vinay and {Rathna Kumar}, S. and {Linder}, Eric and {Mandel}, Kaisey and {Meng}, Xiao-Li and {Meylan}, Georges and {Moustakas}, Leonidas A. and {Prabhu}, Tushar P. and {Romero-Wolf}, Andrew and {Shafieloo}, Arman and {Siemiginowska}, Aneta and {Stalin}, Chelliah S. and {Tak}, Hyungsuk and {Tewes}, Malte and {van Dyk}, David},
        title = "{Strong Lens Time Delay Challenge. II. Results of TDC1}",
      journal = {\apj},
         year = 2015,
        month = feb,
       volume = {800},
       number = {1},
          eid = {11},
        pages = {11},
          doi = {10.1088/0004-637X/800/1/11},
archivePrefix = {arXiv},
       eprint = {1409.1254},
 primaryClass = {astro-ph.IM},
       adsurl = {https://ui.adsabs.harvard.edu/abs/2015ApJ...800...11L}
}

@ARTICLE{oguri:2010,
       author = {{Oguri}, Masamune and {Marshall}, Philip J.},
        title = "{Gravitationally lensed quasars and supernovae in future wide-field optical imaging surveys}",
      journal = {\mnras},
         year = 2010,
        month = jul,
       volume = {405},
       number = {4},
        pages = {2579-2593},
          doi = {10.1111/j.1365-2966.2010.16639.x},
archivePrefix = {arXiv},
       eprint = {1001.2037},
 primaryClass = {astro-ph.CO},
       adsurl = {https://ui.adsabs.harvard.edu/abs/2010MNRAS.405.2579O}
}

@ARTICLE{more:2012,
       author = {{More}, A. and {Cabanac}, R. and {More}, S. and {Alard}, C. and {Limousin}, M. and {Kneib}, J. -P. and {Gavazzi}, R. and {Motta}, V.},
        title = "{The CFHTLS-Strong Lensing Legacy Survey (SL2S): Investigating the Group-scale Lenses with the SARCS Sample}",
      journal = {\apj},
         year = 2012,
        month = apr,
       volume = {749},
       number = {1},
          eid = {38},
        pages = {38},
          doi = {10.1088/0004-637X/749/1/38},
archivePrefix = {arXiv},
       eprint = {1109.1821},
 primaryClass = {astro-ph.CO},
       adsurl = {https://ui.adsabs.harvard.edu/abs/2012ApJ...749...38M}
}

@ARTICLE{weinberg:2013,
       author = {{Weinberg}, David H. and {Mortonson}, Michael J. and {Eisenstein}, Daniel J. and {Hirata}, Christopher and {Riess}, Adam G. and {Rozo}, Eduardo},
        title = "{Observational probes of cosmic acceleration}",
      journal = {\physrep},
         year = 2013,
        month = sep,
       volume = {530},
       number = {2},
        pages = {87-255},
          doi = {10.1016/j.physrep.2013.05.001},
archivePrefix = {arXiv},
       eprint = {1201.2434},
 primaryClass = {astro-ph.CO},
       adsurl = {https://ui.adsabs.harvard.edu/abs/2013PhR...530...87W}
}

@MISC{bajaj:2020,
       author = {{Bajaj}, V. and {Calamida}, A. and {Mack}, J.},
        title = "{Updated WFC3/IR Photometric Calibration}",
 howpublished = {Instrument Science Report WFC3 2020-10, 19 pages},
         year = 2020,
        month = dec,
        pages = {10},
       adsurl = {https://ui.adsabs.harvard.edu/abs/2020wfc..rept...10B}
}

@ARTICLE{bolton:2006,
       author = {{Bolton}, Adam S. and {Burles}, Scott and {Koopmans}, L{\'e}on V.~E. and {Treu}, Tommaso and {Moustakas}, Leonidas A.},
        title = "{The Sloan Lens ACS Survey. I. A Large Spectroscopically Selected Sample of Massive Early-Type Lens Galaxies}",
      journal = {\apj},
         year = 2006,
        month = feb,
       volume = {638},
       number = {2},
        pages = {703-724},
          doi = {10.1086/498884},
archivePrefix = {arXiv},
       eprint = {astro-ph/0511453},
 primaryClass = {astro-ph},
       adsurl = {https://ui.adsabs.harvard.edu/abs/2006ApJ...638..703B}
}

@INPROCEEDINGS{kennedy:1995,
  author={Kennedy, J. and Eberhart, R.},
  booktitle={Proceedings of ICNN'95 - International Conference on Neural Networks}, 
  title={Particle swarm optimization}, 
  year={1995},
  volume={4},
  number={},
  pages={1942-1948 vol.4},
  doi={10.1109/ICNN.1995.488968}}

@ARTICLE{oguri:2005,
       author = {{Oguri}, Masamune and {Inada}, Naohisa and {Hennawi}, Joseph F. and {Richards}, Gordon T. and {Johnston}, David E. and {Frieman}, Joshua A. and {Pindor}, Bartosz and {Strauss}, Michael A. and {Brunner}, Robert J. and {Becker}, Robert H. and {Castander}, Francisco J. and {Gregg}, Michael D. and {Hall}, Patrick B. and {Rix}, Hans-Walter and {Schneider}, Donald P. and {Bahcall}, Neta A. and {Brinkmann}, Jonathan and {York}, Donald G.},
        title = "{Discovery of Two Gravitationally Lensed Quasars with Image Separations of 3'' from the Sloan Digital Sky Survey}",
      journal = {\apj},
         year = 2005,
        month = mar,
       volume = {622},
       number = {1},
        pages = {106-115},
          doi = {10.1086/428087},
archivePrefix = {arXiv},
       eprint = {astro-ph/0411250},
 primaryClass = {astro-ph},
       adsurl = {https://ui.adsabs.harvard.edu/abs/2005ApJ...622..106O}
}

@ARTICLE{richards:2002,
       author = {{Richards}, Gordon T. and {Fan}, Xiaohui and {Newberg}, Heidi Jo and {Strauss}, Michael A. and {Vanden Berk}, Daniel E. and {Schneider}, Donald P. and {Yanny}, Brian and {Boucher}, Adam and {Burles}, Scott and {Frieman}, Joshua A. and {Gunn}, James E. and {Hall}, Patrick B. and {Ivezi{\'c}}, {\v{Z}}eljko and {Kent}, Stephen and {Loveday}, Jon and {Lupton}, Robert H. and {Rockosi}, Constance M. and {Schlegel}, David J. and {Stoughton}, Chris and {SubbaRao}, Mark and {York}, Donald G.},
        title = "{Spectroscopic Target Selection in the Sloan Digital Sky Survey: The Quasar Sample}",
      journal = {\aj},
         year = 2002,
        month = jun,
       volume = {123},
       number = {6},
        pages = {2945-2975},
          doi = {10.1086/340187},
archivePrefix = {arXiv},
       eprint = {astro-ph/0202251},
 primaryClass = {astro-ph},
       adsurl = {https://ui.adsabs.harvard.edu/abs/2002AJ....123.2945R}
}

@ARTICLE{inada:2012,
       author = {{Inada}, Naohisa and {Oguri}, Masamune and {Shin}, Min-Su and {Kayo}, Issha and {Strauss}, Michael A. and {Morokuma}, Tomoki and {Rusu}, Cristian E. and {Fukugita}, Masataka and {Kochanek}, Christopher S. and {Richards}, Gordon T. and {Schneider}, Donald P. and {York}, Donald G. and {Bahcall}, Neta A. and {Frieman}, Joshua A. and {Hall}, Patrick B. and {White}, Richard L.},
        title = "{The Sloan Digital Sky Survey Quasar Lens Search. V. Final Catalog from the Seventh Data Release}",
      journal = {\aj},
         year = 2012,
        month = may,
       volume = {143},
       number = {5},
          eid = {119},
        pages = {119},
          doi = {10.1088/0004-6256/143/5/119},
archivePrefix = {arXiv},
       eprint = {1203.1087},
 primaryClass = {astro-ph.CO},
       adsurl = {https://ui.adsabs.harvard.edu/abs/2012AJ....143..119I}
}

@ARTICLE{dawes:2023,
       author = {{Dawes}, C. and {Storfer}, C. and {Huang}, X. and {Aldering}, G. and {Cikota}, Aleksandar and {Dey}, Arjun and {Schlegel}, D.~J.},
        title = "{Finding Multiply Lensed and Binary Quasars in the DESI Legacy Imaging Surveys}",
      journal = {\apjs},
         year = 2023,
        month = dec,
       volume = {269},
       number = {2},
          eid = {61},
        pages = {61},
          doi = {10.3847/1538-4365/ad015a},
archivePrefix = {arXiv},
       eprint = {2208.06356},
 primaryClass = {astro-ph.CO},
       adsurl = {https://ui.adsabs.harvard.edu/abs/2023ApJS..269...61D}
}

@ARTICLE{dux:2025,
       author = {{Dux}, F. and {Millon}, M. and {Galan}, A. and {Paic}, E. and {Lemon}, C. and {Courbin}, F. and {Bonvin}, V. and {Anguita}, T. and {Auger}, M. and {Birrer}, S. and {Buckley-Geer}, E. and {Fassnacht}, C.~D. and {Frieman}, J. and {McMahon}, R.~G. and {Marshall}, P.~J. and {Melo}, A. and {Motta}, V. and {Neira}, F. and {Sluse}, D. and {Suyu}, S.~H. and {Treu}, T. and {Agnello}, A. and {{\'A}vila}, F. and {Chan}, J. and {Chijani}, M. and {Rojas}, K. and {Hempel}, A. and {Hempel}, M. and {Kim}, S. and {Eigenthaler}, P. and {Lachaume}, R. and {Rabus}, M.},
        title = "{TDCOSMO: XVII. New time delays in 22 lensed quasars from optical monitoring with the ESO-VST 2.6m and MPG 2.2m telescopes}",
      journal = {\aap},
         year = 2025,
        month = may,
       volume = {697},
          eid = {A139},
        pages = {A139},
          doi = {10.1051/0004-6361/202553807},
archivePrefix = {arXiv},
       eprint = {2504.02932},
 primaryClass = {astro-ph.CO},
       adsurl = {https://ui.adsabs.harvard.edu/abs/2025A&A...697A.139D}
}

@ARTICLE{lenstronomy_two,
       author = {{Birrer}, Simon and {Shajib}, Anowar and {Gilman}, Daniel and {Galan}, Aymeric and {Aalbers}, Jelle and {Millon}, Martin and {Morgan}, Robert and {Pagano}, Giulia and {Park}, Ji and {Teodori}, Luca and {Tessore}, Nicolas and {Ueland}, Madison and {Van de Vyvere}, Lyne and {Wagner-Carena}, Sebastian and {Wempe}, Ewoud and {Yang}, Lilan and {Ding}, Xuheng and {Schmidt}, Thomas and {Sluse}, Dominique and {Zhang}, Ming and {Amara}, Adam},
        title = "{lenstronomy II: A gravitational lensing software ecosystem}",
      journal = {The Journal of Open Source Software},
         year = 2021,
        month = jun,
       volume = {6},
       number = {62},
          eid = {3283},
        pages = {3283},
          doi = {10.21105/joss.03283},
archivePrefix = {arXiv},
       eprint = {2106.05976},
 primaryClass = {astro-ph.CO},
       adsurl = {https://ui.adsabs.harvard.edu/abs/2021JOSS....6.3283B}
}

@software{psfr,
       author = {{Birrer}, Simon and {Bhamre}, Vikram and {Nierenberg}, Anna and {Yang}, Lilan and {Van de Vyvere}, Lyne},
        title = "{PSFr: Point Spread Function reconstruction}",
 howpublished = {Astrophysics Source Code Library, record ascl:2210.005},
         year = 2022,
        month = oct,
          eid = {ascl:2210.005},
archivePrefix = {ascl},
       eprint = {2210.005},
       adsurl = {https://ui.adsabs.harvard.edu/abs/2022ascl.soft10005B}
}

@ARTICLE{sluse:2008,
       author = {{Sluse}, D. and {Courbin}, F. and {Eigenbrod}, A. and {Meylan}, G.},
        title = "{A sharp look at the gravitationally lensed quasar SDSS J0806+2006{\enskip}with laser guide star adaptive optics at the VLT}",
      journal = {\aap},
         year = 2008,
        month = dec,
       volume = {492},
       number = {2},
        pages = {L39-L42},
          doi = {10.1051/0004-6361:200810977},
archivePrefix = {arXiv},
       eprint = {0809.2980},
 primaryClass = {astro-ph},
       adsurl = {https://ui.adsabs.harvard.edu/abs/2008A&A...492L..39S}
}

@article{sergeyev:2015,
    author = {Sergeyev, A. V. and Zheleznyak, A. P. and Shalyapin, V. N. and Goicoechea, L. J.},
    title = {Discovery of the optically bright, wide separation double quasar SDSS J1442+4055},
    journal = {Monthly Notices of the Royal Astronomical Society},
    volume = {456},
    number = {2},
    pages = {1948-1954},
    year = {2015},
    month = {12},
    issn = {0035-8711},
    doi = {10.1093/mnras/stv2763},
    url = {https://doi.org/10.1093/mnras/stv2763},
    eprint = {https://academic.oup.com/mnras/article-pdf/456/2/1948/18514279/stv2763.pdf},
}

@ARTICLE{gavazzi:2007,
       author = {{Gavazzi}, Rapha{\"e}l and {Treu}, Tommaso and {Rhodes}, Jason D. and {Koopmans}, L{\'e}on V.~E. and {Bolton}, Adam S. and {Burles}, Scott and {Massey}, Richard J. and {Moustakas}, Leonidas A.},
        title = "{The Sloan Lens ACS Survey. IV. The Mass Density Profile of Early-Type Galaxies out to 100 Effective Radii}",
      journal = {\apj},
         year = 2007,
        month = sep,
       volume = {667},
       number = {1},
        pages = {176-190},
          doi = {10.1086/519237},
archivePrefix = {arXiv},
       eprint = {astro-ph/0701589},
 primaryClass = {astro-ph},
       adsurl = {https://ui.adsabs.harvard.edu/abs/2007ApJ...667..176G}
}

@ARTICLE{auger:2010,
       author = {{Auger}, M.~W. and {Treu}, T. and {Bolton}, A.~S. and {Gavazzi}, R. and {Koopmans}, L.~V.~E. and {Marshall}, P.~J. and {Moustakas}, L.~A. and {Burles}, S.},
        title = "{The Sloan Lens ACS Survey. X. Stellar, Dynamical, and Total Mass Correlations of Massive Early-type Galaxies}",
      journal = {\apj},
         year = 2010,
        month = nov,
       volume = {724},
       number = {1},
        pages = {511-525},
          doi = {10.1088/0004-637X/724/1/511},
archivePrefix = {arXiv},
       eprint = {1007.2880},
 primaryClass = {astro-ph.CO},
       adsurl = {https://ui.adsabs.harvard.edu/abs/2010ApJ...724..511A}
}

@article{inada:2014,
doi = {10.1088/0004-6256/147/6/153},
url = {https://doi.org/10.1088/0004-6256/147/6/153},
year = {2014},
month = {may},
publisher = {The American Astronomical Society},
volume = {147},
number = {6},
pages = {153},
author = {Inada, Naohisa and Oguri, Masamune and Rusu, Cristian E. and Kayo, Issha and Morokuma, Tomoki},
title = {DISCOVERY OF FOUR DOUBLY IMAGED QUASAR LENSES FROM THE SLOAN DIGITAL SKY SURVEY},
journal = {The Astronomical Journal}
}

@article{rusu:2016,
    author = {Rusu, Cristian E. and Oguri, Masamune and Minowa, Yosuke and Iye, Masanori and Inada, Naohisa and Oya, Shin and Kayo, Issha and Hayano, Yutaka and Hattori, Masayuki and Saito, Yoshihiko and Ito, Meguru and Pyo, Tae-Soo and Terada, Hiroshi and Takami, Hideki and Watanabe, Makoto},
    title = {Subaru Telescope adaptive optics observations of gravitationally lensed quasars in the Sloan Digital Sky Survey},
    journal = {Monthly Notices of the Royal Astronomical Society},
    volume = {458},
    number = {1},
    pages = {2-55},
    year = {2016},
    month = {03},
    issn = {0035-8711},
    doi = {10.1093/mnras/stw092},
    url = {https://doi.org/10.1093/mnras/stw092},
    eprint = {https://academic.oup.com/mnras/article-pdf/458/1/2/8180472/stw092.pdf},
}

@ARTICLE{etherington:2024,
       author = {{Etherington}, Amy and {Nightingale}, James W. and {Massey}, Richard and {Tam}, Sut-Ieng and {Cao}, XiaoYue and {Niemiec}, Anna and {He}, Qiuhan and {Robertson}, Andrew and {Li}, Ran and {Amvrosiadis}, Aristeidis and {Cole}, Shaun and {Diego}, Jose M. and {Frenk}, Carlos S. and {Frye}, Brenda L. and {Harvey}, David and {Jauzac}, Mathilde and {Koekemoer}, Anton M. and {Lagattuta}, David J. and {Lange}, Samuel and {Limousin}, Marceau and {Mahler}, Guillaume and {Sirks}, Ellen and {Steinhardt}, Charles L.},
        title = "{Strong gravitational lensing's 'external shear' is not shear}",
      journal = {\mnras},
         year = 2024,
        month = jul,
       volume = {531},
       number = {3},
        pages = {3684-3697},
          doi = {10.1093/mnras/stae1375},
archivePrefix = {arXiv},
       eprint = {2301.05244},
 primaryClass = {astro-ph.CO},
       adsurl = {https://ui.adsabs.harvard.edu/abs/2024MNRAS.531.3684E}
}

@ARTICLE{shajib:2021,
       author = {{Shajib}, Anowar J. and {Molina}, Eden and {Agnello}, Adriano and {Williams}, Peter R. and {Birrer}, Simon and {Treu}, Tommaso and {Fassnacht}, Christopher D. and {Morishita}, Takahiro and {Abramson}, Louis and {Schechter}, Paul L. and {Wisotzki}, Lutz},
        title = "{High-resolution imaging follow-up of doubly imaged quasars}",
      journal = {\mnras},
         year = 2021,
        month = may,
       volume = {503},
       number = {2},
        pages = {1557-1567},
          doi = {10.1093/mnras/stab532},
archivePrefix = {arXiv},
       eprint = {2011.01971},
 primaryClass = {astro-ph.GA},
       adsurl = {https://ui.adsabs.harvard.edu/abs/2021MNRAS.503.1557S}
}

@ARTICLE{kumar:2013,
       author = {{Rathna Kumar}, S. and {Tewes}, M. and {Stalin}, C.~S. and {Courbin}, F. and {Asfandiyarov}, I. and {Meylan}, G. and {Eulaers}, E. and {Prabhu}, T.~P. and {Magain}, P. and {Van Winckel}, H. and {Ehgamberdiev}, Sh.},
        title = "{COSMOGRAIL: the COSmological MOnitoring of GRAvItational Lenses. XIV. Time delay of the doubly lensed quasar SDSS J1001+5027}",
      journal = {\aap},
         year = 2013,
        month = sep,
       volume = {557},
          eid = {A44},
        pages = {A44},
          doi = {10.1051/0004-6361/201322116},
archivePrefix = {arXiv},
       eprint = {1306.5105},
 primaryClass = {astro-ph.CO},
       adsurl = {https://ui.adsabs.harvard.edu/abs/2013A&A...557A..44R}
}

@ARTICLE{treu:2022,
       author = {{Treu}, Tommaso and {Suyu}, Sherry H. and {Marshall}, Philip J.},
        title = "{Strong lensing time-delay cosmography in the 2020s}",
      journal = {\aapr},
         year = 2022,
        month = dec,
       volume = {30},
       number = {1},
          eid = {8},
        pages = {8},
          doi = {10.1007/s00159-022-00145-y},
archivePrefix = {arXiv},
       eprint = {2210.15794},
 primaryClass = {astro-ph.CO},
       adsurl = {https://ui.adsabs.harvard.edu/abs/2022A&ARv..30....8T}
}

@ARTICLE{tdcosmo:2025,
       author = {{TDCOSMO Collaboration} and {Birrer}, Simon and {Buckley-Geer}, Elizabeth J. and {Cappellari}, Michele and {Courbin}, Fr{\'e}d{\'e}ric and {Dux}, Fr{\'e}d{\'e}ric and {Fassnacht}, Christopher D. and {Frieman}, Joshua A. and {Galan}, Aymeric and {Gilman}, Daniel and {Huang}, Xiang-Yu and {Knabel}, Shawn and {Langeroodi}, Danial and {Lin}, Huan and {Millon}, Martin and {Morishita}, Takahiro and {Motta}, Veronica and {Mozumdar}, Pritom and {Paic}, Eric and {Shajib}, Anowar J. and {Sheu}, William and {Sluse}, Dominique and {Sonnenfeld}, Alessandro and {Spiniello}, Chiara and {Stiavelli}, Massimo and {Suyu}, Sherry H. and {Tan}, Chin Yi and {Treu}, Tommaso and {van de Vyvere}, Lyne and {Wang}, Han and {Wells}, Patrick and {Williams}, Devon M. and {Wong}, Kenneth C.},
        title = "{TDCOSMO 2025: Cosmological constraints from strong lensing time delays}",
      journal = {\aap},
         year = 2025,
        month = dec,
       volume = {704},
          eid = {A63},
        pages = {A63},
          doi = {10.1051/0004-6361/202555801},
archivePrefix = {arXiv},
       eprint = {2506.03023},
 primaryClass = {astro-ph.CO},
       adsurl = {https://ui.adsabs.harvard.edu/abs/2025A&A...704A..63T}
}

@ARTICLE{paic:2025,
       author = {{Paic}, Eric and {Courbin}, Fr{\'e}d{\'e}ric and {Fassnacht}, Christopher D. and {Galan}, Aymeric and {Millon}, Martin and {Sluse}, Dominique and {Williams}, Devon M. and {Birrer}, Simon and {Buckley-Geer}, Elizabeth J. and {Cappellari}, Michele and {Dux}, Fr{\'e}d{\'e}ric and {Huang}, Xiang-Yu and {Knabel}, Shawn and {Lemon}, Cameron and {Shajib}, Anowar J. and {Suyu}, Sherry H. and {Treu}, Tommaso and {Wong}, Kenneth C. and {Christensen}, Lise and {Motta}, Veronica and {Sonnenfeld}, Alessandro},
        title = "{TDCOSMO. XXIV. Measurement of the Hubble constant from the doubly lensed quasar HE1104-1805}",
      journal = {arXiv e-prints},
         year = 2025,
        month = dec,
          eid = {arXiv:2512.03178},
        pages = {arXiv:2512.03178},
          doi = {10.48550/arXiv.2512.03178},
archivePrefix = {arXiv},
       eprint = {2512.03178},
 primaryClass = {astro-ph.GA},
       adsurl = {https://ui.adsabs.harvard.edu/abs/2025arXiv251203178P}
}

@ARTICLE{pantheon,
       author = {{Scolnic}, Dan and {Brout}, Dillon and {Carr}, Anthony and {Riess}, Adam G. and {Davis}, Tamara M. and {Dwomoh}, Arianna and {Jones}, David O. and {Ali}, Noor and {Charvu}, Pranav and {Chen}, Rebecca and {Peterson}, Erik R. and {Popovic}, Brodie and {Rose}, Benjamin M. and {Wood}, Charlotte M. and {Brown}, Peter J. and {Chambers}, Ken and {Coulter}, David A. and {Dettman}, Kyle G. and {Dimitriadis}, Georgios and {Filippenko}, Alexei V. and {Foley}, Ryan J. and {Jha}, Saurabh W. and {Kilpatrick}, Charles D. and {Kirshner}, Robert P. and {Pan}, Yen-Chen and {Rest}, Armin and {Rojas-Bravo}, Cesar and {Siebert}, Matthew R. and {Stahl}, Benjamin E. and {Zheng}, WeiKang},
        title = "{The Pantheon+ Analysis: The Full Data Set and Light-curve Release}",
      journal = {\apj},
         year = 2022,
        month = oct,
       volume = {938},
       number = {2},
          eid = {113},
        pages = {113},
          doi = {10.3847/1538-4357/ac8b7a},
archivePrefix = {arXiv},
       eprint = {2112.03863},
 primaryClass = {astro-ph.CO},
       adsurl = {https://ui.adsabs.harvard.edu/abs/2022ApJ...938..113S}
}

@ARTICLE{adnan:2025,
       author = {{Adnan}, S.~M. Rafee and {Hasan}, Muhammad Jobair and {Al-Imtiaz}, Ahmad and {Robin}, Sulyman H. and {Shwadhin}, Fahim R. and {Shajib}, Anowar J. and {Nahid}, Mamun Hossain and {Tanver}, Mehedi Hasan and {Akter}, Tanjela and {Jahan}, Nusrath and {Jafar}, Zareef and {Rashid}, Mamunur and {Biswas}, Anik and {Chowdhury}, Akbar Ahmed and {Feardous}, Jannatul and {Rahaman}, Ajmi and {Ridwan}, Masuk and {Sharma}, Rahul D. and {Chowdhury}, Zannat and {Hossain}, Mir Sazzat},
        title = "{Investigating the relation between the environment and internal structure of massive elliptical galaxies using strong lensing}",
      journal = {\aap},
         year = 2025,
        month = jul,
       volume = {699},
          eid = {A259},
        pages = {A259},
          doi = {10.1051/0004-6361/202453239},
archivePrefix = {arXiv},
       eprint = {2412.00361},
 primaryClass = {astro-ph.GA},
       adsurl = {https://ui.adsabs.harvard.edu/abs/2025A&A...699A.259A}
}

@ARTICLE{Treu:2009,
       author = {{Treu}, Tommaso and {Gavazzi}, Rapha{\"e}l and {Gorecki}, Alexia and {Marshall}, Philip J. and {Koopmans}, L{\'e}on V.~E. and {Bolton}, Adam S. and {Moustakas}, Leonidas A. and {Burles}, Scott},
        title = "{The SLACS Survey. VIII. The Relation between Environment and Internal Structure of Early-Type Galaxies}",
      journal = {\apj},
         year = 2009,
        month = jan,
       volume = {690},
       number = {1},
        pages = {670-682},
          doi = {10.1088/0004-637X/690/1/670},
archivePrefix = {arXiv},
       eprint = {0806.1056},
 primaryClass = {astro-ph},
       adsurl = {https://ui.adsabs.harvard.edu/abs/2009ApJ...690..670T}
}

@ARTICLE{Sluse:2012,
       author = {{Sluse}, D. and {Chantry}, V. and {Magain}, P. and {Courbin}, F. and {Meylan}, G.},
        title = "{COSMOGRAIL: the COSmological MOnitoring of GRAvItational Lenses. X. Modeling based on high-precision astrometry of a sample of 25 lensed quasars: consequences for ellipticity, shear, and astrometric anomalies}",
      journal = {\aap},
         year = 2012,
        month = feb,
       volume = {538},
          eid = {A99},
        pages = {A99},
          doi = {10.1051/0004-6361/201015844},
archivePrefix = {arXiv},
       eprint = {1112.0005},
 primaryClass = {astro-ph.CO},
       adsurl = {https://ui.adsabs.harvard.edu/abs/2012A&A...538A..99S}
}

@ARTICLE{Shajib:2019,
       author = {{Shajib}, A.~J. and {Birrer}, S. and {Treu}, T. and {Auger}, M.~W. and {Agnello}, A. and {Anguita}, T. and {Buckley-Geer}, E.~J. and {Chan}, J.~H.~H. and {Collett}, T.~E. and {Courbin}, F. and {Fassnacht}, C.~D. and {Frieman}, J. and {Kayo}, I. and {Lemon}, C. and {Lin}, H. and {Marshall}, P.~J. and {McMahon}, R. and {More}, A. and {Morgan}, N.~D. and {Motta}, V. and {Oguri}, M. and {Ostrovski}, F. and {Rusu}, C.~E. and {Schechter}, P.~L. and {Shanks}, T. and {Suyu}, S.~H. and {Meylan}, G. and {Abbott}, T.~M.~C. and {Allam}, S. and {Annis}, J. and {Avila}, S. and {Bertin}, E. and {Brooks}, D. and {Carnero Rosell}, A. and {Carrasco Kind}, M. and {Carretero}, J. and {Cunha}, C.~E. and {da Costa}, L.~N. and {De Vicente}, J. and {Desai}, S. and {Doel}, P. and {Flaugher}, B. and {Fosalba}, P. and {Garc{\'\i}a-Bellido}, J. and {Gerdes}, D.~W. and {Gruen}, D. and {Gruendl}, R.~A. and {Gutierrez}, G. and {Hartley}, W.~G. and {Hollowood}, D.~L. and {Hoyle}, B. and {James}, D.~J. and {Kuehn}, K. and {Kuropatkin}, N. and {Lahav}, O. and {Lima}, M. and {Maia}, M.~A.~G. and {March}, M. and {Marshall}, J.~L. and {Melchior}, P. and {Menanteau}, F. and {Miquel}, R. and {Plazas}, A.~A. and {Sanchez}, E. and {Scarpine}, V. and {Sevilla-Noarbe}, I. and {Smith}, M. and {Soares-Santos}, M. and {Sobreira}, F. and {Suchyta}, E. and {Swanson}, M.~E.~C. and {Tarle}, G. and {Walker}, A.~R.},
        title = "{Is every strong lens model unhappy in its own way? Uniform modelling of a sample of 13 quadruply+ imaged quasars}",
      journal = {\mnras},
         year = 2019,
        month = mar,
       volume = {483},
       number = {4},
        pages = {5649-5671},
          doi = {10.1093/mnras/sty3397},
archivePrefix = {arXiv},
       eprint = {1807.09278},
 primaryClass = {astro-ph.GA},
       adsurl = {https://ui.adsabs.harvard.edu/abs/2019MNRAS.483.5649S}
}

@ARTICLE{hu:2023,
       author = {{Hu}, Jian-Ping and {Wang}, Fa-Yin},
        title = "{Hubble Tension: The Evidence of New Physics}",
      journal = {Universe},
         year = 2023,
        month = feb,
       volume = {9},
       number = {2},
          eid = {94},
        pages = {94},
          doi = {10.3390/universe9020094},
archivePrefix = {arXiv},
       eprint = {2302.05709},
 primaryClass = {astro-ph.CO},
       adsurl = {https://ui.adsabs.harvard.edu/abs/2023Univ....9...94H}
}

@ARTICLE{perivolaropoulos:2024,
       author = {{Perivolaropoulos}, Leandros},
        title = "{Hubble tension or distance ladder crisis?}",
      journal = {\prd},
         year = 2024,
        month = dec,
       volume = {110},
       number = {12},
          eid = {123518},
        pages = {123518},
          doi = {10.1103/PhysRevD.110.123518},
archivePrefix = {arXiv},
       eprint = {2408.11031},
 primaryClass = {astro-ph.CO},
       adsurl = {https://ui.adsabs.harvard.edu/abs/2024PhRvD.110l3518P}
}

@ARTICLE{luhtaru:2021,
       author = {{Luhtaru}, Richard and {Schechter}, Paul L. and {de Soto}, Kaylee M.},
        title = "{What Makes Quadruply Lensed Quasars Quadruple?}",
      journal = {\apj},
         year = 2021,
        month = jul,
       volume = {915},
       number = {1},
          eid = {4},
        pages = {4},
          doi = {10.3847/1538-4357/abfda1},
archivePrefix = {arXiv},
       eprint = {2102.08470},
 primaryClass = {astro-ph.GA},
       adsurl = {https://ui.adsabs.harvard.edu/abs/2021ApJ...915....4L}
}

@ARTICLE{keeton:1997,
       author = {{Keeton}, C.~R. and {Kochanek}, C.~S. and {Seljak}, U.},
        title = "{Shear and Ellipticity in Gravitational Lenses}",
      journal = {\apj},
         year = 1997,
        month = jun,
       volume = {482},
       number = {2},
        pages = {604-620},
          doi = {10.1086/304172},
archivePrefix = {arXiv},
       eprint = {astro-ph/9610163},
 primaryClass = {astro-ph},
       adsurl = {https://ui.adsabs.harvard.edu/abs/1997ApJ...482..604K}
}

@ARTICLE{more:2016,
       author = {{More}, Anupreeta and {Oguri}, Masamune and {Kayo}, Issha and {Zinn}, Joel and {Strauss}, Michael A. and {Santiago}, Basilio X. and {Mosquera}, Ana M. and {Inada}, Naohisa and {Kochanek}, Christopher S. and {Rusu}, Cristian E. and {Brownstein}, Joel R. and {da Costa}, Luiz N. and {Kneib}, Jean-Paul and {Maia}, Marcio A.~G. and {Quimby}, Robert M. and {Schneider}, Donald P. and {Streblyanska}, Alina and {York}, Donald G.},
        title = "{The SDSS-III BOSS quasar lens survey: discovery of 13 gravitationally lensed quasars}",
      journal = {\mnras},
         year = 2016,
        month = feb,
       volume = {456},
       number = {2},
        pages = {1595-1606},
          doi = {10.1093/mnras/stv2813},
archivePrefix = {arXiv},
       eprint = {1509.07917},
 primaryClass = {astro-ph.GA},
       adsurl = {https://ui.adsabs.harvard.edu/abs/2016MNRAS.456.1595M}
}

@ARTICLE{paris:2014,
       author = {{P{\^a}ris}, Isabelle and {Petitjean}, Patrick and {Aubourg}, {\'E}ric and {Ross}, Nicholas P. and {Myers}, Adam D. and {Streblyanska}, Alina and {Bailey}, Stephen and {Hall}, Patrick B. and {Strauss}, Michael A. and {Anderson}, Scott F. and {Bizyaev}, Dmitry and {Borde}, Arnaud and {Brinkmann}, J. and {Bovy}, Jo and {Brandt}, William N. and {Brewington}, Howard and {Brownstein}, Joel R. and {Cook}, Benjamin A. and {Ebelke}, Garrett and {Fan}, Xiaohui and {Filiz Ak}, Nurten and {Finley}, Hayley and {Font-Ribera}, Andreu and {Ge}, Jian and {Hamann}, Fred and {Ho}, Shirley and {Jiang}, Linhua and {Kinemuchi}, Karen and {Malanushenko}, Elena and {Malanushenko}, Viktor and {Marchante}, Moses and {McGreer}, Ian D. and {McMahon}, Richard G. and {Miralda-Escud{\'e}}, Jordi and {Muna}, Demitri and {Noterdaeme}, Pasquier and {Oravetz}, Daniel and {Palanque-Delabrouille}, Nathalie and {Pan}, Kaike and {Perez-Fournon}, Isma{\"e}l and {Pieri}, Matthew and {Riffel}, Rog{\'e}rio and {Schlegel}, David J. and {Schneider}, Donald P. and {Simmons}, Audrey and {Viel}, Matteo and {Weaver}, Benjamin A. and {Wood-Vasey}, W. Michael and {Y{\`e}che}, Christophe and {York}, Donald G.},
        title = "{The Sloan Digital Sky Survey quasar catalog: tenth data release}",
      journal = {\aap},
         year = 2014,
        month = mar,
       volume = {563},
          eid = {A54},
        pages = {A54},
          doi = {10.1051/0004-6361/201322691},
archivePrefix = {arXiv},
       eprint = {1311.4870},
 primaryClass = {astro-ph.CO},
       adsurl = {https://ui.adsabs.harvard.edu/abs/2014A&A...563A..54P}
}

@ARTICLE{tdcosmox,
       author = {{Ertl}, S. and {Schuldt}, S. and {Suyu}, S.~H. and {Schmidt}, T. and {Treu}, T. and {Birrer}, S. and {Shajib}, A.~J. and {Sluse}, D.},
        title = "{TDCOSMO. X. Automated modeling of nine strongly lensed quasars and comparison between lens-modeling software}",
      journal = {\aap},
         year = 2023,
        month = apr,
       volume = {672},
          eid = {A2},
        pages = {A2},
          doi = {10.1051/0004-6361/202244909},
archivePrefix = {arXiv},
       eprint = {2209.03094},
 primaryClass = {astro-ph.CO},
       adsurl = {https://ui.adsabs.harvard.edu/abs/2023A&A...672A...2E}
}

@ARTICLE{2021A&A...649A...1G,
       author = {{Gaia Collaboration} and {Brown}, A.~G.~A. and {Vallenari}, A. and {Prusti}, T. and {de Bruijne}, J.~H.~J. and {Babusiaux}, C. and {Biermann}, M. and {Creevey}, O.~L. and {Evans}, D.~W. and {Eyer}, L. and {Hutton}, A. and {Jansen}, F. and {Jordi}, C. and {Klioner}, S.~A. and {Lammers}, U. and {Lindegren}, L. and {Luri}, X. and {Mignard}, F. and {Panem}, C. and {Pourbaix}, D. and {Randich}, S. and {Sartoretti}, P. and {Soubiran}, C. and {Walton}, N.~A. and {Arenou}, F. and {Bailer-Jones}, C.~A.~L. and {Bastian}, U. and {Cropper}, M. and {Drimmel}, R. and {Katz}, D. and {Lattanzi}, M.~G. and {van Leeuwen}, F. and {Bakker}, J. and {Cacciari}, C. and {Casta{\~n}eda}, J. and {De Angeli}, F. and {Ducourant}, C. and {Fabricius}, C. and {Fouesneau}, M. and {Fr{\'e}mat}, Y. and {Guerra}, R. and {Guerrier}, A. and {Guiraud}, J. and {Jean-Antoine Piccolo}, A. and {Masana}, E. and {Messineo}, R. and {Mowlavi}, N. and {Nicolas}, C. and {Nienartowicz}, K. and {Pailler}, F. and {Panuzzo}, P. and {Riclet}, F. and {Roux}, W. and {Seabroke}, G.~M. and {Sordo}, R. and {Tanga}, P. and {Th{\'e}venin}, F. and {Gracia-Abril}, G. and {Portell}, J. and {Teyssier}, D. and {Altmann}, M. and {Andrae}, R. and {Bellas-Velidis}, I. and {Benson}, K. and {Berthier}, J. and {Blomme}, R. and {Brugaletta}, E. and {Burgess}, P.~W. and {Busso}, G. and {Carry}, B. and {Cellino}, A. and {Cheek}, N. and {Clementini}, G. and {Damerdji}, Y. and {Davidson}, M. and {Delchambre}, L. and {Dell'Oro}, A. and {Fern{\'a}ndez-Hern{\'a}ndez}, J. and {Galluccio}, L. and {Garc{\'\i}a-Lario}, P. and {Garcia-Reinaldos}, M. and {Gonz{\'a}lez-N{\'u}{\~n}ez}, J. and {Gosset}, E. and {Haigron}, R. and {Halbwachs}, J.-L. and {Hambly}, N.~C. and {Harrison}, D.~L. and {Hatzidimitriou}, D. and {Heiter}, U. and {Hern{\'a}ndez}, J. and {Hestroffer}, D. and {Hodgkin}, S.~T. and {Holl}, B. and {Jan{\ss}en}, K. and {Jevardat de Fombelle}, G. and {Jordan}, S. and {Krone-Martins}, A. and {Lanzafame}, A.~C. and {L{\"o}ffler}, W. and {Lorca}, A. and {Manteiga}, M. and {Marchal}, O. and {Marrese}, P.~M. and {Moitinho}, A. and {Mora}, A. and {Muinonen}, K. and {Osborne}, P. and {Pancino}, E. and {Pauwels}, T. and {Petit}, J.-M. and {Recio-Blanco}, A. and {Richards}, P.~J. and {Riello}, M. and {Rimoldini}, L. and {Robin}, A.~C. and {Roegiers}, T. and {Rybizki}, J. and {Sarro}, L.~M. and {Siopis}, C. and {Smith}, M. and {Sozzetti}, A. and {Ulla}, A. and {Utrilla}, E. and {van Leeuwen}, M. and {van Reeven}, W. and {Abbas}, U. and {Abreu Aramburu}, A. and {Accart}, S. and {Aerts}, C. and {Aguado}, J.~J. and {Ajaj}, M. and {Altavilla}, G. and {{\'A}lvarez}, M.~A. and {{\'A}lvarez Cid-Fuentes}, J. and {Alves}, J. and {Anderson}, R.~I. and {Anglada Varela}, E. and {Antoja}, T. and {Audard}, M. and {Baines}, D. and {Baker}, S.~G. and {Balaguer-N{\'u}{\~n}ez}, L. and {Balbinot}, E. and {Balog}, Z. and {Barache}, C. and {Barbato}, D. and {Barros}, M. and {Barstow}, M.~A. and {Bartolom{\'e}}, S. and {Bassilana}, J.-L. and {Bauchet}, N. and {Baudesson-Stella}, A. and {Becciani}, U. and {Bellazzini}, M. and {Bernet}, M. and {Bertone}, S. and {Bianchi}, L. and {Blanco-Cuaresma}, S. and {Boch}, T. and {Bombrun}, A. and {Bossini}, D. and {Bouquillon}, S. and {Bragaglia}, A. and {Bramante}, L. and {Breedt}, E. and {Bressan}, A. and {Brouillet}, N. and {Bucciarelli}, B. and {Burlacu}, A. and {Busonero}, D. and {Butkevich}, A.~G. and {Buzzi}, R. and {Caffau}, E. and {Cancelliere}, R. and {C{\'a}novas}, H. and {Cantat-Gaudin}, T. and {Carballo}, R. and {Carlucci}, T. and {Carnerero}, M.~I. and {Carrasco}, J.~M. and {Casamiquela}, L. and {Castellani}, M. and {Castro-Ginard}, A. and {Castro Sampol}, P. and {Chaoul}, L. and {Charlot}, P. and {Chemin}, L. and {Chiavassa}, A. and {Cioni}, M.-R.~L. and {Comoretto}, G. and {Cooper}, W.~J. and {Cornez}, T. and {Cowell}, S. and {Crifo}, F. and {Crosta}, M. and {Crowley}, C. and {Dafonte}, C. and {Dapergolas}, A. and {David}, M. and {David}, P.},
        title = "{Gaia Early Data Release 3. Summary of the contents and survey properties}",
      journal = {\aap},
         year = 2021,
        month = may,
       volume = {649},
          eid = {A1},
        pages = {A1},
          doi = {10.1051/0004-6361/202039657},
archivePrefix = {arXiv},
       eprint = {2012.01533},
 primaryClass = {astro-ph.GA},
       adsurl = {https://ui.adsabs.harvard.edu/abs/2021A&A...649A...1G}
}

@ARTICLE{gaia:dr2,
       author = {{Gaia Collaboration} and {Brown}, A.~G.~A. and {Vallenari}, A. and {Prusti}, T. and {de Bruijne}, J.~H.~J. and {Babusiaux}, C. and {Bailer-Jones}, C.~A.~L. and {Biermann}, M. and {Evans}, D.~W. and {Eyer}, L. and {Jansen}, F. and {Jordi}, C. and {Klioner}, S.~A. and {Lammers}, U. and {Lindegren}, L. and {Luri}, X. and {Mignard}, F. and {Panem}, C. and {Pourbaix}, D. and {Randich}, S. and {Sartoretti}, P. and {Siddiqui}, H.~I. and {Soubiran}, C. and {van Leeuwen}, F. and {Walton}, N.~A. and {Arenou}, F. and {Bastian}, U. and {Cropper}, M. and {Drimmel}, R. and {Katz}, D. and {Lattanzi}, M.~G. and {Bakker}, J. and {Cacciari}, C. and {Casta{\~n}eda}, J. and {Chaoul}, L. and {Cheek}, N. and {De Angeli}, F. and {Fabricius}, C. and {Guerra}, R. and {Holl}, B. and {Masana}, E. and {Messineo}, R. and {Mowlavi}, N. and {Nienartowicz}, K. and {Panuzzo}, P. and {Portell}, J. and {Riello}, M. and {Seabroke}, G.~M. and {Tanga}, P. and {Th{\'e}venin}, F. and {Gracia-Abril}, G. and {Comoretto}, G. and {Garcia-Reinaldos}, M. and {Teyssier}, D. and {Altmann}, M. and {Andrae}, R. and {Audard}, M. and {Bellas-Velidis}, I. and {Benson}, K. and {Berthier}, J. and {Blomme}, R. and {Burgess}, P. and {Busso}, G. and {Carry}, B. and {Cellino}, A. and {Clementini}, G. and {Clotet}, M. and {Creevey}, O. and {Davidson}, M. and {De Ridder}, J. and {Delchambre}, L. and {Dell'Oro}, A. and {Ducourant}, C. and {Fern{\'a}ndez-Hern{\'a}ndez}, J. and {Fouesneau}, M. and {Fr{\'e}mat}, Y. and {Galluccio}, L. and {Garc{\'\i}a-Torres}, M. and {Gonz{\'a}lez-N{\'u}{\~n}ez}, J. and {Gonz{\'a}lez-Vidal}, J.~J. and {Gosset}, E. and {Guy}, L.~P. and {Halbwachs}, J.-L. and {Hambly}, N.~C. and {Harrison}, D.~L. and {Hern{\'a}ndez}, J. and {Hestroffer}, D. and {Hodgkin}, S.~T. and {Hutton}, A. and {Jasniewicz}, G. and {Jean-Antoine-Piccolo}, A. and {Jordan}, S. and {Korn}, A.~J. and {Krone-Martins}, A. and {Lanzafame}, A.~C. and {Lebzelter}, T. and {L{\"o}ffler}, W. and {Manteiga}, M. and {Marrese}, P.~M. and {Mart{\'\i}n-Fleitas}, J.~M. and {Moitinho}, A. and {Mora}, A. and {Muinonen}, K. and {Osinde}, J. and {Pancino}, E. and {Pauwels}, T. and {Petit}, J.-M. and {Recio-Blanco}, A. and {Richards}, P.~J. and {Rimoldini}, L. and {Robin}, A.~C. and {Sarro}, L.~M. and {Siopis}, C. and {Smith}, M. and {Sozzetti}, A. and {S{\"u}veges}, M. and {Torra}, J. and {van Reeven}, W. and {Abbas}, U. and {Abreu Aramburu}, A. and {Accart}, S. and {Aerts}, C. and {Altavilla}, G. and {{\'A}lvarez}, M.~A. and {Alvarez}, R. and {Alves}, J. and {Anderson}, R.~I. and {Andrei}, A.~H. and {Anglada Varela}, E. and {Antiche}, E. and {Antoja}, T. and {Arcay}, B. and {Astraatmadja}, T.~L. and {Bach}, N. and {Baker}, S.~G. and {Balaguer-N{\'u}{\~n}ez}, L. and {Balm}, P. and {Barache}, C. and {Barata}, C. and {Barbato}, D. and {Barblan}, F. and {Barklem}, P.~S. and {Barrado}, D. and {Barros}, M. and {Barstow}, M.~A. and {Bartholom{\'e} Mu{\~n}oz}, S. and {Bassilana}, J.-L. and {Becciani}, U. and {Bellazzini}, M. and {Berihuete}, A. and {Bertone}, S. and {Bianchi}, L. and {Bienaym{\'e}}, O. and {Blanco-Cuaresma}, S. and {Boch}, T. and {Boeche}, C. and {Bombrun}, A. and {Borrachero}, R. and {Bossini}, D. and {Bouquillon}, S. and {Bourda}, G. and {Bragaglia}, A. and {Bramante}, L. and {Breddels}, M.~A. and {Bressan}, A. and {Brouillet}, N. and {Br{\"u}semeister}, T. and {Brugaletta}, E. and {Bucciarelli}, B. and {Burlacu}, A. and {Busonero}, D. and {Butkevich}, A.~G. and {Buzzi}, R. and {Caffau}, E. and {Cancelliere}, R. and {Cannizzaro}, G. and {Cantat-Gaudin}, T. and {Carballo}, R. and {Carlucci}, T. and {Carrasco}, J.~M. and {Casamiquela}, L. and {Castellani}, M. and {Castro-Ginard}, A. and {Charlot}, P. and {Chemin}, L. and {Chiavassa}, A. and {Cocozza}, G. and {Costigan}, G. and {Cowell}, S. and {Crifo}, F. and {Crosta}, M. and {Crowley}, C. and {Cuypers}, J. and {Dafonte}, C. and {Damerdji}, Y. and {Dapergolas}, A. and {David}, P. and {David}, M. and {de Laverny}, P. and {De Luise}, F.},
        title = "{Gaia Data Release 2. Summary of the contents and survey properties}",
      journal = {\aap},
         year = 2018,
        month = aug,
       volume = {616},
          eid = {A1},
        pages = {A1},
          doi = {10.1051/0004-6361/201833051},
archivePrefix = {arXiv},
       eprint = {1804.09365},
 primaryClass = {astro-ph.GA},
       adsurl = {https://ui.adsabs.harvard.edu/abs/2018A&A...616A...1G}
}

@ARTICLE{suyu:2009,
       author = {{Suyu}, S.~H. and {Marshall}, P.~J. and {Blandford}, R.~D. and {Fassnacht}, C.~D. and {Koopmans}, L.~V.~E. and {McKean}, J.~P. and {Treu}, T.},
        title = "{Dissecting the Gravitational Lens B1608+656. I. Lens Potential Reconstruction}",
      journal = {\apj},
         year = 2009,
        month = jan,
       volume = {691},
       number = {1},
        pages = {277-298},
          doi = {10.1088/0004-637X/691/1/277},
archivePrefix = {arXiv},
       eprint = {0804.2827},
 primaryClass = {astro-ph},
       adsurl = {https://ui.adsabs.harvard.edu/abs/2009ApJ...691..277S}
}

@ARTICLE{birrer:2015,
       author = {{Birrer}, Simon and {Amara}, Adam and {Refregier}, Alexandre},
        title = "{Gravitational Lens Modeling with Basis Sets}",
      journal = {\apj},
         year = 2015,
        month = nov,
       volume = {813},
       number = {2},
          eid = {102},
        pages = {102},
          doi = {10.1088/0004-637X/813/2/102},
archivePrefix = {arXiv},
       eprint = {1504.07629},
 primaryClass = {astro-ph.CO},
       adsurl = {https://ui.adsabs.harvard.edu/abs/2015ApJ...813..102B}
}

@ARTICLE{cosmograil:xix,
       author = {{Millon}, M. and {Courbin}, F. and {Bonvin}, V. and {Paic}, E. and {Meylan}, G. and {Tewes}, M. and {Sluse}, D. and {Magain}, P. and {Chan}, J.~H.~H. and {Galan}, A. and {Joseph}, R. and {Lemon}, C. and {Tihhonova}, O. and {Anderson}, R.~I. and {Marmier}, M. and {Chazelas}, B. and {Lendl}, M. and {Triaud}, A.~H.~M.~J. and {Wyttenbach}, A.},
        title = "{COSMOGRAIL. XIX. Time delays in 18 strongly lensed quasars from 15 years of optical monitoring}",
      journal = {\aap},
         year = 2020,
        month = aug,
       volume = {640},
          eid = {A105},
        pages = {A105},
          doi = {10.1051/0004-6361/202037740},
archivePrefix = {arXiv},
       eprint = {2002.05736},
 primaryClass = {astro-ph.CO},
       adsurl = {https://ui.adsabs.harvard.edu/abs/2020A&A...640A.105M}
}

@ARTICLE{tdcosmo:ii,
       author = {{Millon}, M. and {Courbin}, F. and {Bonvin}, V. and {Buckley-Geer}, E. and {Fassnacht}, C.~D. and {Frieman}, J. and {Marshall}, P.~J. and {Suyu}, S.~H. and {Treu}, T. and {Anguita}, T. and {Motta}, V. and {Agnello}, A. and {Chan}, J.~H.~H. and {Chao}, D.~C.-Y. and {Chijani}, M. and {Gilman}, D. and {Gilmore}, K. and {Lemon}, C. and {Lucey}, J.~R. and {Melo}, A. and {Paic}, E. and {Rojas}, K. and {Sluse}, D. and {Williams}, P.~R. and {Hempel}, A. and {Kim}, S. and {Lachaume}, R. and {Rabus}, M.},
        title = "{TDCOSMO. II. Six new time delays in lensed quasars from high-cadence monitoring at the MPIA 2.2 m telescope}",
      journal = {\aap},
         year = 2020,
        month = oct,
       volume = {642},
          eid = {A193},
        pages = {A193},
          doi = {10.1051/0004-6361/202038698},
archivePrefix = {arXiv},
       eprint = {2006.10066},
 primaryClass = {astro-ph.CO},
       adsurl = {https://ui.adsabs.harvard.edu/abs/2020A&A...642A.193M}
}

@ARTICLE{desi_legacy_imaging,
       author = {{Dey}, Arjun and {Schlegel}, David J. and {Lang}, Dustin and {Blum}, Robert and {Burleigh}, Kaylan and {Fan}, Xiaohui and {Findlay}, Joseph R. and {Finkbeiner}, Doug and {Herrera}, David and {Juneau}, St{\'e}phanie and {Landriau}, Martin and {Levi}, Michael and {McGreer}, Ian and {Meisner}, Aaron and {Myers}, Adam D. and {Moustakas}, John and {Nugent}, Peter and {Patej}, Anna and {Schlafly}, Edward F. and {Walker}, Alistair R. and {Valdes}, Francisco and {Weaver}, Benjamin A. and {Y{\`e}che}, Christophe and {Zou}, Hu and {Zhou}, Xu and {Abareshi}, Behzad and {Abbott}, T.~M.~C. and {Abolfathi}, Bela and {Aguilera}, C. and {Alam}, Shadab and {Allen}, Lori and {Alvarez}, A. and {Annis}, James and {Ansarinejad}, Behzad and {Aubert}, Marie and {Beechert}, Jacqueline and {Bell}, Eric F. and {BenZvi}, Segev Y. and {Beutler}, Florian and {Bielby}, Richard M. and {Bolton}, Adam S. and {Brice{\~n}o}, C{\'e}sar and {Buckley-Geer}, Elizabeth J. and {Butler}, Karen and {Calamida}, Annalisa and {Carlberg}, Raymond G. and {Carter}, Paul and {Casas}, Ricard and {Castander}, Francisco J. and {Choi}, Yumi and {Comparat}, Johan and {Cukanovaite}, Elena and {Delubac}, Timoth{\'e}e and {DeVries}, Kaitlin and {Dey}, Sharmila and {Dhungana}, Govinda and {Dickinson}, Mark and {Ding}, Zhejie and {Donaldson}, John B. and {Duan}, Yutong and {Duckworth}, Christopher J. and {Eftekharzadeh}, Sarah and {Eisenstein}, Daniel J. and {Etourneau}, Thomas and {Fagrelius}, Parker A. and {Farihi}, Jay and {Fitzpatrick}, Mike and {Font-Ribera}, Andreu and {Fulmer}, Leah and {G{\"a}nsicke}, Boris T. and {Gaztanaga}, Enrique and {George}, Koshy and {Gerdes}, David W. and {Gontcho}, Satya Gontcho A. and {Gorgoni}, Claudio and {Green}, Gregory and {Guy}, Julien and {Harmer}, Diane and {Hernandez}, M. and {Honscheid}, Klaus and {Huang}, Lijuan Wendy and {James}, David J. and {Jannuzi}, Buell T. and {Jiang}, Linhua and {Joyce}, Richard and {Karcher}, Armin and {Karkar}, Sonia and {Kehoe}, Robert and {Kneib}, Jean-Paul and {Kueter-Young}, Andrea and {Lan}, Ting-Wen and {Lauer}, Tod R. and {Le Guillou}, Laurent and {Le Van Suu}, Auguste and {Lee}, Jae Hyeon and {Lesser}, Michael and {Perreault Levasseur}, Laurence and {Li}, Ting S. and {Mann}, Justin L. and {Marshall}, Robert and {Mart{\'\i}nez-V{\'a}zquez}, C.~E. and {Martini}, Paul and {du Mas des Bourboux}, H{\'e}lion and {McManus}, Sean and {Meier}, Tobias Gabriel and {M{\'e}nard}, Brice and {Metcalfe}, Nigel and {Mu{\~n}oz-Guti{\'e}rrez}, Andrea and {Najita}, Joan and {Napier}, Kevin and {Narayan}, Gautham and {Newman}, Jeffrey A. and {Nie}, Jundan and {Nord}, Brian and {Norman}, Dara J. and {Olsen}, Knut A.~G. and {Paat}, Anthony and {Palanque-Delabrouille}, Nathalie and {Peng}, Xiyan and {Poppett}, Claire L. and {Poremba}, Megan R. and {Prakash}, Abhishek and {Rabinowitz}, David and {Raichoor}, Anand and {Rezaie}, Mehdi and {Robertson}, A.~N. and {Roe}, Natalie A. and {Ross}, Ashley J. and {Ross}, Nicholas P. and {Rudnick}, Gregory and {Safonova}, Sasha and {Saha}, Abhijit and {S{\'a}nchez}, F. Javier and {Savary}, Elodie and {Schweiker}, Heidi and {Scott}, Adam and {Seo}, Hee-Jong and {Shan}, Huanyuan and {Silva}, David R. and {Slepian}, Zachary and {Soto}, Christian and {Sprayberry}, David and {Staten}, Ryan and {Stillman}, Coley M. and {Stupak}, Robert J. and {Summers}, David L. and {Sien Tie}, Suk and {Tirado}, H. and {Vargas-Maga{\~n}a}, Mariana and {Vivas}, A. Katherina and {Wechsler}, Risa H. and {Williams}, Doug and {Yang}, Jinyi and {Yang}, Qian and {Yapici}, Tolga and {Zaritsky}, Dennis and {Zenteno}, A. and {Zhang}, Kai and {Zhang}, Tianmeng and {Zhou}, Rongpu and {Zhou}, Zhimin},
        title = "{Overview of the DESI Legacy Imaging Surveys}",
      journal = {\aj},
         year = 2019,
        month = may,
       volume = {157},
       number = {5},
          eid = {168},
        pages = {168},
          doi = {10.3847/1538-3881/ab089d},
archivePrefix = {arXiv},
       eprint = {1804.08657},
 primaryClass = {astro-ph.IM},
       adsurl = {https://ui.adsabs.harvard.edu/abs/2019AJ....157..168D}
}

@article{kess_raftery:1995,
author = {Robert E. Kass and Adrian E. Raftery},
title = {Bayes Factors},
journal = {Journal of the American Statistical Association},
volume = {90},
number = {430},
pages = {773--795},
year = {1995},
publisher = {Taylor \& Francis},
doi = {10.1080/01621459.1995.10476572},


URL = { 
    
        https://doi.org/10.1080/01621459.1995.10476572
    
    

},
eprint = { 
    
        https://doi.org/10.1080/01621459.1995.10476572
    
    

}
}

@ARTICLE{divalentino:2025,
       author = {{Di Valentino}, Eleonora and {Said}, Jackson Levi and {Riess}, Adam and {Pollo}, Agnieszka and {Poulin}, Vivian and {G{\'o}mez-Valent}, Adri{\`a} and {Weltman}, Amanda and {Palmese}, Antonella and {Huang}, Caroline D. and {van de Bruck}, Carsten and {Saraf}, Chandra Shekhar and {Kuo}, Cheng-Yu and {Uhlemann}, Cora and {Grand{\'o}n}, Daniela and {Paz}, Dante and {Eckert}, Dominique and {Teixeira}, Elsa M. and {Saridakis}, Emmanuel N. and {Colg{\'a}in}, Eoin {\'O}. and {Beutler}, Florian and {Niedermann}, Florian and {Bajardi}, Francesco and {Barenboim}, Gabriela and {Gubitosi}, Giulia and {Musella}, Ilaria and {Banik}, Indranil and {Szapudi}, Istvan and {Singal}, Jack and {Cases}, Jaume Haro and {Chluba}, Jens and {Torrado}, Jes{\'u}s and {Mifsud}, Jurgen and {Jedamzik}, Karsten and {Said}, Khaled and {Dialektopoulos}, Konstantinos and {Herold}, Laura and {Perivolaropoulos}, Leandros and {Zu}, Lei and {Galbany}, Llu{\'\i}s and {Breuval}, Louise and {Visinelli}, Luca and {Escamilla}, Luis A. and {Anchordoqui}, Luis A. and {Sheikh-Jabbari}, M.~M. and {Lembo}, Margherita and {Dainotti}, Maria Giovanna and {Vincenzi}, Maria and {Asgari}, Marika and {Gerbino}, Martina and {Forconi}, Matteo and {Cantiello}, Michele and {Moresco}, Michele and {Benetti}, Micol and {Sch{\"o}neberg}, Nils and {Akarsu}, {\"O}zg{\"u}r and {Nunes}, Rafael C. and {Bernardo}, Reginald Christian and {Ch{\'a}vez}, Ricardo and {Anderson}, Richard I. and {Watkins}, Richard and {Capozziello}, Salvatore and {Li}, Siyang and {Vagnozzi}, Sunny and {Pan}, Supriya and {Treu}, Tommaso and {Irsic}, Vid and {Handley}, Will and {Giar{\`e}}, William and {Murakami}, Yukei and {Banihashemi}, Abdolali and {Poudou}, Ad{\`e}le and {Heavens}, Alan and {Kogut}, Alan and {Domi}, Alba and {Lenart}, Aleksander {\L}ukasz and {Melchiorri}, Alessandro and {Vadal{\`a}}, Alessandro and {Amon}, Alexandra and {Rivera}, Alexander Bonilla and {Reeves}, Alexander and {Zhuk}, Alexander and {Bonanno}, Alfio and {{\"O}vg{\"u}n}, Ali and {Pisani}, Alice and {Talebian}, Alireza and {Abebe}, Amare and {Aboubrahim}, Amin and {Gonz{\'a}lez Mor{\'a}n}, Ana Luisa and {Kov{\'a}cs}, Andr{\'a}s and {Lymperis}, Andreas and {Papatriantafyllou}, Andreas and {Liddle}, Andrew R. and {Paliathanasis}, Andronikos and {Borowiec}, Andrzej and {Yadav}, Anil Kumar and {Yadav}, Anita and {Sen}, Anjan Ananda and {William}, Anjitha John and {Davis}, Anne Christine and {Shajib}, Anowar J. and {Walters}, Anthony and {Lonappan}, Anto Idicherian and {Chudaykin}, Anton and {Capodagli}, Antonio and {da Silva}, Antonio and {De Felice}, Antonio and {Racioppi}, Antonio and {Oficial}, Araceli Soler and {Montiel}, Ariadna and {Favale}, Arianna and {Bernui}, Armando and {Velasco}, Arrianne Crystal and {Heinesen}, Asta and {Bakopoulos}, Athanasios and {Chatzistavrakidis}, Athanasios and {Khanpour}, Bahman and {Sathyaprakash}, Bangalore S. and {Zgirski}, Bartek and {L'Huillier}, Benjamin and {Famaey}, Benoit and {Jain}, Bhuvnesh and {Zhang}, Bing and {Karmakar}, Biswajit and {Dragovich}, Branko and {Thomas}, Brooks and {Correa}, Carlos and {Boiza}, Carlos G. and {Marques}, Catarina and {Escamilla-Rivera}, Celia and {Tzerefos}, Charalampos and {Zhang}, Chi and {De Leo}, Chiara and {Pfeifer}, Christian and {Lee}, Christine and {Venter}, Christo and {Gomes}, Cl{\'a}udio and {Roque De bom}, Clecio and {Moreno-Pulido}, Cristian and {Iosifidis}, Damianos and {Grin}, Dan and {Blixt}, Daniel and {Scolnic}, Dan and {Oriti}, Daniele and {Dobrycheva}, Daria and {Bettoni}, Dario and {Benisty}, David and {Fern{\'a}ndez-Arenas}, David and {Wiltshire}, David L. and {Sanchez Cid}, David and {Tamayo}, David and {Valls-Gabaud}, David and {Pedrotti}, Davide and {Wang}, Deng and {Staicova}, Denitsa and {Totolou}, Despoina and {Rubiera-Garcia}, Diego and {Milakovi{\'c}}, Dinko and {Pesce}, Dominic W. and {Sluse}, Dominique and {Borka}, Du{\v{s}}ko and {Yusofi}, Ebrahim and {Giusarma}, Elena and {Terlevich}, Elena and {Tomasetti}, Elena and {Vagenas}, Elias C. and {Fazzari}, Elisa and {Ferreira}, Elisa G.~M. and {Barakovic}, Elvis and {Dimastrogiovanni}, Emanuela and {Holm}, Emil Brinch and {Mottola}, Emil and {{\"O}z{\"u}lker}, Emre and {Specogna}, Enrico and {Brocato}, Enzo and {Jensko}, Erik and {Enriquez}, Erika Antonette and {Bhatia}, Esha and {Bresolin}, Fabio and {Avila}, Felipe and {Bouch{\`e}}, Filippo and {Bombacigno}, Flavio and {Anagnostopoulos}, Fotios K. and {Pace}, Francesco and {Sorrenti}, Francesco and {Lobo}, Francisco S.~N. and {Courbin}, Fr{\'e}d{\'e}ric and {Hansen}, Frode K. and {Sloan}, Greg and {Farrugia}, Gabriel and {Lynch}, Gabriel and {Garcia-Arroyo}, Gabriela and {Raimondo}, Gabriella and {Lambiase}, Gaetano and {Anand}, Gagandeep S. and {Poulot}, Gaspard and {Leon}, Genly and {Kouniatalis}, Gerasimos and {Nardini}, Germano and {Cs{\"o}rnyei}, G{\'e}za and {Galloni}, Giacomo},
        title = "{The CosmoVerse White Paper: Addressing observational tensions in cosmology with systematics and fundamental physics}",
      journal = {Physics of the Dark Universe},
         year = 2025,
        month = sep,
       volume = {49},
          eid = {101965},
        pages = {101965},
          doi = {10.1016/j.dark.2025.101965},
archivePrefix = {arXiv},
       eprint = {2504.01669},
 primaryClass = {astro-ph.CO},
       adsurl = {https://ui.adsabs.harvard.edu/abs/2025PDU....4901965D}
}

@ARTICLE{gabor:2009,
       author = {{Gabor}, J.~M. and {Impey}, C.~D. and {Jahnke}, K. and {Simmons}, B.~D. and {Trump}, J.~R. and {Koekemoer}, A.~M. and {Brusa}, M. and {Cappelluti}, N. and {Schinnerer}, E. and {Smol{\v{c}}i{\'c}}, V. and {Salvato}, M. and {Rhodes}, J.~D. and {Mobasher}, B. and {Capak}, P. and {Massey}, R. and {Leauthaud}, A. and {Scoville}, N.},
        title = "{Active Galactic Nucleus Host Galaxy Morphologies in COSMOS}",
      journal = {\apj},
         year = 2009,
        month = jan,
       volume = {691},
       number = {1},
        pages = {705-722},
          doi = {10.1088/0004-637X/691/1/705},
archivePrefix = {arXiv},
       eprint = {0809.0309},
 primaryClass = {astro-ph},
       adsurl = {https://ui.adsabs.harvard.edu/abs/2009ApJ...691..705G}
}

@ARTICLE{ding:2021,
       author = {{Ding}, Xuheng and {Treu}, Tommaso and {Birrer}, Simon and {Agnello}, Adriano and {Sluse}, Dominique and {Fassnacht}, Chris and {Auger}, Matthew W. and {Wong}, Kenneth C. and {Suyu}, Sherry H. and {Morishita}, Takahiro and {Rusu}, Cristian E. and {Galan}, Aymeric},
        title = "{Testing the evolution of correlations between supermassive black holes and their host galaxies using eight strongly lensed quasars}",
      journal = {\mnras},
         year = 2021,
        month = feb,
       volume = {501},
       number = {1},
        pages = {269-280},
          doi = {10.1093/mnras/staa2992},
archivePrefix = {arXiv},
       eprint = {2005.13550},
 primaryClass = {astro-ph.GA},
       adsurl = {https://ui.adsabs.harvard.edu/abs/2021MNRAS.501..269D}
}

@ARTICLE{ding:2017,
       author = {{Ding}, Xuheng and {Liao}, Kai and {Treu}, Tommaso and {Suyu}, Sherry H. and {Chen}, Geoff C.-F. and {Auger}, Matthew W. and {Marshall}, Philip J. and {Agnello}, Adriano and {Courbin}, Frederic and {Nierenberg}, Anna M. and {Rusu}, Cristian E. and {Sluse}, Dominique and {Sonnenfeld}, Alessandro and {Wong}, Kenneth C.},
        title = "{H0LiCOW. VI. Testing the fidelity of lensed quasar host galaxy reconstruction}",
      journal = {\mnras},
         year = 2017,
        month = mar,
       volume = {465},
       number = {4},
        pages = {4634-4649},
          doi = {10.1093/mnras/stw3078},
archivePrefix = {arXiv},
       eprint = {1610.08504},
 primaryClass = {astro-ph.GA},
       adsurl = {https://ui.adsabs.harvard.edu/abs/2017MNRAS.465.4634D}
}

@ARTICLE{ding:2017b,
       author = {{Ding}, Xuheng and {Treu}, Tommaso and {Suyu}, Sherry H. and {Wong}, Kenneth C. and {Morishita}, Takahiro and {Park}, Daeseong and {Sluse}, Dominique and {Auger}, Matthew W. and {Agnello}, Adriano and {Bennert}, Vardha N. and {Collett}, Thomas E.},
        title = "{H0LiCOW VII: cosmic evolution of the correlation between black hole mass and host galaxy luminosity}",
      journal = {\mnras},
         year = 2017,
        month = nov,
       volume = {472},
       number = {1},
        pages = {90-103},
          doi = {10.1093/mnras/stx1972},
archivePrefix = {arXiv},
       eprint = {1703.02041},
 primaryClass = {astro-ph.GA},
       adsurl = {https://ui.adsabs.harvard.edu/abs/2017MNRAS.472...90D}
}

@ARTICLE{silervamn:2019,
       author = {{Silverman}, John D. and {Treu}, Tommaso and {Ding}, Xuheng and {Jahnke}, Knud and {Bennert}, Vardha N. and {Birrer}, Simon and {Schramm}, Malte and {Schulze}, Andreas and {Kartaltepe}, Jeyhan S. and {Sanders}, David B. and {Cen}, Renyue},
        title = "{Where Do Quasar Hosts Lie with Respect to the Size-Mass Relation of Galaxies?}",
      journal = {\apjl},
         year = 2019,
        month = dec,
       volume = {887},
       number = {1},
          eid = {L5},
        pages = {L5},
          doi = {10.3847/2041-8213/ab5851},
archivePrefix = {arXiv},
       eprint = {1910.14242},
 primaryClass = {astro-ph.GA},
       adsurl = {https://ui.adsabs.harvard.edu/abs/2019ApJ...887L...5S}
}

@ARTICLE{wolf:2008,
       author = {{Wolf}, Marsha J. and {Sheinis}, Andrew I.},
        title = "{Host Galaxies of Luminous Quasars: Structural Properties and the Fundamental Plane}",
      journal = {\aj},
         year = 2008,
        month = oct,
       volume = {136},
       number = {4},
        pages = {1587-1606},
          doi = {10.1088/0004-6256/136/4/1587},
archivePrefix = {arXiv},
       eprint = {0808.0918},
 primaryClass = {astro-ph},
       adsurl = {https://ui.adsabs.harvard.edu/abs/2008AJ....136.1587W}
}

@ARTICLE{teodori:2022,
       author = {{Teodori}, Luca and {Blum}, Kfir and {Castorina}, Emanuele and {Simonovi{\'c}}, Marko and {Soreq}, Yotam},
        title = "{Comments on the mass sheet degeneracy in cosmography analyses}",
      journal = {\jcap},
         year = 2022,
        month = jul,
       volume = {2022},
       number = {7},
          eid = {027},
        pages = {027},
          doi = {10.1088/1475-7516/2022/07/027},
archivePrefix = {arXiv},
       eprint = {2201.05111},
 primaryClass = {astro-ph.CO},
       adsurl = {https://ui.adsabs.harvard.edu/abs/2022JCAP...07..027T}
}

@ARTICLE{falco:1985,
       author = {{Falco}, E.~E. and {Gorenstein}, M.~V. and {Shapiro}, I.~I.},
        title = "{On model-dependent bounds on H 0 from gravitational images : application to Q 0957+561 A, B.}",
      journal = {\apjl},
         year = 1985,
        month = feb,
       volume = {289},
        pages = {L1-L4},
          doi = {10.1086/184422},
       adsurl = {https://ui.adsabs.harvard.edu/abs/1985ApJ...289L...1F}
}

@ARTICLE{saha:2000,
       author = {{Saha}, Prasenjit},
        title = "{Lensing Degeneracies Revisited}",
      journal = {\aj},
         year = 2000,
        month = oct,
       volume = {120},
       number = {4},
        pages = {1654-1659},
          doi = {10.1086/301581},
archivePrefix = {arXiv},
       eprint = {astro-ph/0006432},
 primaryClass = {astro-ph},
       adsurl = {https://ui.adsabs.harvard.edu/abs/2000AJ....120.1654S}
}

@ARTICLE{saha:2006,
       author = {{Saha}, Prasenjit and {Coles}, Jonathan and {Macci{\`o}}, Andrea V. and {Williams}, Liliya L.~R.},
        title = "{The Hubble Time Inferred from 10 Time Delay Lenses}",
      journal = {\apjl},
         year = 2006,
        month = oct,
       volume = {650},
       number = {1},
        pages = {L17-L20},
          doi = {10.1086/507583},
archivePrefix = {arXiv},
       eprint = {astro-ph/0607240},
 primaryClass = {astro-ph},
       adsurl = {https://ui.adsabs.harvard.edu/abs/2006ApJ...650L..17S}
}

@ARTICLE{suyu:2012,
       author = {{Suyu}, S.~H.},
        title = "{Cosmography from two-image lens systems: overcoming the lens profile slope degeneracy}",
      journal = {\mnras},
         year = 2012,
        month = oct,
       volume = {426},
       number = {2},
        pages = {868-879},
          doi = {10.1111/j.1365-2966.2012.21661.x},
archivePrefix = {arXiv},
       eprint = {1202.0287},
 primaryClass = {astro-ph.CO},
       adsurl = {https://ui.adsabs.harvard.edu/abs/2012MNRAS.426..868S}
}

@dataset{brady_2026_19421439,
  author       = {Brady, Ryan},
  title        = {Metadata for TDCOSMO XXVI: Uniform Lens Modeling of
                   Eight Doubly Imaged Quasars
                  },
  month        = apr,
  year         = 2026,
  publisher    = {Zenodo},
  doi          = {10.5281/zenodo.19421439},
  url          = {https://doi.org/10.5281/zenodo.19421439},
}

@ARTICLE{tdcosmo:i,
       author = {{Millon}, M. and {Galan}, A. and {Courbin}, F. and {Treu}, T. and {Suyu}, S.~H. and {Ding}, X. and {Birrer}, S. and {Chen}, G.~C.-F. and {Shajib}, A.~J. and {Sluse}, D. and {Wong}, K.~C. and {Agnello}, A. and {Auger}, M.~W. and {Buckley-Geer}, E.~J. and {Chan}, J.~H.~H. and {Collett}, T. and {Fassnacht}, C.~D. and {Hilbert}, S. and {Koopmans}, L.~V.~E. and {Motta}, V. and {Mukherjee}, S. and {Rusu}, C.~E. and {Sonnenfeld}, A. and {Spiniello}, C. and {Van de Vyvere}, L.},
        title = "{TDCOSMO. I. An exploration of systematic uncertainties in the inference of H$_{0}$ from time-delay cosmography}",
      journal = {\aap},
         year = 2020,
        month = jul,
       volume = {639},
          eid = {A101},
        pages = {A101},
          doi = {10.1051/0004-6361/201937351},
archivePrefix = {arXiv},
       eprint = {1912.08027},
 primaryClass = {astro-ph.CO},
       adsurl = {https://ui.adsabs.harvard.edu/abs/2020A&A...639A.101M}
}

@ARTICLE{tdcosmo:xix,
       author = {{Knabel}, Shawn and {Mozumdar}, Pritom and {Shajib}, Anowar J. and {Treu}, Tommaso and {Cappellari}, Michele and {Spiniello}, Chiara and {Birrer}, Simon},
        title = "{TDCOSMO: XIX. Measuring stellar velocity dispersion with sub-percent accuracy for cosmography}",
      journal = {\aap},
         year = 2025,
        month = nov,
       volume = {703},
          eid = {A117},
        pages = {A117},
          doi = {10.1051/0004-6361/202554229},
archivePrefix = {arXiv},
       eprint = {2502.16034},
 primaryClass = {astro-ph.GA},
       adsurl = {https://ui.adsabs.harvard.edu/abs/2025A&A...703A.117K}
}

@ARTICLE{tdcosmo:ix,
       author = {{Shajib}, A.~J. and {Wong}, K.~C. and {Birrer}, S. and {Suyu}, S.~H. and {Treu}, T. and {Buckley-Geer}, E.~J. and {Lin}, H. and {Rusu}, C.~E. and {Poh}, J. and {Palmese}, A. and {Agnello}, A. and {Auger-Williams}, M.~W. and {Galan}, A. and {Schuldt}, S. and {Sluse}, D. and {Courbin}, F. and {Frieman}, J. and {Millon}, M.},
        title = "{TDCOSMO. IX. Systematic comparison between lens modelling software programs: Time-delay prediction for WGD 2038‒4008}",
      journal = {\aap},
         year = 2022,
        month = nov,
       volume = {667},
          eid = {A123},
        pages = {A123},
          doi = {10.1051/0004-6361/202243401},
archivePrefix = {arXiv},
       eprint = {2202.11101},
 primaryClass = {astro-ph.CO},
       adsurl = {https://ui.adsabs.harvard.edu/abs/2022A&A...667A.123S}
}
\bibliographystyle{aasjournal}

\appendix

\section{\textbf{A. Direct Comparison with the Literature}}
\label{sec:appendA}

\begin{figure}[t]
    \centering
    \includegraphics[width=0.999\textwidth]{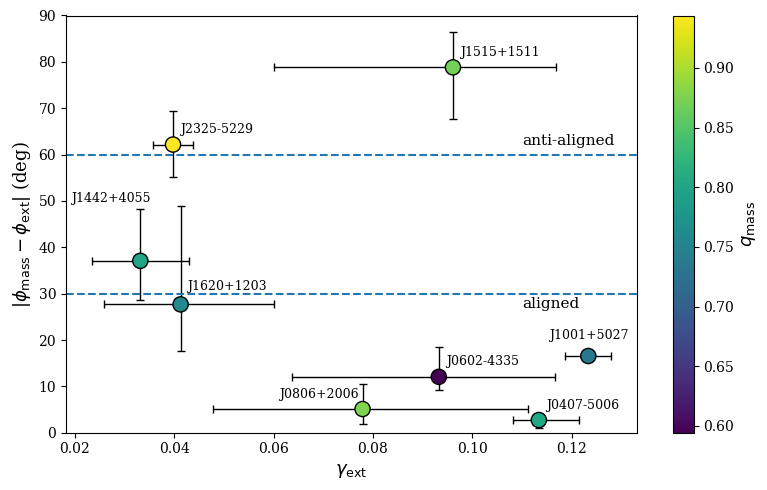}
    \caption{Misalignment angle between the best-fit mass position angle $\phi_\mathrm{mass}$ and external shear position angle $\phi_\mathrm{ext}$ as a function of shear magnitude $\gamma_\mathrm{ext}$, color-coded by mass axis ratio $q_\mathrm{mass}$. Dashed blue lines at $30$ degrees and $60$ degrees delineate aligned, intermediate, and anti-aligned regimes. Error bars reflect $1\sigma$ posterior uncertainties from the full image modeling. Seven of eight systems fall in the aligned or anti-aligned regime, which is consistent with the behavior identified by \citet{etherington:2024} for power-law models, wherein external shear tends to compensate for residual mass model complexity rather than tracing 
    a physical line-of-sight perturbation.}
    \label{fig:mass_shear_alignment}
\end{figure}

This appendix provides a direct comparison between our multi-band HST lens model parameters and those reported in previous studies for each system in our sample. We focus on the image separations, Einstein radii, mass ellipticity, and external shear magnitude, and assess the level of agreement in regard to differing data quality and modeling choices. Figure \ref{fig:mass_shear_alignment} shows the misalignment angle between the inferred mass position angle and the external shear position angle as a function of shear magnitude from our full image modeling. This diagnostic follows directly from \citet{etherington:2024}, who demonstrated that in both mock and real HST data, best-fit external shear in power-law lens models is predominantly aligned ($|\phi_\text{mass} - \phi_\text{ext}| \leq 30^\circ$) or anti-aligned ($|\phi_\text{mass} - \phi_\text{ext}| \geq 60^\circ$) with the lens mass distribution, rather than randomly oriented as would be expected if $\gamma_\mathrm{ext}$ were tracing a physical line-of-sight perturbation. Our sample of doubles is consistent with this behavior. Seven systems reside in the aligned or anti-aligned regimes, with only one system occupying the intermediate range. While our sample size is small, this bimodal distribution is suggestive of the external shear term absorbing internal complexities of the lens mass distribution that are not captured by the adopted SIE profile. Deviations in inferred $\gamma_\text{ext}$ and $q_\text{mass}$ across studies thus may indicate movement along this degeneracy as opposed to physical differences.

\textbf{J0407-5006.} Our modeled Einstein radius, $\theta_\mathrm{E} = 0\farcs787^{+0.006}_{-0.006}$, is in $4.5\sigma$ tension with the value reported by \citet{anguita:2018} derived from SIE mass modeling from Magellan-IMACS and Dark Energy Survey photometry. The aforementioned work also reports an image separation of $\Delta \theta = 1\farcs69$, while we measure $\Delta \theta = 1\farcs704$, which lies between their value and the Gaia DR2 measurement of $\Delta \theta = 1\farcs721$. This work does not explicitly state their inferred value of $q_\text{mass}$. As discussed in Section \ref{sec:results}, the use of multi-band HST imaging likely improves our ability to deblend Image B from the lens light, particularly given their small separation of $0\farcs28$, which may contribute to the observed difference in $\theta_\mathrm{E}$.

\textbf{J0602-4335.} No previous works have performed lens modeling on this system. Through our \lenstronomy~multi-band HST modeling, we found $\theta_\text{E}=0\farcs914^{+0.014}_{-0.019}$, $\Delta \theta = 1\farcs826$, $q_\text{mass}=0.59^{+0.06}_{-0.05}$, and $\gamma_\text{ext}=0.093^{+0.023}_{-0.029}$.

\textbf{J0806+2006.} Our measured image separation, $\Delta \theta = 1\farcs484$, is 16 mas smaller than the value reported by \citet{sluse:2008} from data obtained at the ESO Paranal Observatory and 84 mas larger than that of \citet{inada:2006} with photometric data taken at the University of Hawaii 2.2m telescope and the W. M. Keck Observatory’s Keck I. Our measured value of $\theta_\text{E}$ is consistent within $1.2\sigma$ with the result of \citet{inada:2006}, who adopted an SIS + shear model. Our measurement of $\theta_\text{E}$ is in closer agreement with \citet{sluse:2008}, who found $\theta_\mathrm{E} = 0\farcs74$ (SIS + shear) and $\theta_\mathrm{E} = 0\farcs76$ (SIE without shear, $q_\text{mass}=0.96$). We also find consistency within $1\sigma$ for $\theta_\text{E}$ and $q_\text{mass}$ with \citet{shajib:2021}, who reported $\theta_\mathrm{E} = 0\farcs76^{+0.10}_{-0.07}$ and $q_\text{mass}=0.85$ using an SIE model without shear from data acquired at the W.M. Keck observatory. Despite this overall agreement in $\theta_\text{E}$, the inferred external shear differs across studies, with \citet{inada:2006} and \citet{sluse:2008} finding $\gamma_\text{ext}=0.03$ and $\gamma_\text{ext}=0.017$, respectively, both lower than our value of $\gamma_\text{ext}=0.078 \pm 0.03$. In contrast to systems with bright extended arcs, the lensed host emission in J0806+2006 is relatively faint, leading to broader mass posteriors and allowing a wider range of $(q_\text{mass}, \gamma_\text{ext})$ combinations to reproduce the observed image configuration. This is shown in Figure \ref{fig:four_conj_posteriors}, where the $\gamma_\text{ext}$ posterior spans a large fraction of the allowed parameter space. The differences in shear amplitude across studies likely reflect different positions along this degeneracy rather than a physical discrepancy.

\textbf{J1001+5027.} For this system, our measured image separation is 4 mas larger than that reported by \citet{rusu:2016}. Our inferred Einstein radius is closest to their SIE + shear model, but remains in $2.9\sigma$ tension with that result. That work also reports relatively large values of $q_\text{mass}=0.90^{+0.08}_{-0.05}$ and $\gamma_\text{ext}=0.181^{+0.017}_{-0.040}$. In contrast, we find good agreement in $\theta_\text{E}$ with \citet{shajib:2021} at the $0.3\sigma$ level, who modeled the system using an SIE framework and report $q_\text{mass}=0.59^{+0.13}_{-0.17}$, which is also consistent with our result within $1\sigma$. J1001+5027 also exhibits the brightest lensed arcs in our sample, which provides strong constraints on the lens potential and significantly reduces the degeneracy between mass ellipticity and external shear, as reflected in Figure \ref{fig:four_conj_posteriors}. The combination of high shear and a relatively round mass profile inferred by \citet{rusu:2016} may be consistent with solutions that move along this degeneracy. Our multi-band HST modeling, with improved resolution of the extended arcs, more effectively separates these contributions, yielding a lower shear and a correspondingly more elliptical mass distribution. We further note that these studies utilize different observational datasets, with adaptive optics imaging from the Subaru Telescope in \citet{rusu:2016} compared to W.M. Keck Near-Infrared Camera2 imaging in \citet{shajib:2021}. Such differences can introduce systematics in PSF reconstruction and astrometric precision that propagate into the inferred lens parameters. However, the ability of our HST data to resolve the extended host arcs likely plays the dominant role in tightening constraints and mitigating degeneracy-driven parameter shifts.

\textbf{J1442+4055.} Our predicted $\theta_\text{E}=1\farcs032^{+0.008}_{-0.007}$ agrees at the $1.1 \sigma$ level with the reported value from \citet{more:2016}, derived from an SIS mass model with data taken at the  Hiltner Telescope. The aforementioned work reports an image separation of $\Delta \theta=2\farcs13$, which differs from our work by 1 mas. Both \citet{sergeyev:2015} and \citet{shalyapin:2019} report significantly higher values of $\theta_\text{E}=1\farcs08$ and $\theta_\text{E} = 1\farcs078$, respectively, from SIS models. This discrepancy likely reflects the increased sensitivity of our multi-band HST imaging to extended lensed arcs, together with the inclusion of $\gamma_\text{ext}$, which can account for perturbations from the nearby secondary galaxy visible in Figure \ref{fig:rgbs} that are not captured with an SIS model.

\textbf{J1515+1511.} We determined the image separation to be $\Delta \theta = 2\farcs008$, which differs by 19 mas from the value reported in \citet{inada:2014} and is in precise agreement with the value presented in \citet{rusu:2016}. Additionally, our determination of $\theta_\text{E}$ is consistent at the $0.1\sigma$ level with the value determined by \citet{inada:2014} via SIE mass modeling with photometric data from the University of Hawaii 2.2m telescope. When comparing our mass model results to \citet{rusu:2016} and \citet{shalyapin:2017}, our valuation of $\theta_\text{E}$ differs significantly from their reported value of $\theta_\text{E}=1\farcs21$ derived from SIE + shear models. One possible explanation is that these models favor solutions with unusually large external shear, with $\gamma_\mathrm{ext} = 0.283$ and $\gamma_\mathrm{ext} = 0.286$ reported by \citet{rusu:2016} and \citet{shalyapin:2017}, respectively. Data for \citet{rusu:2016} was acquired at the Subaru Telescope, while data for \citet{shalyapin:2017} was obtained at the Liverpool Telescope. As discussed in Section \ref{sec:modeling_procedure}, external shear can compensate for insufficient model complexity rather than reflect true environmental effects. While a secondary galaxy is present at a separation of $\sim$$16\farcs0$, such large shear values may be difficult to reconcile with a physical interpretation. In contrast, we find a more moderate value of $\gamma_\mathrm{ext} = 0.096^{+0.021}_{-0.036}$. Both works also report $q_\text{mass} = 0.81$, which matches our result within $1\sigma$. Conversely, \citet{inada:2006} report a significantly lower $q_\text{mass} = 0.60$.

\textbf{J1620+1203.} Our measured image separation differs from the results of \citet{rusu:2016} by 6 mas. We found $\theta_\text{E}$ to be within $2.9\sigma$ of the value found by \citet{kayo:2010} through observations conducted at the University of Hawaii 2.2m telescope. In that work, the mass was modeled via SIE without external shear. That work reports $q_\text{mass} = 0.77 \pm 0.06$, which is within $1\sigma$ of our result. We find similar consistency in $\theta_\text{E}$ with the SIE + shear measurement of \citet{rusu:2016}, whose data was obtained by the Subaru Telescope and who report $q_\text{mass} = 0.76 \pm 0.02$, as well as a larger $\gamma_\text{ext}=0.08^{+0.006}_{-0.025}$, both of which are within $1\sigma$ of our results.

\textbf{J2325-5229.} We find an image separation of $\Delta \theta = 2\farcs82$, which differs from the value measured in \citet{ostrovski:2017} by 80 mas. This is expected, as the aforementioned observations were obtained from the ground-based Dark Energy Survey, VISTA Hemisphere Survey, and WISE photometry. Given that our image separation agrees with Gaia DR2 at the 1 mas level, we do not consider this discrepancy to be cause for concern. Despite the astrometric differences, our inferred Einstein radius remains in excellent agreement with their result at the $0.2\sigma$ level. We note that their mass model adopts an SIE profile without external shear. This work also reports a mass ellipticity of $q_\text{mass} = 0.77 \pm 0.06$, which is significantly lower than our measurement of $q_\text{mass} = 0.94 \pm 0.02$. Given that this system exhibits some of the brightest lensed arcs in our sample, the degeneracy between mass ellipticity and external shear is significantly reduced, as evidenced by the well-constrained posteriors in Figure \ref{fig:three_conj_posteriors}. We infer a modest $\gamma_\text{ext} = 0.04 \pm 0.004$, and the difference with \citet{ostrovski:2017} likely reflects residual trade-offs between these components. In their case, the absence of shear may lead to a more elliptical mass distribution, while our modeling supports a correspondingly rounder mass profile.

In summary, we find no evidence for systematic discrepancies that would compromise the reliability of our lens models. Instead, the observed differences are consistent with observational and modeling heterogeneity, as well as the improved resolution of extended host galaxy arcs in multi-band HST imaging, which may lead to shifts in the inferred mass profiles. The differences in $q_\mathrm{mass}$ and $\gamma_\mathrm{ext}$ across studies can potentially be attributed to the mass-shear degeneracy. As demonstrated across our sample, systems with bright, well-resolved lensed arcs exhibit a significantly reduced degeneracy between these parameters, leading to tighter overall constraints on the mass profiles than would be obtainable with ground-based datasets.

\section{\textbf{B. Visualization of Remaining Models and Inferred Source Parameters}}
\label{append::b}

\begin{figure*}[h]
    \centering
    \includegraphics[width=.44\textwidth]{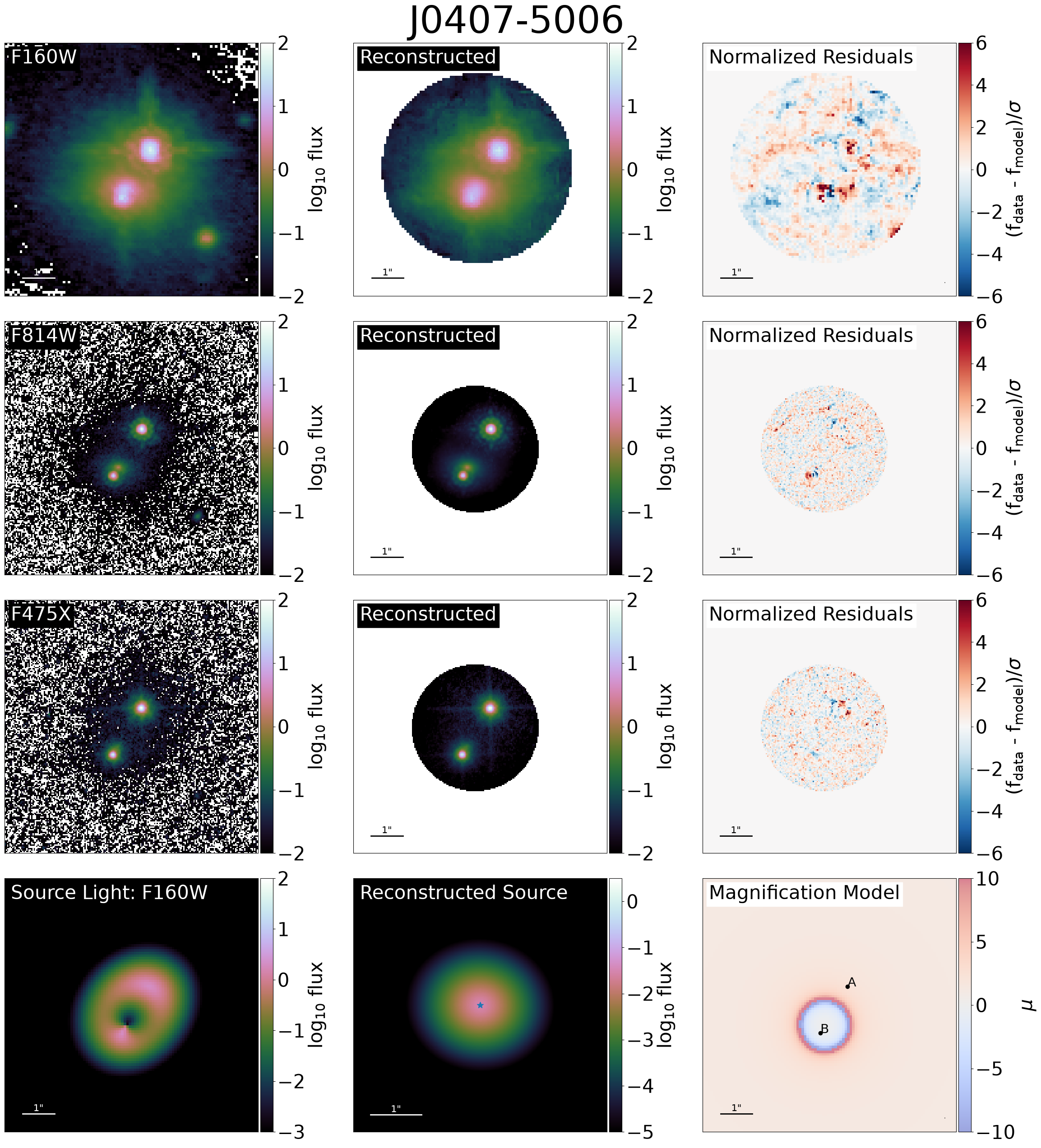} \hspace{.3in}
\includegraphics[width=.44\textwidth]{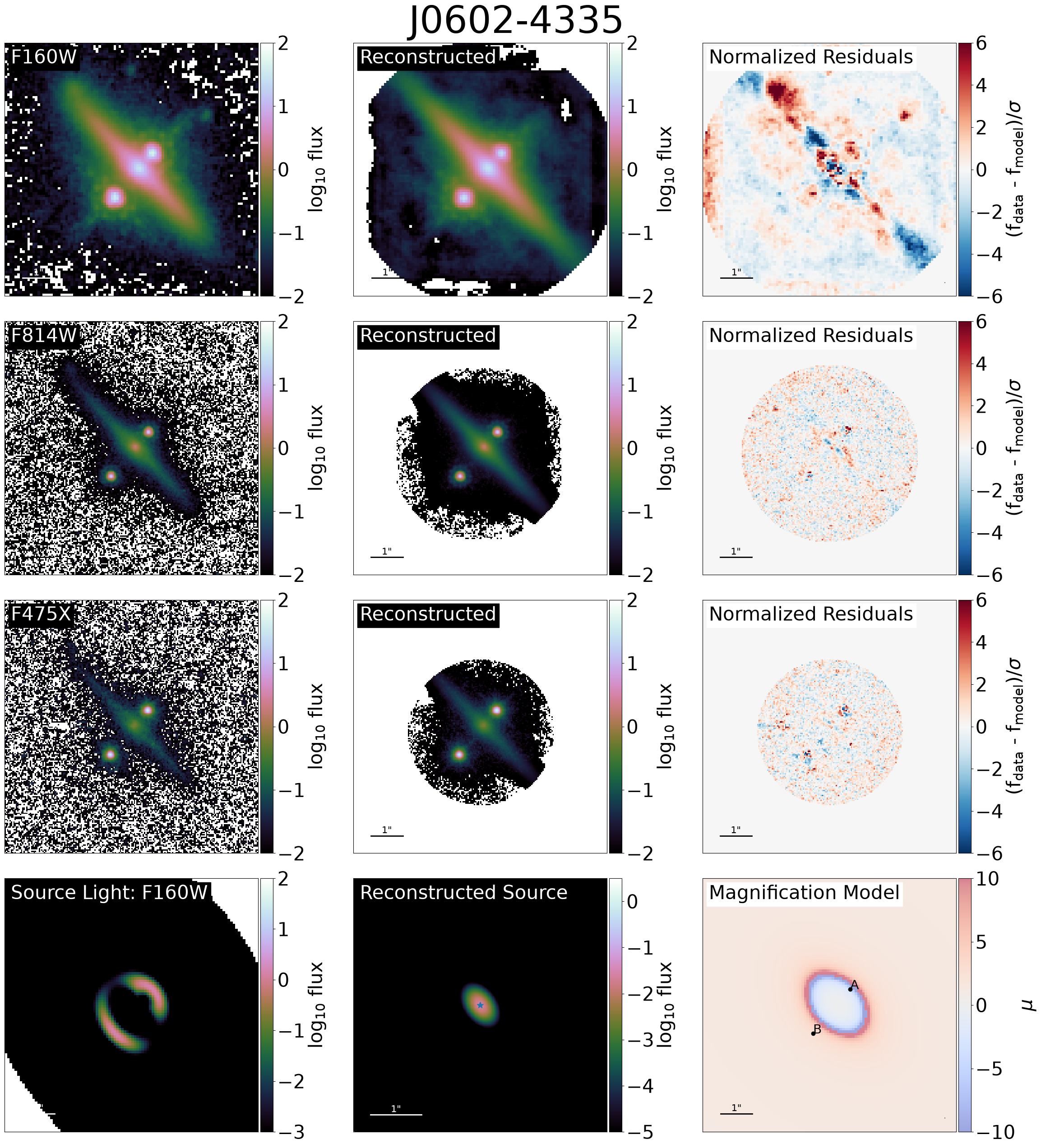}
\includegraphics[width=.44\textwidth]{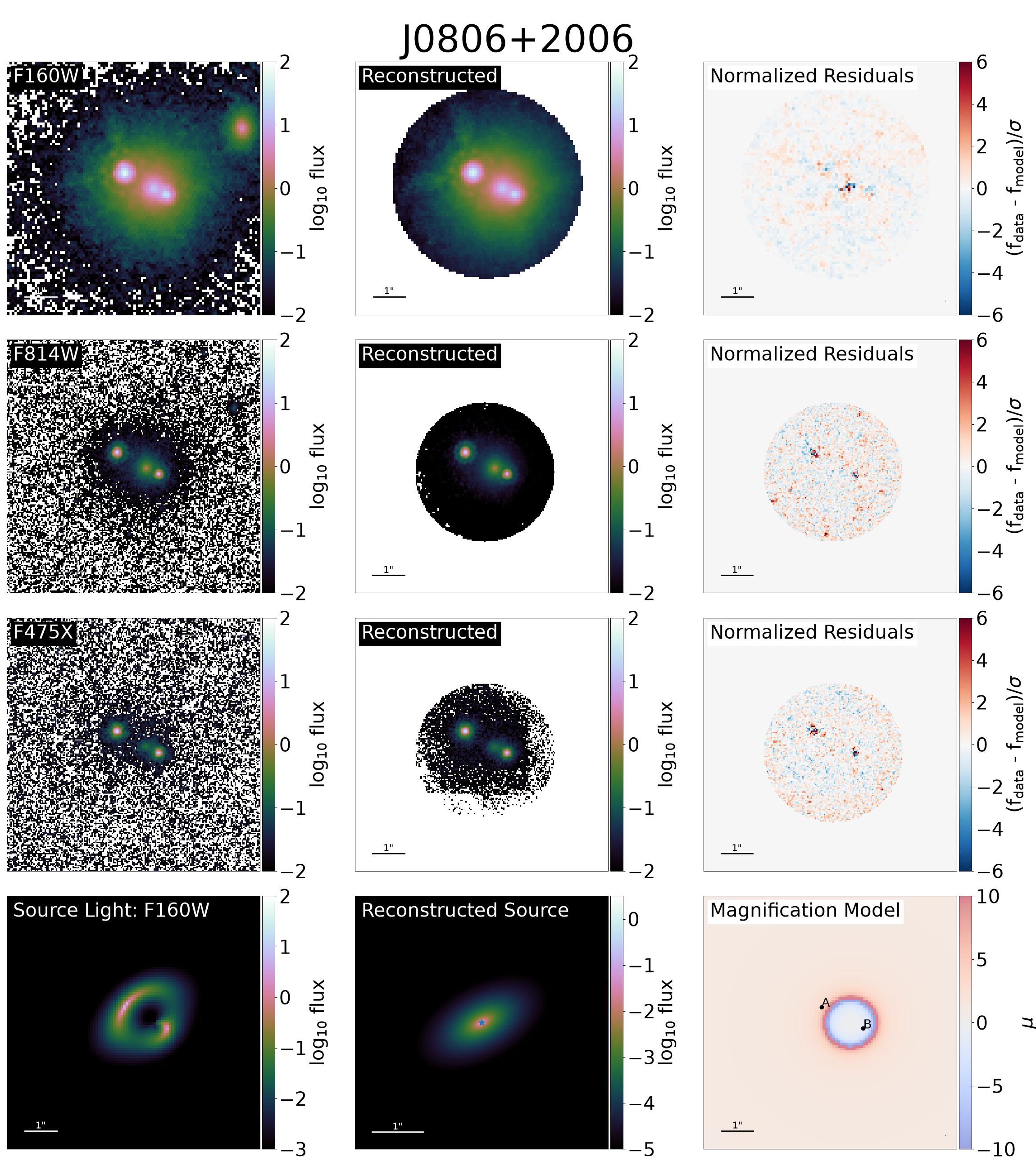} \hspace{.3in}
\includegraphics[width=.44\textwidth]{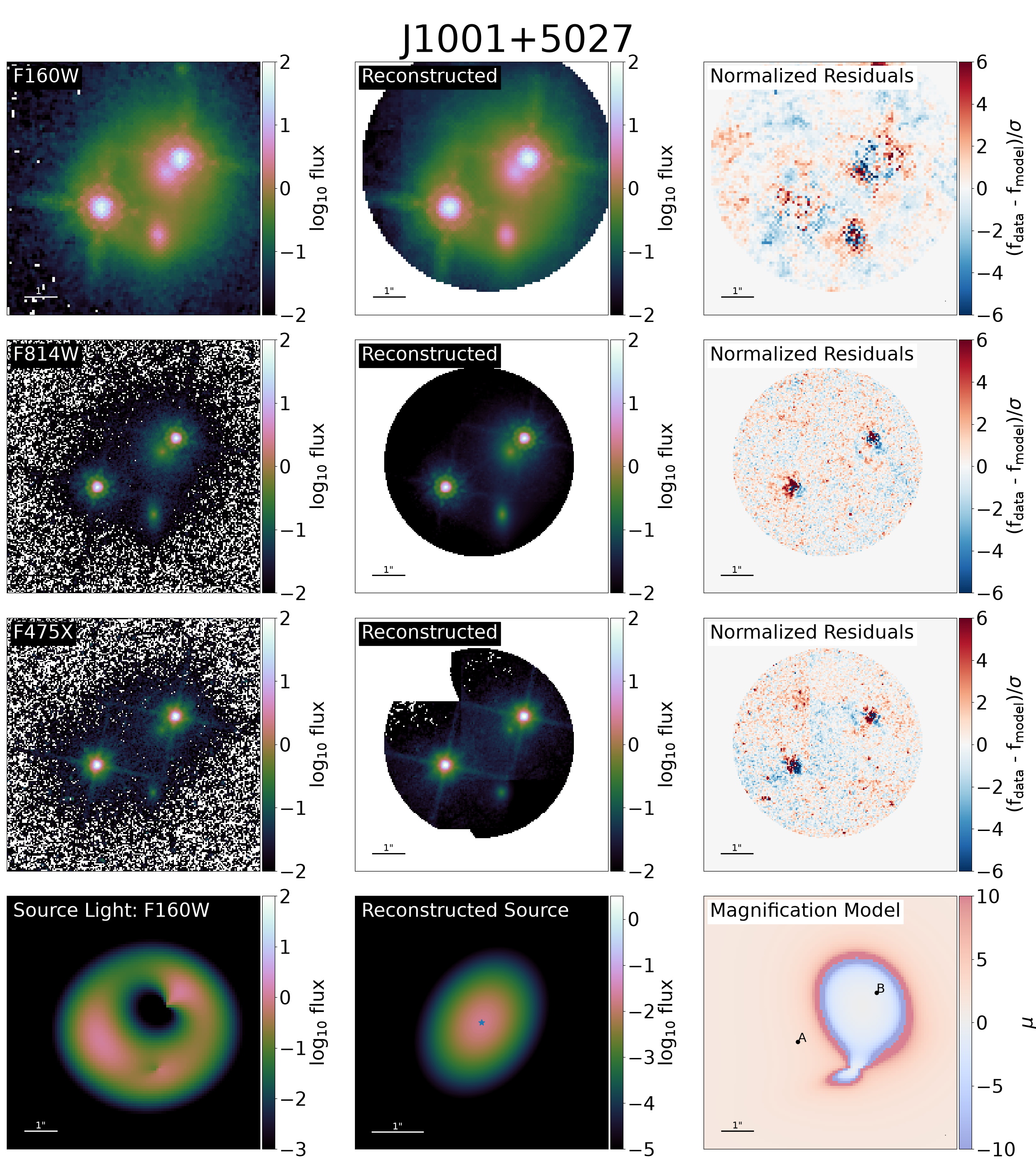}
    \caption{\lenstronomy~model visualizations for the systems J0407-5006 (top left), J0602-4335 (top right), J0806+2006 (lower left), and J1001+5027 (lower right). \textbf{Top Rows.} HST IR science image cutout, followed by the IR best-fit reconstruction, followed by residuals of the fit normalized by the pixel noise level. \textbf{Second Rows.} The same components as the top row, but for the F814W data. \textbf{Third Rows.} The same components as the top row, but for the F475X data. \textbf{Bottom Rows.} Reconstructed source arcs modeled in the image plane for the IR data, followed by the reconstructed source in the source plane. Here, the blue star represents the position of the quasar. The last panel showcases the magnification model displaying spatial variations in magnification.}
    \label{fig:4_models}
\end{figure*}

\begin{figure*}[h]
    \centering
    \includegraphics[width=.44\textwidth]{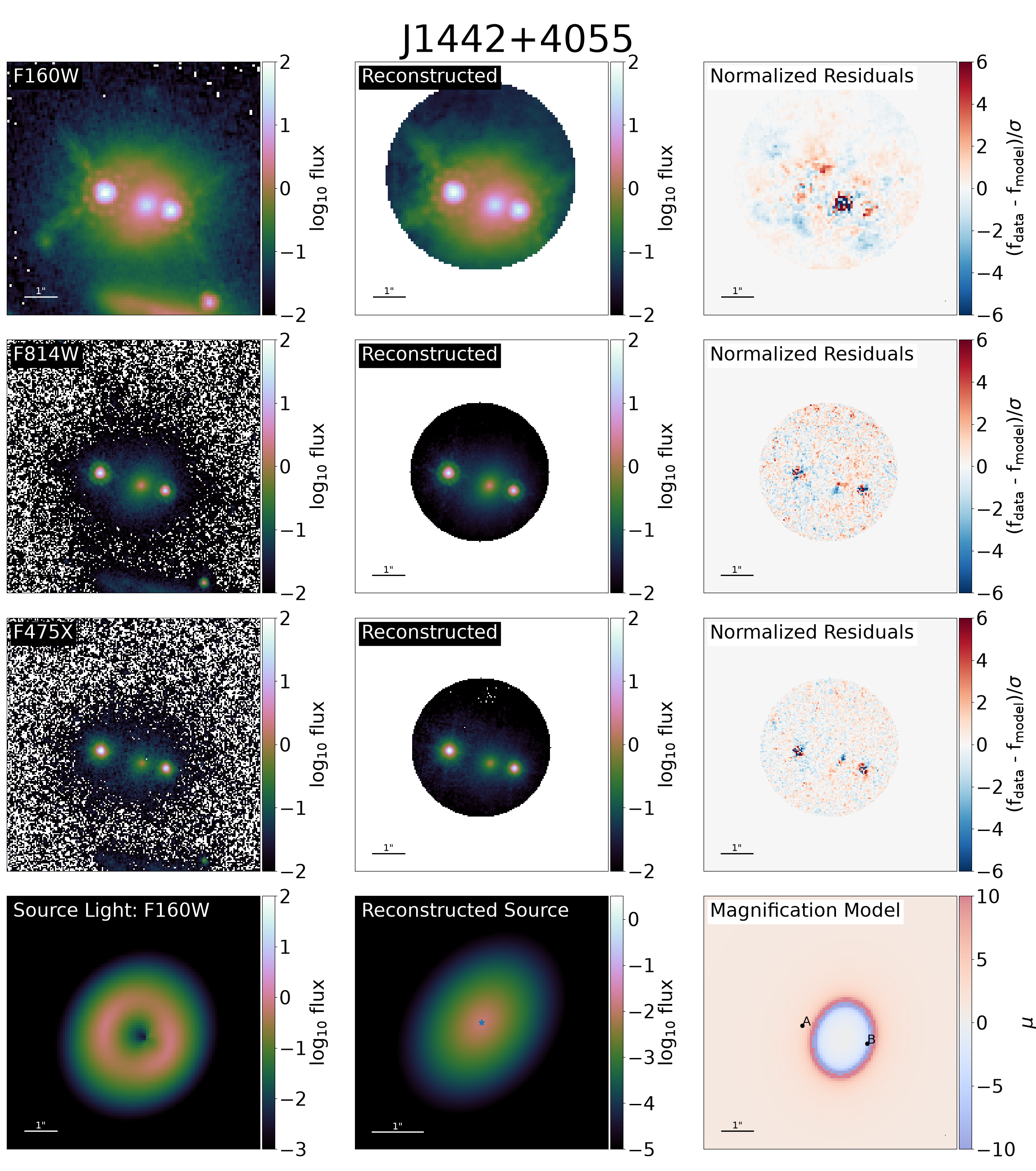} \hspace{0.3in}
    \includegraphics[width=.44\textwidth]{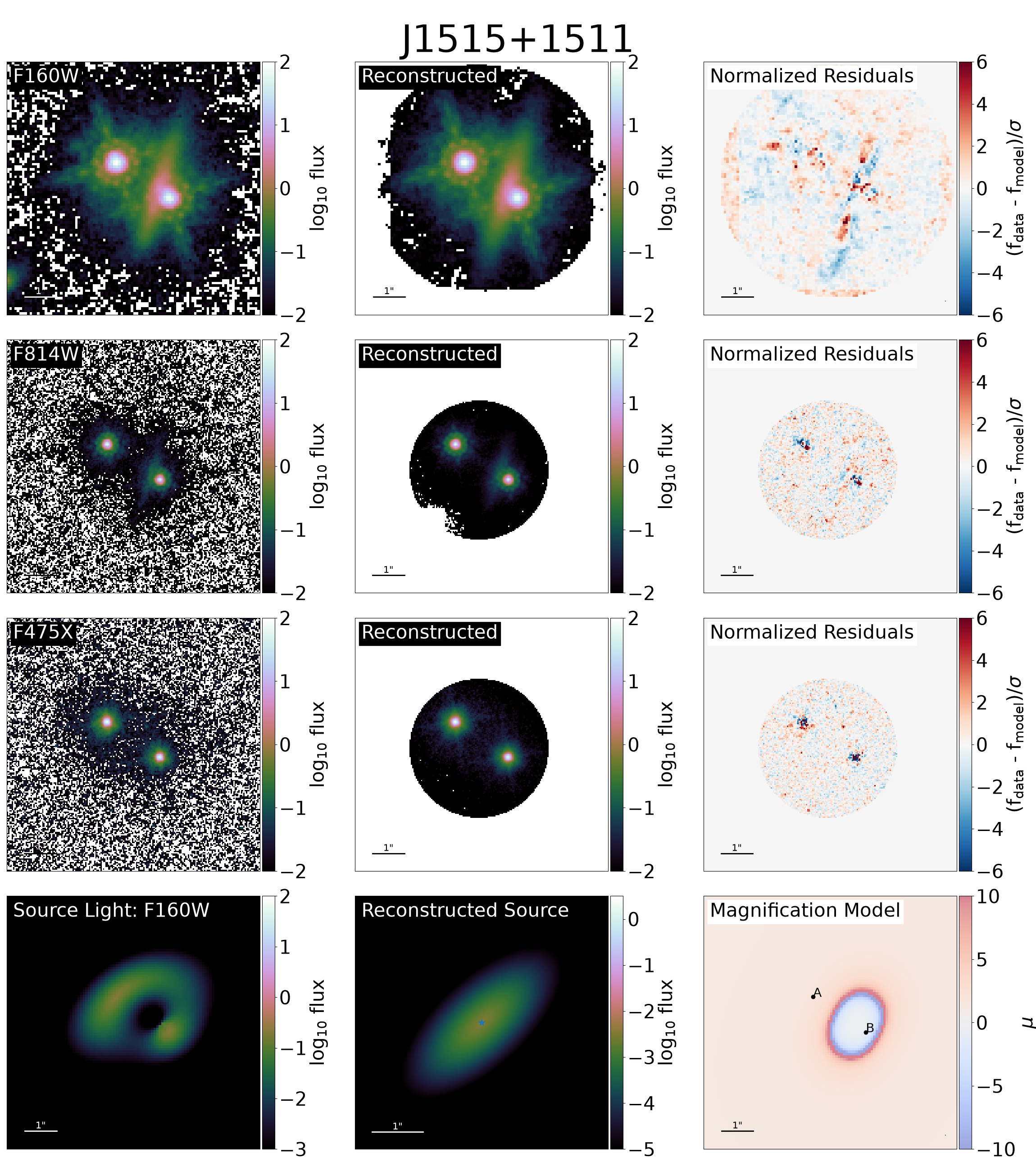}
    \includegraphics[width=.44\textwidth]{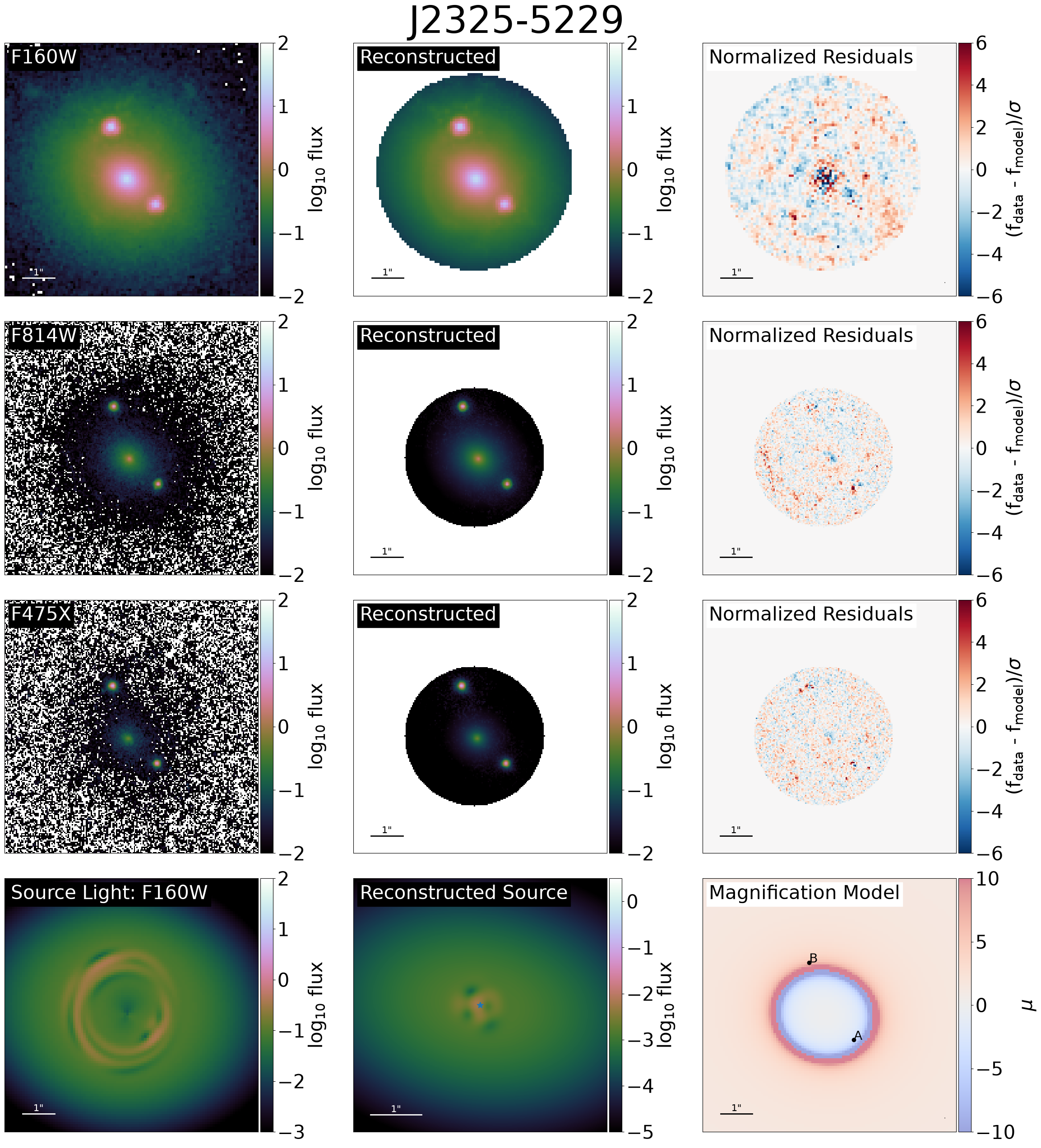}
    
    \caption{\lenstronomy~model visualizations for the systems J1442+4055 (top left), J1515+1511 (top right), and J2325-5229 (lower middle). \textbf{Top Rows.} HST IR science image cutout, followed by the IR best-fit reconstruction, followed by residuals of the fit normalized by the pixel noise level. \textbf{Second Rows.} The same components as the top row, but for the F814W data. \textbf{Third Rows.} The same components as the top row, but for the F475X data. \textbf{Bottom Rows.} Reconstructed source arcs modeled in the image plane for the IR data, followed by the reconstructed source in the source plane. Here, the blue star represents the position of the quasar. The last panel showcases the magnification model displaying spatial variations in magnification.}
    \label{fig:2_models}
\end{figure*}

Figure \ref{fig:4_models} and Figure \ref{fig:2_models} showcase visualizations of the models of the systems not presented in the main text. A few features are worthy of note in Figure \ref{fig:4_models}. J0602-4335 is of particular interest due to its edge-on lens morphology, which is the most visually distinctive in the sample. The two-component S\'ersic model reproduces the elongated light distribution reasonably well across all three bands, though structured residuals persist along the disk axis in the F160W panel, reflecting the inherent difficulty of fitting a smooth parametric profile to a sharply peaked edge-on system. The IR residuals for J1001+5027 show mild structured features, and notably the UVIS fits display noticeable residuals at the quasar image positions that is likely attributable to PSF mismodeling. As noted in Section~\ref{sec:modeling_procedure}, the UVIS PSFs for this system were constructed from only the two quasar images and one isolated star due to the absence of additional stars in the frame, limiting the quality of the PSF reconstruction relative to other systems in the sample and likely driving the structured residuals seen at the point source locations.

\begin{figure*}
    \centering
    \includegraphics[width=0.49\textwidth]{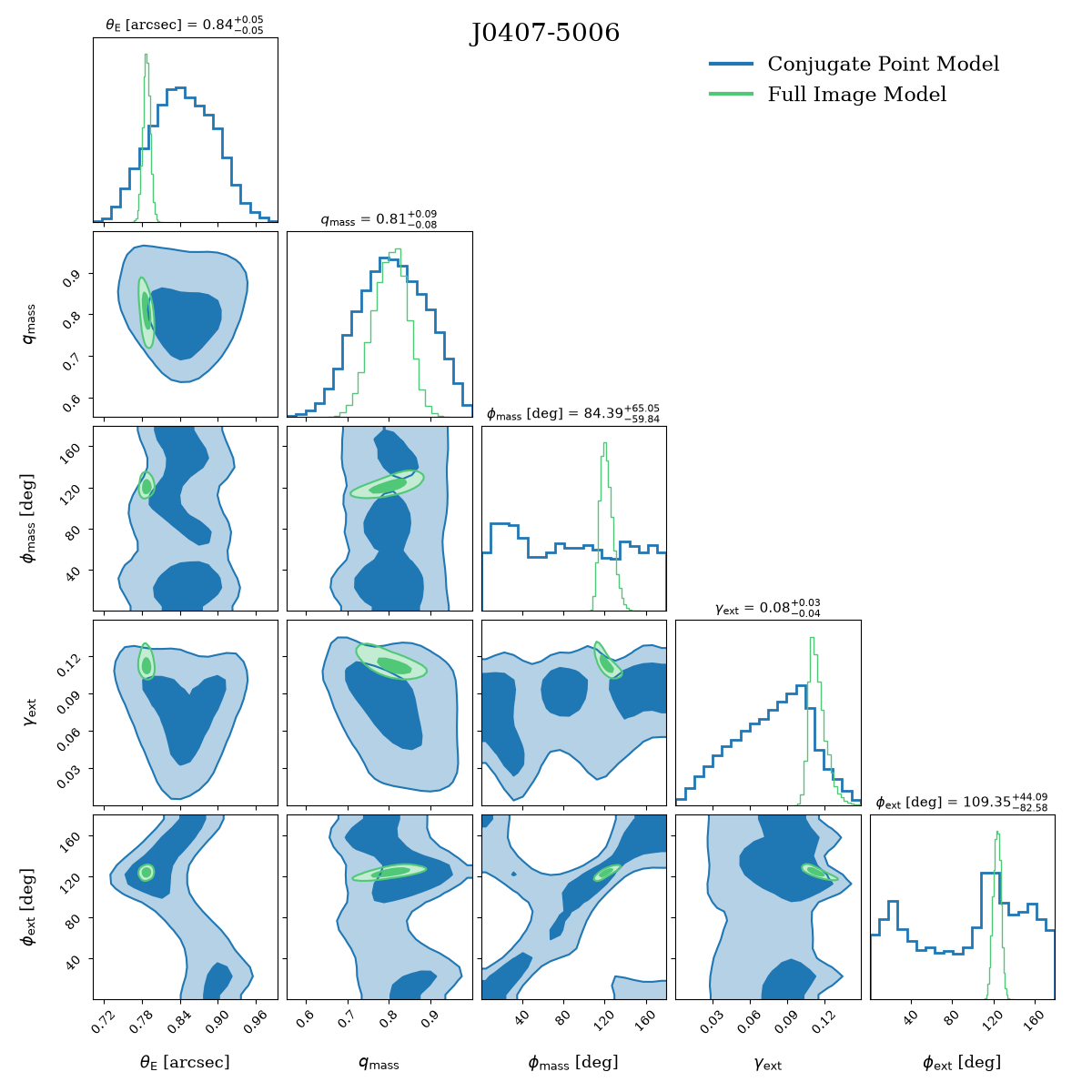}
    \hfill
    \includegraphics[width=0.49\textwidth]{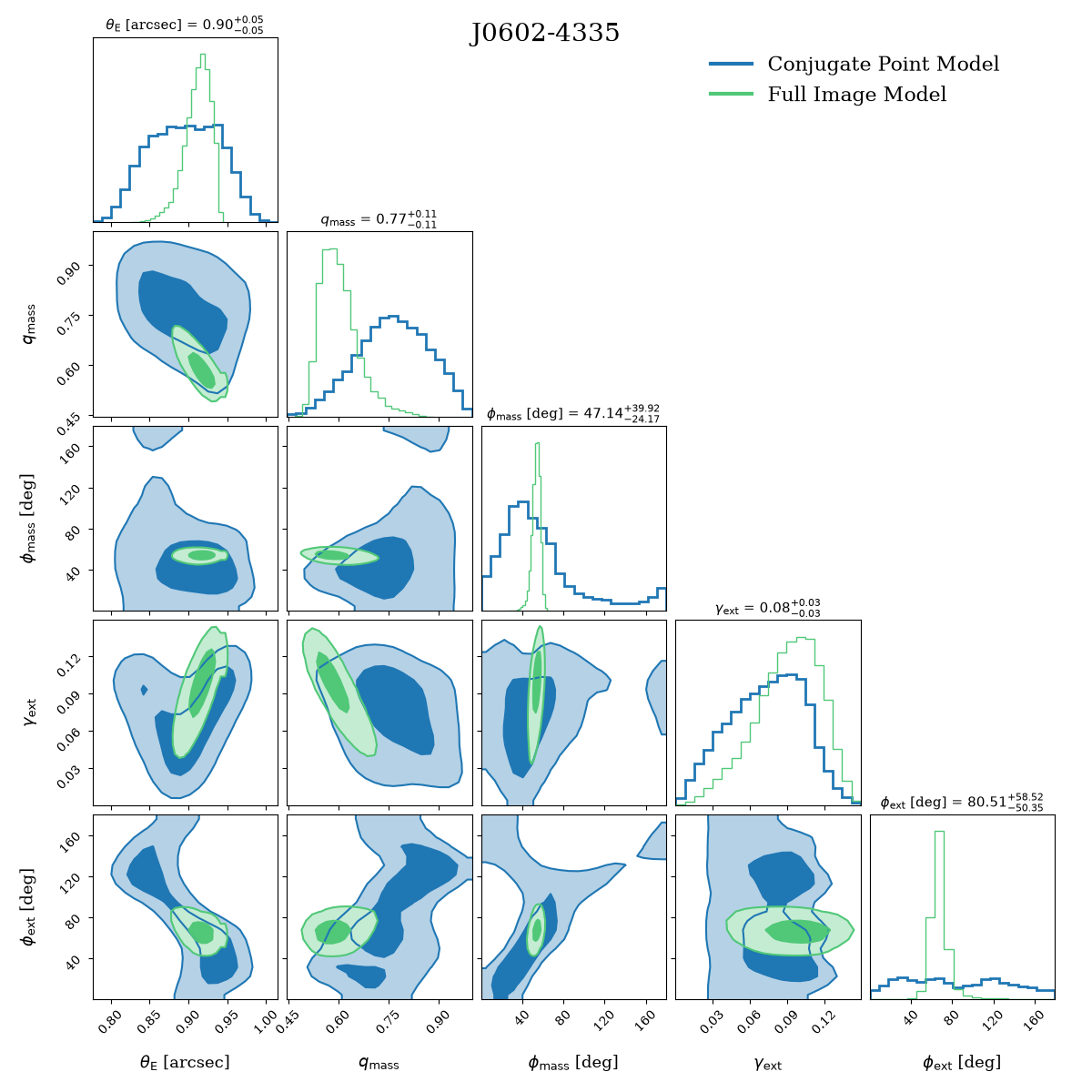}
    \includegraphics[width=0.49\textwidth]{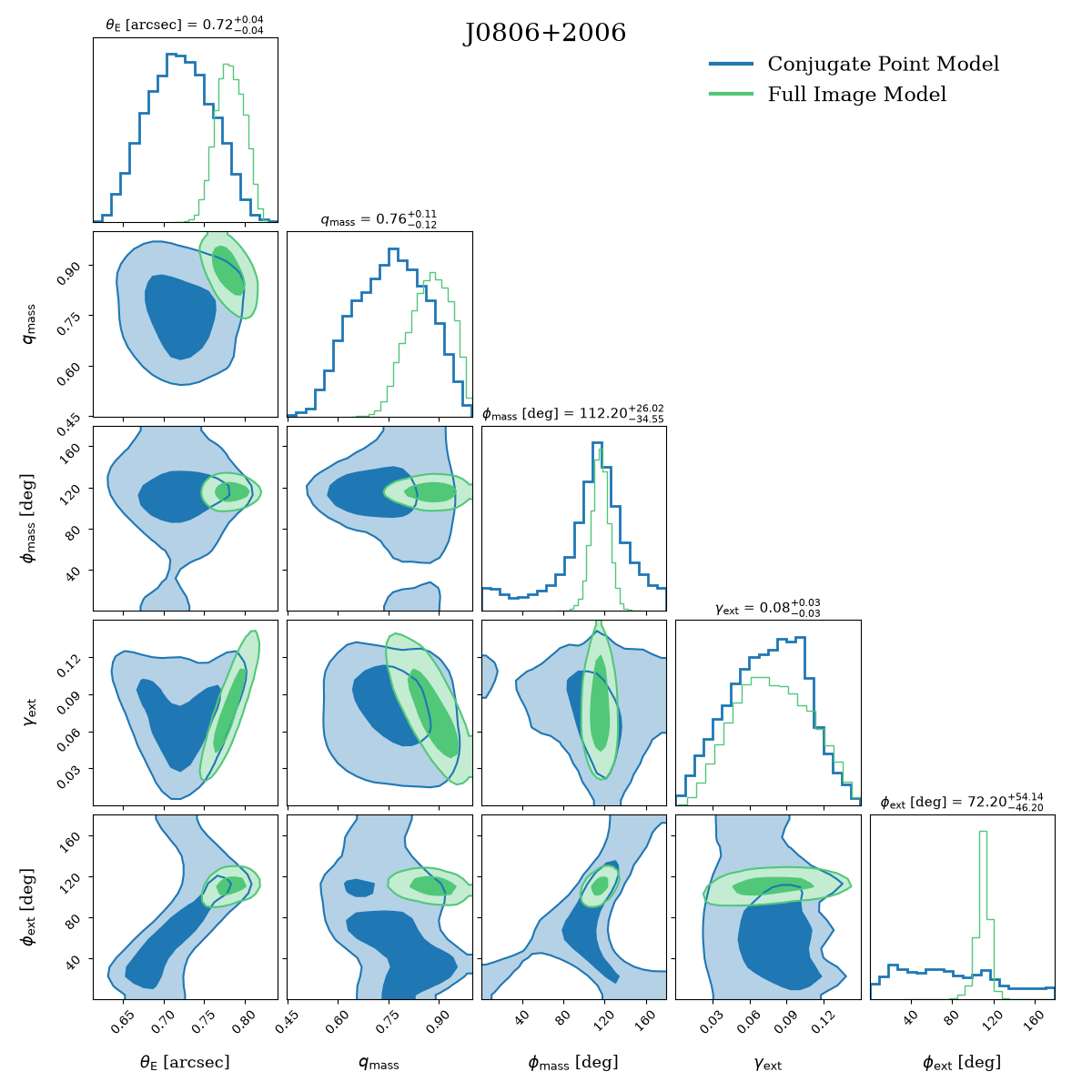}
    \includegraphics[width=0.49\textwidth]{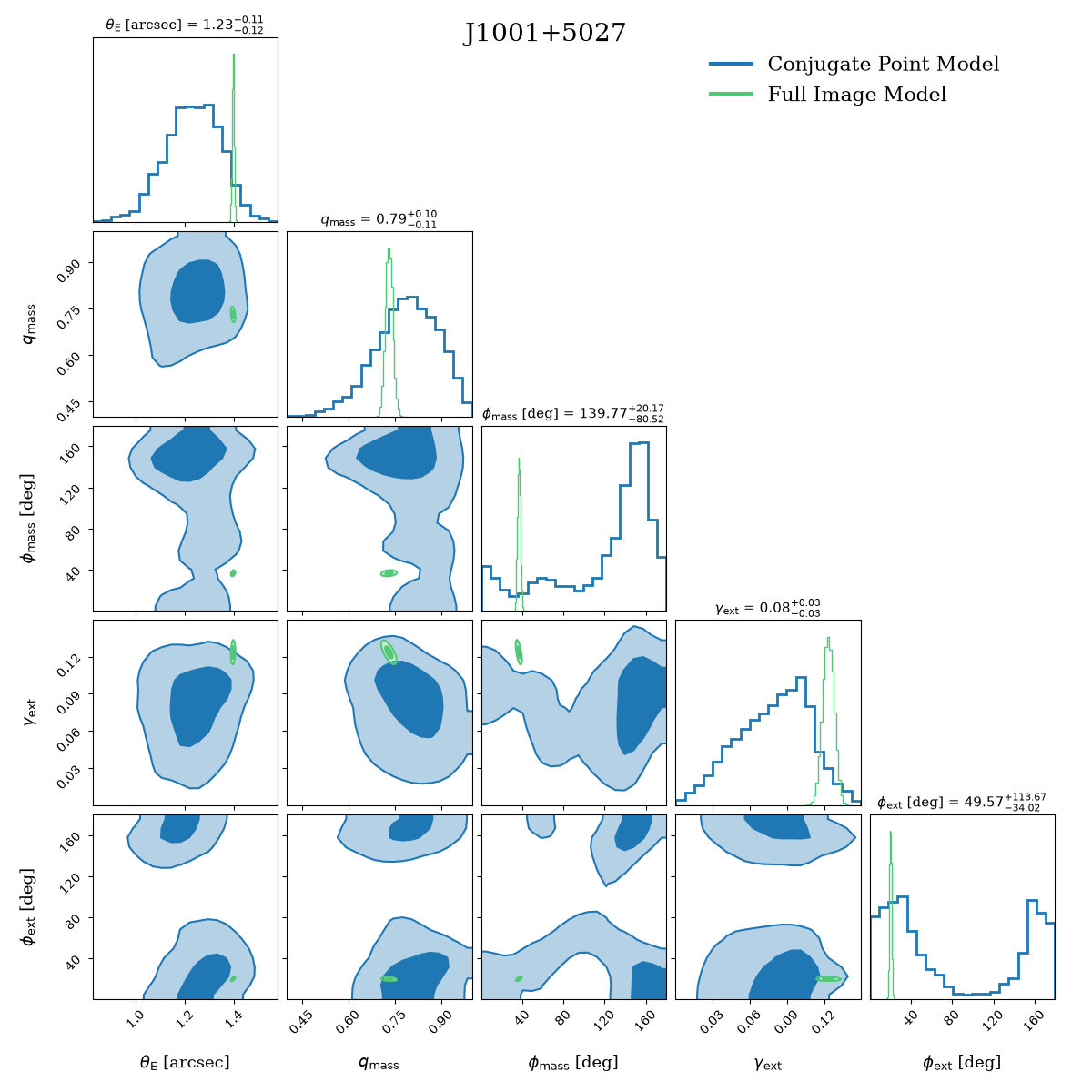}
    \caption{Overlayed posterior probability distributions for the conjugate point model (blue) and the full image model (green) for the systems J0407-5006 (top left), J0602-4335 (top right), J0806+2006 (lower left), and J1001+5027 (lower right). Contours reflect the 1- and 2$\sigma$ credible regions, and the quoted parameter values correspond to the conjugate point posteriors.}
    \label{fig:four_conj_posteriors}
\end{figure*}

\begin{figure*}
    \centering
    
    \includegraphics[width=0.49\textwidth]{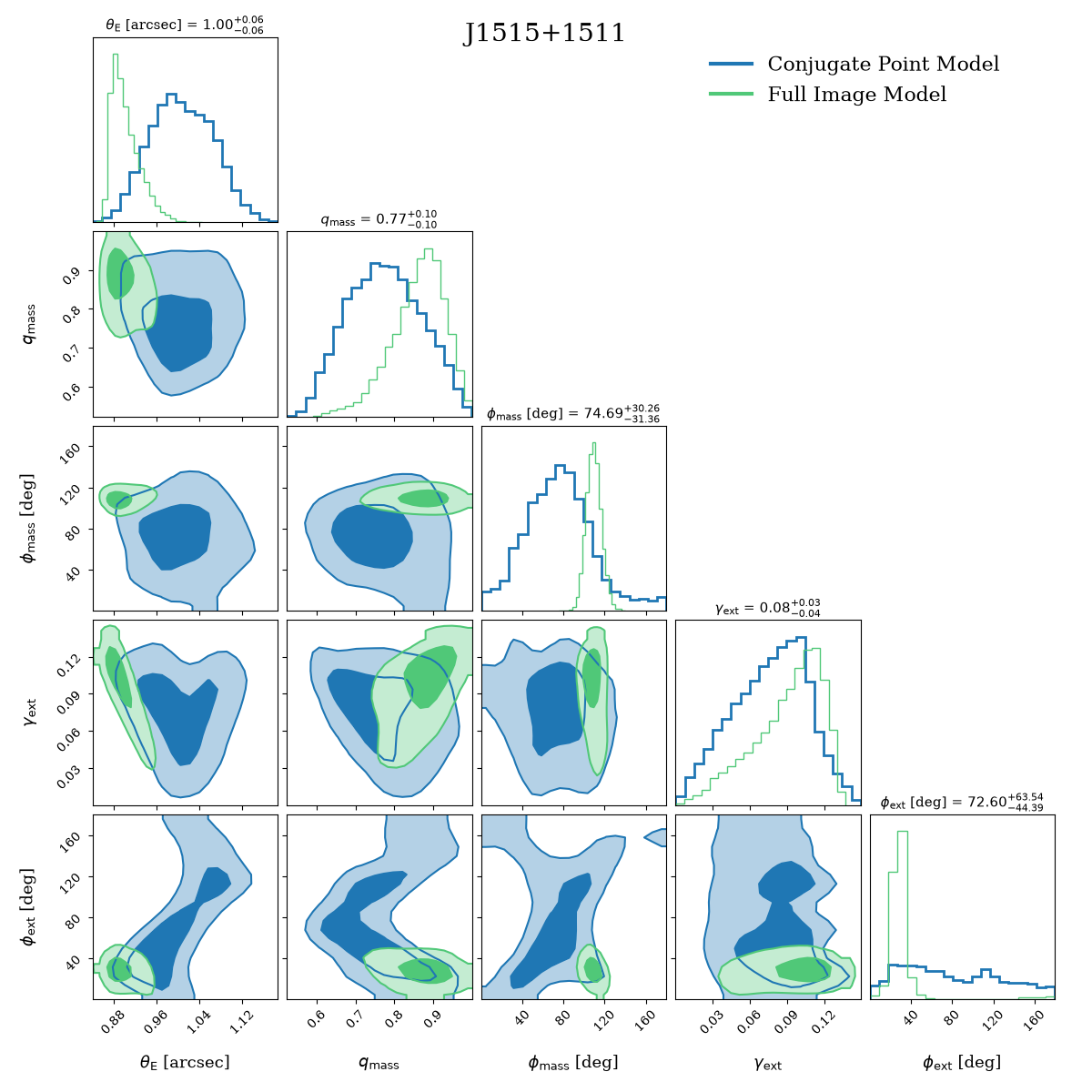}
    \includegraphics[width=0.49\textwidth]{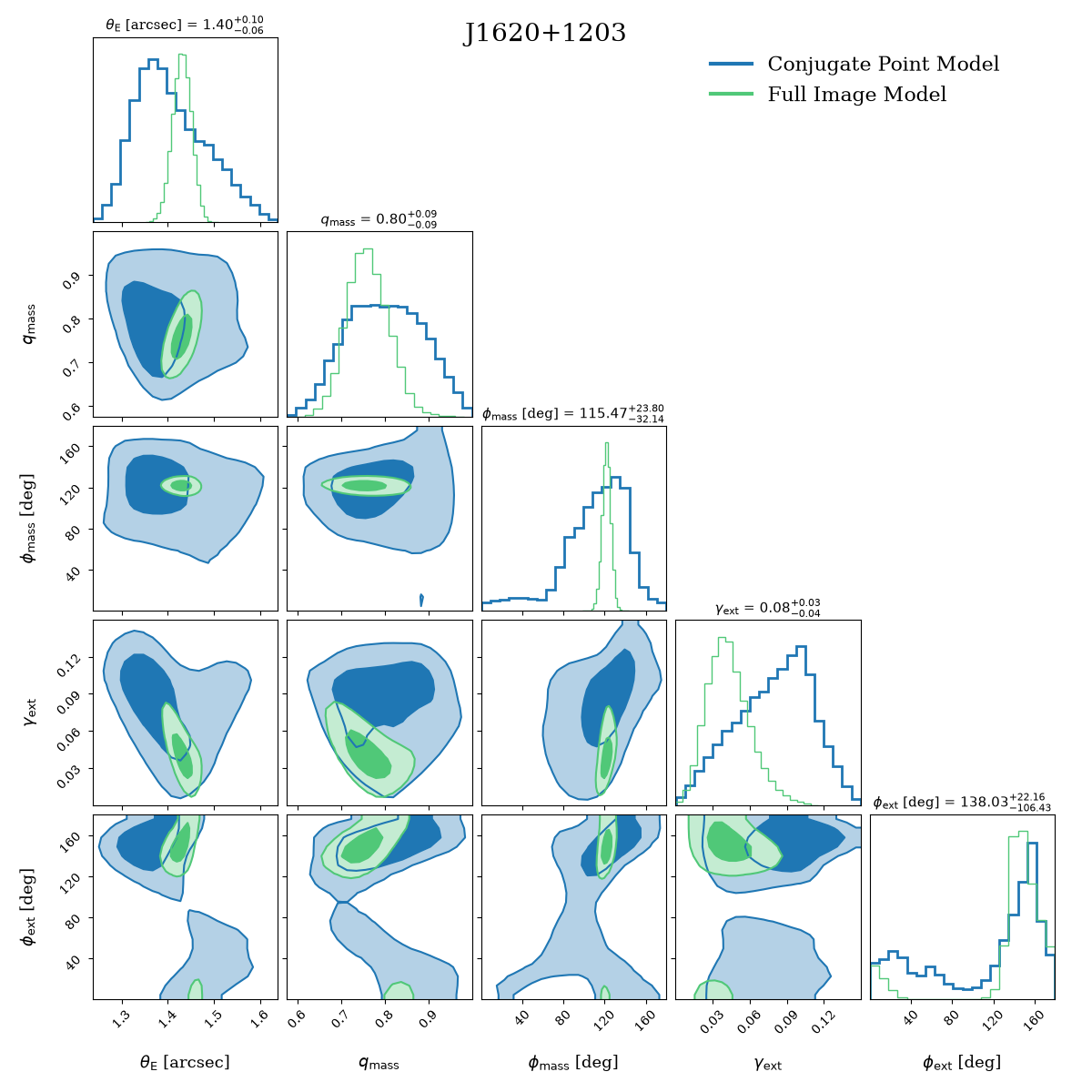}
    \hfill
    \includegraphics[width=0.49\textwidth]{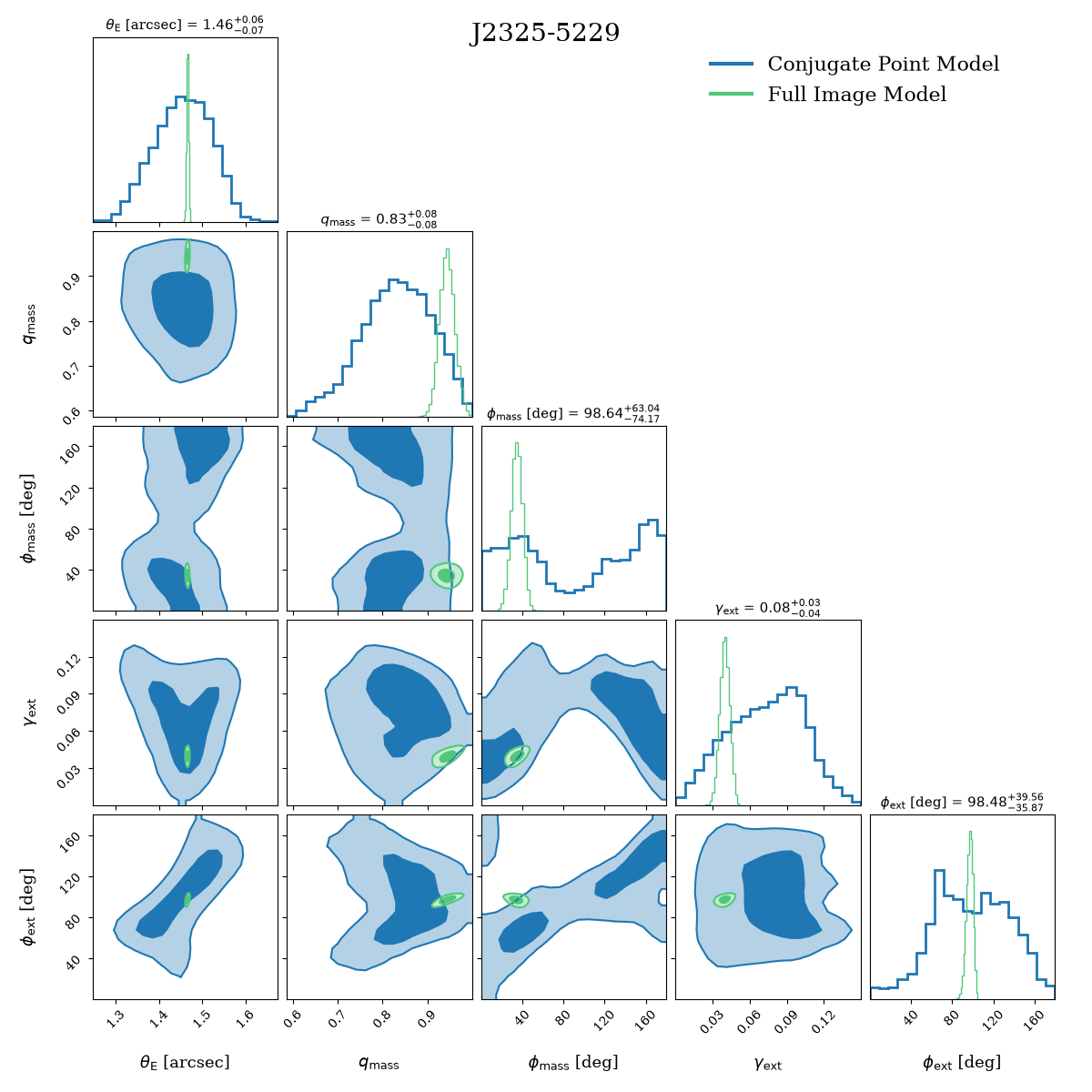}
    \caption{Overlayed posterior probability distributions for the conjugate point model (blue) and the full image model (green) for the systems J1515+1511 (top left), J1620+1203 (top right), and J2325-5229 (bottom). Contours reflect the 1- and 2$\sigma$ credible regions, and the quoted parameter values correspond to the conjugate point posteriors.}
    \label{fig:three_conj_posteriors}
\end{figure*}

Figure \ref{fig:four_conj_posteriors} and Figure \ref{fig:three_conj_posteriors} display the overlayed mass model posteriors for the remaining conjugate point and full image models. The posterior comparisons are consistent with the trends discussed in Section~\ref{sec:conjugate}. In every system, the full image modeling posteriors (green) appear as constrained regions fully within the much broader conjugate point distributions (blue), confirming that the arc surface brightness information systematically tightens the mass parameter constraints. The degree of tightening varies across the sample in a manner that tracks arc brightness. J1001+5027 and J2325-5229, which host the two brightest arcs in the sample, show the most dramatic compression of the full image posteriors relative to the conjugate point contours, while J0806+2006 and J1515+1511, with fainter arcs, show comparatively less compression and broader residual degeneracies between $q_\text{mass}$ and $\gamma_\text{ext}$ in the full image posteriors as well. The mass-shear degeneracy is visible in the conjugate point contours of every system, manifesting as elongated joint distributions, and it is precisely this degeneracy that the extended arc morphology helps to constrain. 

\begin{table}[t]
\centering
\caption{Best-fit source light model parameters. The associated uncertainties are statistical and were calculated using the 84th and 16th percentiles. We adopted a fiducial flat cosmology of $\Hzero = 70 ~\rm km~s^{-1}~Mpc^{-1}$ and $\Omega_\text{m} = 0.3$ to calculate the angular diameter distance in the conversion of $R_\text{eff}$ from angular size to physical size. The position angle $\phi_\text{source}$ is measured in degrees north of East. Refer to Section \ref{sec:hst_sampls} for a discussion on system discoveries and redshift measurements.}
\label{tab:source_light_params_f814w}
\begin{tabular}{lcccccc}
\hline
Lens System & $z_\text{source}$ & $R_{\text{eff}}$ [arcsec] & $R_{\text{eff}}$ [kpc] & $n_{\text{S\'ersic}}$ & $q_\text{source}$ & $\phi_\text{source}$ [deg] \\
\hline
J0407-5006 & 1.515 & $0.330^{+0.008}_{-0.008}$ & $2.80^{+0.07}_{-0.07}$ & $0.63^{+0.06}_{-0.05}$ & $0.91^{+0.03}_{-0.03}$ & $13.1^{+163.0}_{-9.7}$ \\
J0602-4335 & 2.920 & $0.115^{+0.006}_{-0.006}$ & $0.89^{+0.04}_{-0.05}$ & $0.53^{+0.04}_{-0.02}$ & $0.65^{+0.01}_{-0.03}$ & $55.4^{+5.5}_{-6.4}$ \\
J0806+2006 & 1.540 & $0.083^{+0.005}_{-0.002}$ & $0.70^{+0.04}_{-0.02}$ & $3.71^{+0.22}_{-0.46}$ & $0.57^{+0.07}_{-0.07}$ & $156.7^{+12.3}_{-9.4}$ \\
J1001+5027 & 1.841 & $0.398^{+0.003}_{-0.003}$ & $3.35^{+0.02}_{-0.02}$ & $0.56^{+0.02}_{-0.02}$ & $0.75^{+0.01}_{-0.01}$ & $121.1^{+1.3}_{-1.4}$ \\
J1442+4055 & 2.575 & $0.436^{+0.039}_{-0.020}$ & $3.50^{+0.31}_{-0.16}$ & $0.79^{+0.15}_{-0.08}$ & $0.73^{+0.03}_{-0.03}$ & $126.7^{+3.9}_{-3.7}$ \\
J1515+1511 & 2.049 & $0.394^{+0.019}_{-0.018}$ & $3.29^{+0.16}_{-0.15}$ & $0.78^{+0.17}_{-0.12}$ & $0.47^{+0.06}_{-0.04}$ & $140.0^{+3.7}_{-2.6}$ \\
J1620+1203 & 1.158 & $0.315^{+0.016}_{-0.016}$ & $2.60^{+0.13}_{-0.13}$ & $2.03^{+0.30}_{-0.23}$ & $0.85^{+0.06}_{-0.06}$ & $23.6^{+10.7}_{-11.0}$ \\
J2325-5229 & 2.739 & $1.300^{+0.094}_{-0.073}$ & $10.27^{+0.74}_{-0.58}$ & $0.51^{+0.02}_{-0.01}$ & $0.67^{+0.01}_{-0.01}$ & $7.8^{+1.5}_{-1.6}$ \\
\hline
\end{tabular}
\end{table}

The reconstructed source light parameters are presented in Table~\ref{tab:source_light_params_f814w}. Six of eight systems exhibit low S\'ersic indices, which align with disk-like morphologies commonly observed for lensed quasar hosts \citep{gabor:2009}. However, previous lens modeling studies have shown that S\'ersic indices can be biased toward lower values due to regularization and limited signal-to-noise, which tend to smooth centrally concentrated light profiles \citep{ding:2017, ding:2017b}. In addition, the effective radius has been shown to be correlated with the S\'ersic index, introducing further systematic uncertainty in $R_{\rm eff}$ \citep{ding:2021}. Therefore, the low S\'ersic index values and corresponding range of effective radii reported here may partially reflect modeling systematics.

The source light parameters recovered from our lens models can be compared to previous measurements of quasar host galaxy structural properties. \citet{silervamn:2019} studied 32 broad-line AGN hosts at $1.2 < \rm z < 1.7$ using HST/WFC3 imaging, finding effective radii generally between 1 and 6 kpc, with S\'ersic indices spanning the full allowed range of $0.5 < n_\text{S\'ersic} < 6$, with a mean value of $n_\text{S\'ersic} = 2$. The low-$n_\text{S\'ersic}$ majority of our sample thus aligns with the disk-dominated character of AGN hosts at comparable redshifts. In order to convert our modeled angular source size to physical size, we adopted a fiducial flat cosmology with $\Hzero = 70 ~\rm km~s^{-1}~Mpc^{-1}$ and $\Omega_\text{m} = 0.3$. Five of eight systems fall between $2~ \text{kpc} < R_\text{eff} < 4 ~\text{kpc}$, consistent with the expectations reported by \citet{silervamn:2019}. J0602-4335 and J0806+2006 are noticeably more compact, with $R_\text{eff} < 1~\text{kpc}$. The inferred half-light radius of J2325-5229 ($R_\text{eff} = 10.27^{+0.74}_{-0.58} ~\text{kpc}$) is anomalously large relative to the rest of our sample, but is aligned with the regime occupied by radio loud QSO host galaxies, which \citet{wolf:2008} find to have a mean effective radius of 11.4 kpc based on HST and ground-based adaptive optics imaging of 10 quasars. Overall, our inferred source parameters are qualitatively in accordance with the significant size dispersion at fixed stellar mass documented by \citet{silervamn:2019}, who interpret this scatter as evidence that AGN hosts occupy an intermediate structural state between star forming and quiescent galaxies.

\section{\textbf{C. Additional Observational Information}}
\label{append::c}

We summarize here the observational properties of the HST GO-17199 dataset used in this work. Table~\ref{tab:observations} provides the filter, instrument, exposure time, and observation date for each lens system in our sample.

\begin{table*}[b]
\centering
\caption{HST GO-17199 (PI: Lemon) WFC3 observation details for each system presented in this study. At each dither position, both short and long exposures were taken to capture the large dynamic range between the bright AGN and its much fainter host galaxy. The Exposure Time therefore reflects the temporal combination of individual short and long exposure times.}
\begin{tabular}{lcccc}
\hline
System & Filter & Instrument & Exposure Time (s) & Observation Date \\
\hline
J0407-5006 & F160W & WFC3/IR & 4393.9 & 2024-02-28 \\
 & F814W & WFC3/UVIS & 1428.0 & 2024-02-28 \\
 & F475X & WFC3/UVIS & 988.0 & 2024-02-28 \\
\hline
J0602-4335 & F160W & WFC3/IR & 4393.9 & 2023-08-01 \\
 & F814W & WFC3/UVIS & 1428.0 & 2023-08-01 \\
 & F475X & WFC3/UVIS & 793.0 & 2023-08-01 \\
\hline
J0806+2006 & F160W & WFC3/IR & 4393.9 & 2023-04-04 \\
 & F814W & WFC3/UVIS & 1428.0 & 2023-05-16 \\
 & F475X & WFC3/UVIS & 676.0 & 2023-04-04 \\
\hline
J1001+5027 & F160W & WFC3/IR & 1597.7 & 2024-12-20 \\
 & F814W & WFC3/UVIS & 1428.0 & 2023-12-08 \\
 & F475X & WFC3/UVIS & 1397.0 & 2023-12-08 \\
\hline
J1442+4055 & F160W & WFC3/IR & 4393.9 & 2023-07-05 \\
 & F814W & WFC3/UVIS & 1428.0 & 2023-11-14 \\
 & F475X & WFC3/UVIS & 771.0 & 2023-07-05 \\
\hline
J1515+1511 & F160W & WFC3/IR & 4393.9 & 2023-06-28 \\
 & F814W & WFC3/UVIS & 1428.0 & 2023-08-23 \\
 & F475X & WFC3/UVIS & 674.0 & 2023-06-28 \\
\hline
J1620+1203 & F160W & WFC3/IR & 4393.9 & 2023-06-27 \\
 & F814W & WFC3/UVIS & 1428.0 & 2023-06-27 \\
 & F475X & WFC3/UVIS & 615.0 & 2023-06-27 \\
\hline
J2325-5229 & F160W & WFC3/IR & 4393.9 & 2023-08-19 \\
 & F814W & WFC3/UVIS & 1428.0 & 2023-08-19 \\
 & F475X & WFC3/UVIS & 988.0 & 2023-08-19 \\
\hline
\end{tabular}
\label{tab:observations}
\end{table*}

\vspace{1in}

\end{document}